\PassOptionsToPackage{colorlinks=true,urlcolor=blue,linkcolor=black,citecolor=blue}{hyperref} 

\makeatletter
\newcommand{\multihat}[1]{%
  \begingroup
  \@tfor\next\equiv#1\do{\hat{\next}}%
  \endgroup
}
\makeatother

\PassOptionsToPackage{table, dvipsnames}{xcolor} 

\documentclass[submission, Phys]{SciPost}
\usepackage{amsmath,amssymb,mathtools}

\usepackage{bm} 
\usepackage{soul} 
\usepackage{braket}
\usepackage{amsmath}
\usepackage{dsfont} 
\usepackage[table]{xcolor}
\usepackage{subcaption}

\usepackage[utf8]{inputenc}
\usepackage[english]{babel}
\usepackage{physics}
\usepackage{epsfig}
\usepackage{epstopdf}
\usepackage{epsf}
\usepackage{graphicx,amsmath,amssymb}
\usepackage{epsfig,multicol}
\usepackage{epsfig}
\usepackage{amssymb}
\usepackage{mathtools}
\usepackage{framed}
\usepackage[none]{hyphenat}
\usepackage{subcaption}
\usepackage{mathrsfs} 
\usepackage{array}
\usepackage[makeroom]{cancel}
\usepackage{comment}
\usepackage{bm}
\usepackage{bbm}

\usepackage{booktabs}
\newcommand{\thickline}{\specialrule{1pt}{0pt}{0pt}}
\newcommand{\thinline}{\specialrule{0.001pt}{0pt}{0pt}} 

\usepackage{longtable} 

\definecolor{ClassI}{HTML}{FF3333} 
\definecolor{ClassIIa}{HTML}{3344FF}
\definecolor{ClassIIb}{HTML}{3344FF}
\definecolor{ClassIIc}{HTML}{3344FF}
\definecolor{ClassIIIa}{HTML}{50B254}
\definecolor{ClassIIIb}{HTML}{50B254} 
\definecolor{ClassIV}{HTML}{F8FF33} 
\definecolor{DU}{HTML}{000000} 
\definecolor{?}{HTML}{000000}

\newcommand{\defeq}{\overset{\text{def}}{=}} 

\graphicspath{{Figures/}}

\def\CC{\mathcal{C}}

\def\CO{\mathcal{O}}
\def\CP{\mathcal{P}}

\def\CT{\mathcal{T}}
\def\CV{\mathcal{V}}

\def\bea{\begin{eqnarray}}
\def\eea{\end{eqnarray}}

\usepackage{comment}
\begin{document}
\hyphenpenalty=1000 

\begin{center}{\Large \textbf{
Ergodic behaviors in reversible 3-state cellular automata
}}\end{center}

\begin{center}
Rustem Sharipov\textsuperscript{ \raisebox{0.6ex}{\fcolorbox{black}{red}{\rule{0pt}{0.1pt}\rule{0.1pt}{0pt}}}}${}^\star$,
Matija Koterle\textsuperscript{ \raisebox{0.6ex}{\fcolorbox{black}{red}{\rule{0pt}{0.1pt}\rule{0.1pt}{0pt}}} },
Sašo Grozdanov\textsuperscript{ \raisebox{0.6ex}{\fcolorbox{black}{red}{\rule{0pt}{0.1pt}\rule{0.1pt}{0pt}}}\raisebox{0.6ex}{\fcolorbox{black}{white}{\rule{0pt}{0.1pt}\rule{0.1pt}{0pt}}}} and
Tomaž Prosen\textsuperscript{ \raisebox{0.6ex}{\fcolorbox{black}{red}{\rule{0pt}{0.1pt}\rule{0.1pt}{0pt}}}\raisebox{0.6ex}{\fcolorbox{black}{blue}{\rule{0pt}{0.1pt}\rule{0.1pt}{0pt}}}}
\end{center}

\begin{center}
\raisebox{0.6ex}{\fcolorbox{black}{red}{\rule{0pt}{0.1pt}\rule{0.1pt}{0pt}}} \textit{Department of Physics, Faculty of Mathematics and Physics,}

\textit{University of Ljubljana, Jadranska 19, SI-1000 Ljubljana, Slovenia}
\\
\raisebox{0.6ex}{\fcolorbox{black}{blue}{\rule{0pt}{0.1pt}\rule{0.1pt}{0pt}}} \textit{Institute for Mathematics, Physics, and Mechanics,} 

\textit{Jadranska 19, SI-1000 Ljubljana, Slovenia}
\\
\raisebox{0.6ex}{\fcolorbox{black}{white}{\rule{0pt}{0.1pt}\rule{0.1pt}{0pt}}} \textit{Higgs Centre for Theoretical Physics, University of Edinburgh,}

\textit{Edinburgh, EH8 9YL, Scotland}
\\
${}^\star$ {\small \sf rustem.sharipov@fmf.uni-lj.si}
\end{center}

\begin{center}
\today
\end{center}


\section*{Abstract}
{\bf
Classical cellular automata represent a class of explicit discrete spacetime lattice models in which complex large-scale phenomena emerge from simple deterministic rules. With the goal to uncover different physically distinct classes of ergodic behavior, we perform a systematic study of three-state cellular automata (with a stable `vacuum' state and `particles' with $\pm$ charges). The classification is aided by the automata's different transformation properties under discrete symmetries: charge conjugation, spatial parity and time reversal. In particular, we propose a simple classification that distinguishes between types and levels of ergodic behavior in such system as quantified by the following observables:
the mean return time, the number of conserved quantities, and the scaling of correlation functions. In each of the physically distinct classes, we present examples and discuss some of their phenomenology. This includes chaotic or ergodic dynamics, phase-space fragmentation, Ruelle-Pollicott resonances, existence of quasilocal charges, and anomalous transport with a variety of dynamical exponents.
}
\vspace{50pt}
\begin{figure}[h!]
\centering
  \includegraphics[width=1.0\linewidth]{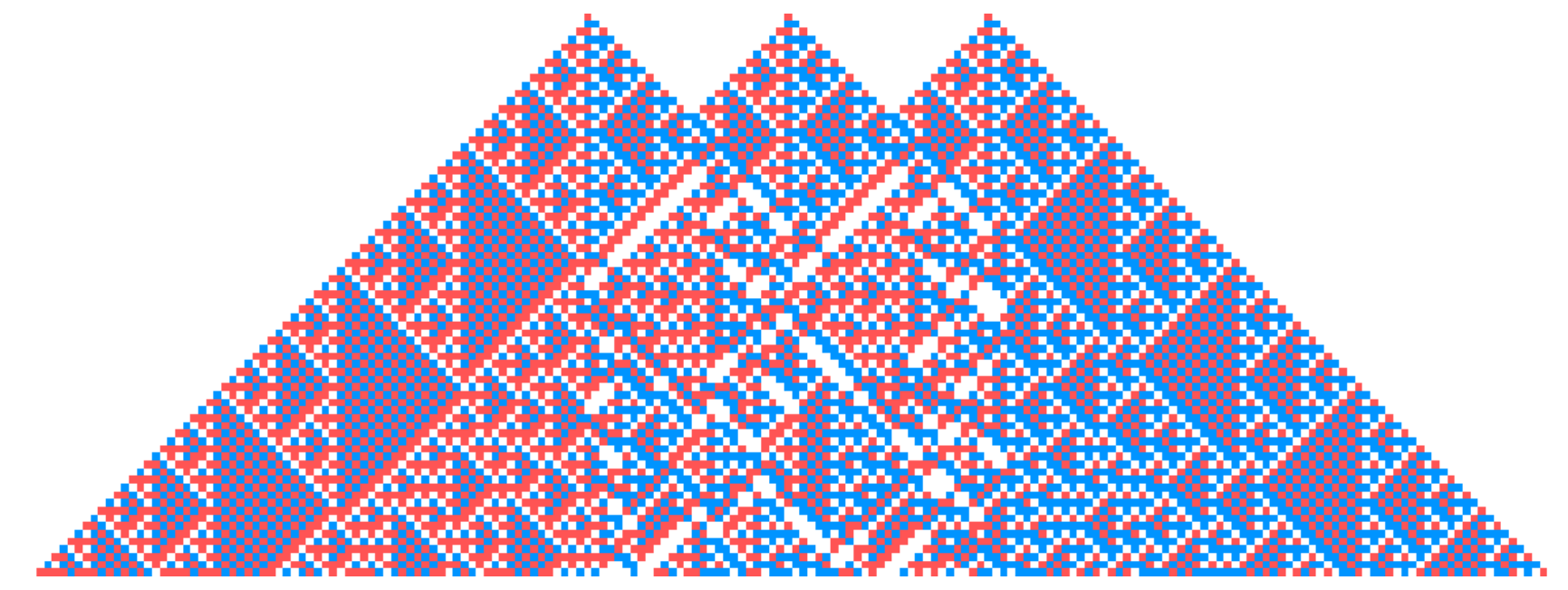}
\end{figure}

\newpage
\vspace{30pt}
\noindent\rule{\textwidth}{1pt}
\setcounter{tocdepth}{1}
\tableofcontents\thispagestyle{fancy}
\noindent\rule{\textwidth}{1pt}
\vspace{10pt}

\section{Introduction}

Finding minimal microscopic models to describe the emergence of complex phenomena is an overarching goal of theoretical physics. Moreover, one may aim at finding all possible emerging dynamical phenomena that can be derived from a reasonable set of local laws of motion that are constrained by physical principles, such as symmetries. 
While identification of minimal models with universality classes of statistical behavior seems well established in the context of equilibrium physics or imaginary time dynamics, the field has only begun developing an analogous understanding of the real-time dynamics of interacting many-particle systems. 

In the general context of interacting systems with few degrees of freedom described by the Hamiltonian formalism, the mathematical apparatus of ergodic theory provides a rather clear picture of a plethora of possible dynamical behaviors, ranging from (Liouville) integrable dynamics, to ergodic, mixing and chaotic dynamics, as quantified by precise notions, such as dynamical correlation functions, dynamical (Kolmogorov) entropies and Lyapunov exponents. The gray area between integrable, ergodic and chaotic dynamics is covered by the so-called Kolmogorov-Arnold-Moser (KAM) theorem, which provides at least a qualitatively satisfactory picture. 

The situation becomes increasingly richer and more complicated when one studies systems with many interacting degrees of freedom, even if the interaction only acts locally (say between nearest neighbors or spatially closest degrees of freedom). 
There is no known classification of all possible dynamical behaviors in either the thermodynamic (IR) or continuum (field theory, UV) limits. There are no general results on the stability of non-ergodic dynamics (say in the vicinity of integrable regimes) under IR and UV limits. An important historical playground has been the Fermi-Pasta-Ulam-Tsingou (FPUT) model \cite{FPUT} which provided key motivation for developing the theories of Hamiltonian chaos and the theory of integrable nonlinear PDEs (also called the soliton theory). Nevertheless, understanding stability thresholds separating ergodic and non-ergodic behavior in FPUT and related models has been a daunting task, despite decades of research~\cite{bermanFPUT,gallavottiFPUT}.
Recent related studies \cite{FlachPRL2019,FlachPRL2022} provided first hints on underlying principles that explain the diversity of stability behaviors.

The state of affairs becomes even more difficult in the quantum domain. Whereas a comprehensive phenomenological study of dynamics of generic interacting quantum many-body systems is out of reach due to exponential proliferation of Hilbert space (at least without a clear utility that is promised by quantum simulators), one may consider classical many-particle systems with discrete state space which in many ways mimic quantum lattice systems~\cite{tHooft}. These are the reversible cellular automata (RCA), where reversibility of local update rule on a discrete state space is an analogue of unitary dynamics over a finite local Hilbert space. We note that a more general class of (non-reversible) cellular automata~\cite{wolfram1983statistical,li1990structure} is known to be able to perform a universal computation~\cite{cook2004universality}, however we here insist on reversible rules as only they are the analogue of the notion of conservative (Hamiltonian) or quantum dynamics.
Even though RCA can be efficiently simulated by sampling over classical trajectories, its large space-time correlation functions are in many ways similar to those of quantum lattice systems. In other words, hydrodynamic descriptions of (most of) quantum many-body lattice systems can be phrased in terms of a classical field theory\footnote{For discussions of classical hydrodynamics, see e.g.~Refs.~\cite{Kovtun:2012rj,Liu:2018kfw,Grozdanov:2019kge}.}, so, for starters, it may be enough to understand all possible emergent behaviors of a sufficiently rich class of RCA.

The first attempt in this direction has been a classification of 2-state (binary) RCA with `lightcone' or `relativistic' local rules by Takesue~\cite{takesue} and Bobenko et al.~\cite{bobenko}. Remarkably, among $2^8=256$ rules one can identify a few exactly solvable rules which describe generic physical phenomena, such as diffusion, convection, etc. Examples are the so-called rules 54, 150, and 201~\cite{RCA54review} (labeling the automata according to the binary code of the rule~\cite{bobenko}). However, it has soon become clear that the set of effective dynamics of RCA can
be much richer than those described by binary local rules (see e.g.,~\cite{Klobas_2022,Gambor}). 
In this paper, we fill this observational gap by systematically exploring a set of one-dimensional 3-state RCA with all possible local rules (physical interactions)
respecting certain physical symmetries.
Specifically, we will focus on the so-called brickwork RCA, sometimes referred to as block cellular automata, where at each instant of time, a reversible mapping is applied to a block (brick) of two neighboring degrees of freedom
 $U: \mathbb{Z}_k \times \mathbb{Z}_k \rightarrow \mathbb{Z}_k \times \mathbb{Z}_k$, while in the following time instant, the reference point is shifted by one site
 (so after two time steps the reference point is again back at the original).
 The number of all reversible block cellular automata is equal to the number of reversible local maps, i.e., $(k^2)!$. In the binary case, $k=2$, it turns out that dynamics generated by all such rules is rather uninteresting, with the least trivial example being perhaps the so-called deterministic East model~\cite{East}. Block RCA rules with $k>2$ have also been studied in specific cases, for example, see Ref.~\cite{Gambor}, while the entire set of block RCA is already prohibitively large to allow for a systematic study even for $k=3$.
 
 Moreover, in order to interpret RCA evolution in terms of particle dynamics, it is  meaningful to identify a special state, called a vacancy or a vacuum state $\varnothing$, 
 such that the latter is stable under local dynamics 
 $U(\varnothing ,\varnothing)=(\varnothing,\varnothing)$.
 This reduces the number of all reversible vacuum-preserving block RCA to $(k^2-1)!$.
 
 In the case $k=3$, which we systematically study in this work, this amounts to a total of $8!=40320$ rules.
 Identifying the remaining two states as particles with charge $\pm 1$, we shall
 focus on particular sets of symmetric block RCA sharing time-reversal, spatial-reversal (reflection) and charge-conjugation symmetries, or various combinations thereof.
 This then amounts to studying only a few hundred distinct symmetric RCA which we classify into 4 groups (classes) based on their fundamental dynamical (ergodic) and complexity behavior. Due to a fundamental difficulty with a proper definition of Lyapunov exponents and dynamical entropies for systems with discrete degrees of freedom, we determine the ergodicity class by a combination of the following empirical indicators: (i) the volume scaling of (average/typical) return time (starting from a random initial condition), (ii) temporal decay of spatio-temporal correlation functions of local observables (or densities of conserved charges), and (iii) leading eigenmodes of a 
 dynamical transfer matrix of local (translationally invariant) observables. The latter correspond to either local (or quasilocal) conserved quantities or Ruelle-Pollicott resonances (chaotic modes), in gapless or gapped cases, respectively.
 We stress that this may be the first clear observation of the existence of Ruelle-Pollicott resonances in discrete state space dynamical systems and can serve as a definition of deterministic chaos.

As a result of our empirical study, we find an amazingly rich set of dynamical behaviors in this set of RCA, which contain simple examples (perhaps minimal models) of various types of ergodic and chaotic behavior. Many of these models warrant further analytical studies  whereby one may hope to uncover new underlying algebraic structures and formalize ergodic theory of reversible discrete state space dynamical systems.

The paper is organized as follows. In Section~\ref{sec:Summary}, we summarize our main findings together with the proposed classification. In Section~\ref{sec:ThreeColorCA}, we then precisely define our class of models,
their symmetries and other elementary features.
In Section~\ref{sec:dynobs}, we define the main dynamical observables used in our phenomenological classification, illustrated with a few examples of RCA models. In the remaining sections we then discuss particular dynamical features of each of the four classes and conclude. The appendices contain, among several other useful technical details, also the complete table of block RCA that have been studied with a list of all dynamical features.
 
{\em We note that while this manuscript was being finalized, a preprint~\cite{kim2025circuitssimpleplatformemergence} appeared in which certain findings overlap with the results of this paper.}

\section{Summary of results}\label{sec:Summary}

The goal of this work is to propose a classification of different types of ergodic behavior in classical, time-reversible three-state cellular automata in one spatial dimension. We introduce this family of models by considering a periodic lattice of even length $L$. The microscopic state of the model is
given by a configuration $s=(s_1,s_2,s_3,\ldots,s_L)$, where each $s_j$ takes a value from the local configuration space $X=\{ +,-,\varnothing \}$. The setup can be seen as a toy model for a system of charged particles and anti-particles (with opposite charges $+$ and $-$) and vacuum $\varnothing$. We depict the sites with $s_i = +$ in red and the sites with $s_i = -$ in blue color. Absence of a `particle' with $s_i = \varnothing$ (vacuum) is represented in white. 

\begin{figure}[h!]
    \centering{
        \includegraphics[width=0.9\linewidth]{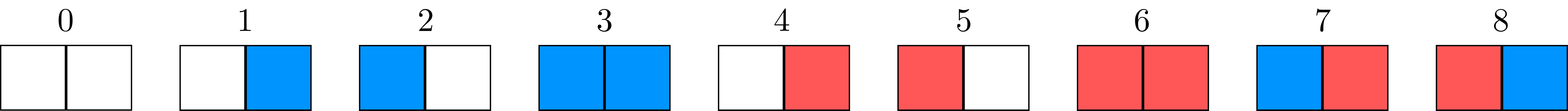}
        \caption{The configuration-labeling scheme used in this work. From left to right, the sequence of integers from $0$ to $8$ designates neighboring sites with all possible configurations from $(\varnothing,\varnothing)$ to $(+,-)$. White, blue and red colors are used to represent $\varnothing$, $-$ and $+$, respectively. 
        }
        \label{fig:rules_definition}
    }
\end{figure}

Time evolution then acts on the configuration $s$ in two consecutive time-steps by updating the sites in a pair-wise manner both times by the same rule. The first (and every odd) step maps $(s_1,s_2) \to (s'_1,s'_2)$, $(s_3,s_4) \to (s'_3,s'_4)$, and so on. Then, the second (and every even) step further maps $(s'_2,s'_3) \to (s''_2, s''_3)$, $(s'_4,s'_5) \to (s''_4, s''_5)$, and due to the periodicity of the lattice, $(s'_L,s'_1) \to (s''_L, s''_1)$. In order to facilitate a shorthand notation that will be used to specify each such rule, we label the $9$ possible pairs of neighboring configurations as in Figure~\ref{fig:rules_definition}.

\begin{figure}[h!]
\centering
  \includegraphics[width=0.6\linewidth]{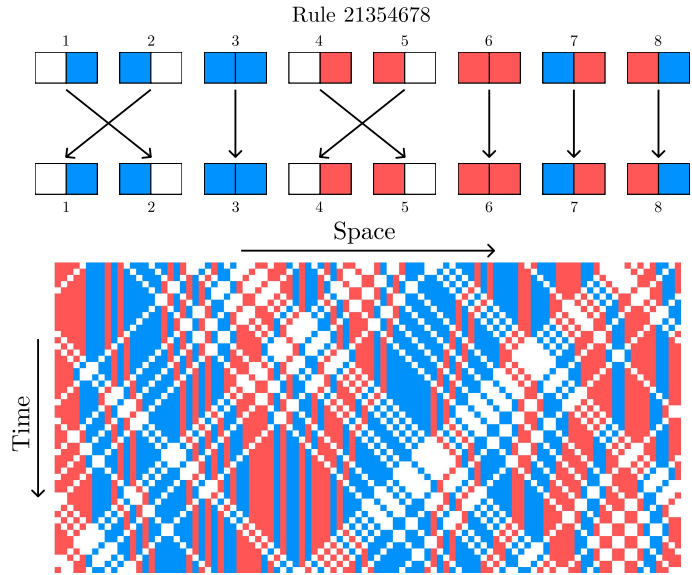}
  \caption{An example of the cellular automaton with rule 21354678 studied previously in \cite{ziga_rule,ziga_rule2,ziga_rule3,ziga_rule4} is used to show the labeling sequence conventions along with an example of its time evolution for dynamics on $L=100$ lattice sites.
  }
  \label{fig:ziga_rule_example_1}
\end{figure}

A rule that defines the evolution of the automaton is then given by a set of $8$ numbers specifying what other configuration each 2-site configuration in Figure~\ref{fig:rules_definition} maps to. We assume that $(\varnothing,\varnothing)$ always remains $(\varnothing,\varnothing)$ and that the mapping rules at even and odd time steps are the same. Then, we use a shorthand notation with the sequence of eight numbers $N_1 N_2\ldots N_8$ that designates what each configuration from Figure~\ref{fig:rules_definition} (with the exception of the trivial 0, which always maps to 0) maps to. That is, $1\to N_1$, $2 \to N_2$, and so on, until $8 \to N_8$. For example, the rule $21354678$ means that under every step of discrete time evolution, $1\to 2$, $2\to 1$, $3\to 3$, $4\to 5$, $5\to 4$, $6\to 6$, $7\to 7$ and $8 \to 8$. We show this example in Figure~\ref{fig:ziga_rule_example_1}. Formal details of the setup and more mathematically precise definitions are given in Section~\ref{sec:ThreeColorCA}.

The classification of ergodic behaviors in this family of cellular automata models is then done by analyzing three different observables: the mean return time computed as a function of the system size, two-point correlation functions and the scaling of the number of (local, translationally invariant) conserved charges with the system size. 

What are the possible types of behavior of these three observables? The mean return time computed as a function of the system size $L$ is found to either grow exponentially $\sim e^{\kappa L}$ or as a power-law $\sim L^p$. For the second measure, we analyze the decay of the `longest-lived' two-point correlation function in the model. In models with conservation laws, such longest-lived correlators are those of densities of conserved quantities, which can be used to diagnose late-time transport. As expected, correlators are found to decay either exponentially $\sim e^{-\alpha t}$ or algebraically $\sim t^{-1/z}$ with time $t$. Finally, we also study the conserved charges which we assume are local and translationally invariant observables with a finite support $r$. Specifically, we study the 
scaling of the number of charges with $r$. Different models exhibit either absence of such conserved charges or, in their presence, the number can be constant, or grow linearly or exponentially.

Due to the large number of different models (there exist $8!=40320$ different rules with the properties defined above), we will simplify our analysis by restricting it to only certain subclasses of models as organized by their different transformation properties under the discrete charge conjugation ($\mathcal{C}$), spatial parity $(\mathcal{P})$, and time-reversal $(\mathcal{T})$ symmetries.

\subsection{Classification}

\begin{figure}[h!]
    \centering{
        \includegraphics[width=1\linewidth]{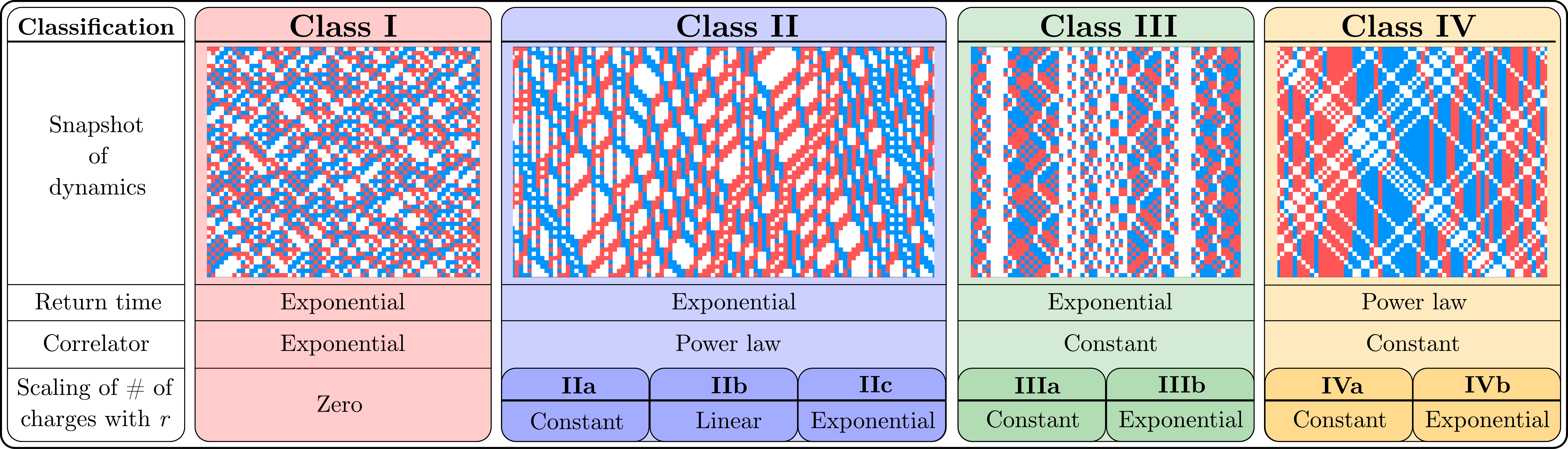}
        \caption{Proposed classification of three-state, reversible cellular automata based on the typical behavior of three observables: the mean return time, decay of the slowest correlation function, and scaling of the number of charges with the support $r$. In each of the classes, we show a representative example of time evolution. From left to right, the examples of rules are 56231487,  43162578, 45321687, 21354678.}
        \label{fig:figure_1}
    }
\end{figure}

We summarize our findings and present the hierarchy of different types of ergodic time-evolution, separated into four general classes, in Figure~\ref{fig:figure_1}.

Generically, we observe that the models fitting into the four above classes exhibit the following pattern of behavior: 

\begin{itemize}
    \item \textbf{Class I}. For models within this class, the average return time scales exponentially, the decay of correlation functions is exponential, and the rules posses no local conserved quantities. The cellular automata in this class are therefore the most chaotic. Further details are discussed in  Section~\ref{sec:class_1}. 

    \item \textbf{Class II:} The models exhibit exponential mean return time, but the correlation function decays algebraically. In this case, we can distinguish three subclasses of ergodic behavior based on the number of local conserved charges that the model possesses. These subclasses has a constant, linearly or exponentially growing number of charges with increasing the maximal support $r$. For details, see Section~\ref{sec:class_2}.

    An interesting feature of several models in this class is that they posses quasilocal charges (in addition or instead of strictly local ones). From the point of view of correlation functions, the models can exhibit diffusive or anomalous (subdiffusive and superdiffusive) transport of conserved quantities. Finally, the three patterns of scaling of the number of conserved charges suggests that the class includes models which are not integrable, integrable and super-integrable.

    \item \textbf{Class III:} In this class, the mean return time scales exponentially whereas the correlation functions decay to a constant value. This is possible with the number of conserved charges either constant or scaling exponentially with maximal support size $r$. The latter two options delineate two subclasses. For details, see Section~\ref{sec:class_3}.

    An interesting feature of these models is that they exhibit domain walls, which are preserved under dynamics. They exist in both subclasses characterized by the scaling of the number of conserved charges, and therefore both in chaotic and (super)integrable models.

    \item \textbf{Class IV:} The last class contains models with a mean return time that grows as a power-law $\sim L^p$ with integer $p \leq 4$. Moreover the models exhibit a constant correlation function, and an exponential growth of the number of charges with maximal support $r$. This class contains some of the simplest models, include those with `free' and `trivial' dynamics. Further details are discussed in Section~\ref{sec:class_4}. 

\end{itemize}

We note that there is a small number of models (in total 35 in all considered symmetry classes) for which it was not possible to clearly establish either one or more of the empirical criteria
in a finite amount of accessible computation time. It is also possible that some
of these models are fundamentally unclassifiable according to our scheme, e.g., their mean return time may be an erratic (or number theoretic) function of the system size. See Appendix~\ref{app:uc} for details.

\subsection{Features of RCA dynamics}

Here, we summarize the certain interesting properties of dynamics exhibited by the 3-state RCA, highlighting behavior that goes beyond what has previously been observed in quantum spin chains and deterministic cellular automata.
\begin{itemize}

\item\textbf{Anomalous transport.} We observe a rich variety of transport phenomena, including anomalous transport that goes beyond conventional diffusive and ballistic behavior. In particular, we identify models exhibiting  subdiffusive transport with the dynamical exponent $z = 3$, which, to our knowledge, has not been observed previously in many-body systems. In addition, we observe many superdiffusive models with different dynamical exponents. Some of them exhibit a  dynamical exponent $z = 3/2$, but with a scaling function different from the Prähofer–Spohn one predicted by the KPZ equation.

\item \textbf{Quasilocal charges in chaotic systems.} We demonstrate the existence of quasilocal charges in models that strictly lack simple local conserved quantities. These quasilocal charges play a crucial role in the dynamics of typical observables, governing the polynomial decay of their correlation functions, which would otherwise be exponential in their absence. Some models exhibit an extensive growth of quasilocal conserved quantities with the support --- this may extend the common notion of integrability with the scaling of local charges.

\item \textbf{Integrable models.} We found a number of new models, which show the phenomenology of integrable models, including an infinite number of conservation laws in the thermodynamic limit, as well as ballistic transport. Moreover, some of them exhibit polynomial return times, which has previously yielded models which can be solved exactly (cf.~\cite{RCA54review,ziga_rule2}).
However, it remains a nontrivial future challange to extend the methods of integrability to solve these models.
\end{itemize}

\vspace{-10pt}
\section{Three-state vacuum preserving cellular automata: models and symmetries}\label{sec:ThreeColorCA}

We proceed with a more formal discussion of the cellular automata models introduced in Section~\ref{sec:Summary}. We first define the models, their local time-evolution (or update) rules and discuss how discrete $\CC$, $\CP$ and $\CT$ symmetries of those rules help us divide the models into smaller subsets of all models. We also discuss two separate properties of the time evolution: dual reversibility and unitarity.   

Considering a local configuration space $X=\{ +,-,\varnothing \}$ and a local two-site update rule (map) $U:X^2 \rightarrow X^2$, as introduced in Section~\ref{sec:Summary} (Figs.~\ref{fig:rules_definition},\ref{fig:ziga_rule_example_1}),
we take periodic one-dimensional lattice of even length $L$ and define a reversible dynamical system generated by a many-body dynamical map
$\mathcal{V}: X^L \rightarrow X^L$ over the space of configurations (aka phase space) $X^L\ni(s_1,s_2,\ldots,s_L)$:
\begin{equation}
    s(t+2)=\mathcal{V} s(t),  \quad t\;\textrm{even},
\end{equation}
where $s(t)$ denotes the configuration at some (integer) time $t$, $s(0)$ is the initial configuration and we use the `Cartesian product' notation for $X^L\equiv X\times X\times...\times X$. The two-step dynamical map $\CV$ is composed of `odd' $\CV_o$ and `even' $\CV_e$  one-time-step maps $\mathcal{V}=\mathcal{V}_e\mathcal{V}_o$, that have the following form 
\begin{equation}
    \mathcal{V}_o=U_{2,3}~U_{4,5}~...~U_{L,1}, \qquad \mathcal{V}_e=U_{1,2}~U_{3,4}~...~U_{L-1,L},
\end{equation}
where $U_{i,j}$ denotes the map $U$ applied to variables $s_i$ and $s_j$ and acting trivially (as identity) on the remaining variables. More precisely, we then define dynamical orbits for all integer times as
\begin{equation}
s(2t+1) = \mathcal V_o s(2t),\quad
s(2t+2) = \mathcal V_e s(2t+1).   \label{eq:orbit} 
\end{equation}
The choice of local dynamical map $U$ thus completely specifies the many-body dynamics of such a reversible block cellular automaton.

We can represent the action of $\CV$ diagrammatically as follows:

\begin{figure}[h!]
\centering
  \includegraphics[width=0.75\linewidth]{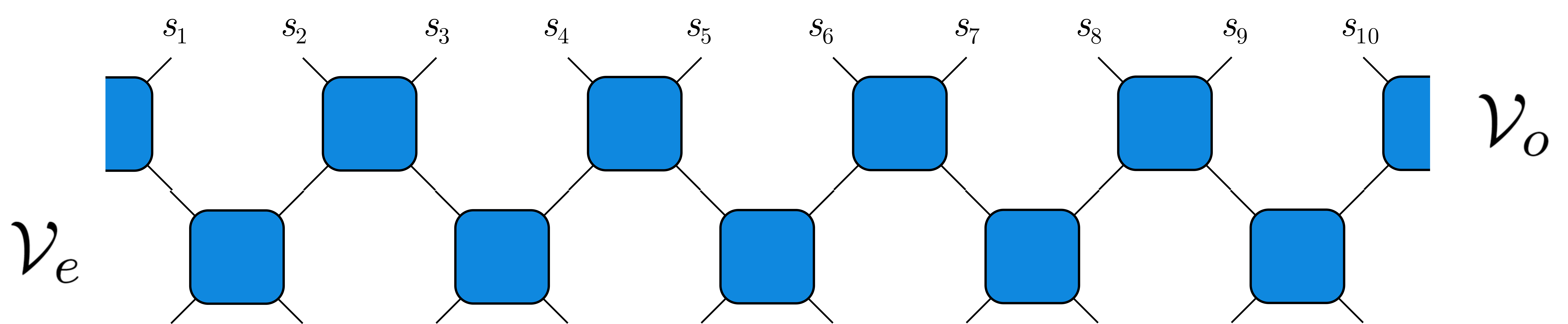}
  \label{fig:floquet}
\end{figure}

where lines denote the discrete variables and boxes the local maps ($U$).

We also impose the time-reversibility of the local rule $U:X^2 \rightarrow X^2$, i.e., we demand that there exists an inverse $U^{-1}$ (hence the term, reversible cellular automaton). Additionally, $U$ must preserve the vacuum configuration: $U(\varnothing,\varnothing)=(\varnothing,\varnothing)$. With these requirements, each update rule, which we label with a sequence of eight numbers as explained in Section~\ref{sec:Summary} (see Figures~\ref{fig:rules_definition} and \ref{fig:ziga_rule_example_1}), is isomorphic to an element of the permutation group $S_8$. We will use the one-line notation for the permutation elements, e.g., for the element $\sigma$ of $S_3$, we have
$\sigma = (2, 3, 1)$ meaning that $ \sigma(1) = 2$, $\sigma(2) = 3$, and $\sigma(3) = 1$. The isomorphism between a map $U$ and an element $\sigma$ of the permutation group $S_8$ is then given by
\begin{equation}
    U=n^{-1} \circ \sigma \circ n.
\end{equation} 
Due to the isomorphism with the permutation group $S_8$, the total number of different local evolution rules is $\text{dim}(S_8)=8!=40320$. Instead of investigating each individual rule, we restrict our attention to subsets of all models characterized by the transformation properties of their rules under discrete $\CC$, $\CP$ and $\CT$ transformations that we define below. It is worth noting that we believe that the classes of cellular automata studied and classified in this work represent a sufficiently broad sample to uncover the breadth of interesting types of ergodic behavior in such models.

\subsection{Symmetries}\label{sec:Symmetries}
To work with discrete symmetries, we consider a map (local symmetry transformation) $\mathcal{S}$ that acts on a configuration of two neighboring cells $\mathcal{S}: X^2 \rightarrow X^2$. We call a rule $\mathcal{S}$-symmetric if it satisfies the following operator identity:
\begin{equation}
  \mathcal{S}\circ U=U \circ \mathcal{S}.
\end{equation}
Note that we will encounter examples in which $\mathcal{S}$ is itself composed of multiple symmetry operators, e.g., $\mathcal{S} = \mathcal{S}_1 \circ \mathcal{S}_2$. We can now define what we mean by the $\CC$, $\CP$ and $\CT$ symmetries in this work where it is important to stress that, in particular, $\CP$ and $\CT$ should be thought as the transformation properties of the update rule and not the entire system. 

\begin{itemize}
\item \textbf{Charge conjugation $\mathcal{C}$}:

\begin{equation}
    \mathcal{C}(s_1,s_2) \defeq
(s_1^*,s_2^*),~\text{where}~s^*=
    \begin{cases}
    -, ~~\text{if}~s=+\\
     +, ~~\text{if}~s=-\\
      \varnothing, ~~\text{if}~s=\varnothing
    \end{cases}.
\end{equation}

\item \textbf{Parity $\mathcal{P}$}:
\begin{equation}
    \mathcal{P}(s_1,s_2) \defeq (s_2,s_1).
\end{equation}
\item \textbf{Time reversal $\mathcal{T}$}:
\begin{equation} 
\mathcal T U \defeq U^{-1}.
\end{equation} 
\end{itemize}
We note that the rule $U$ is called $\mathcal T$-symmetric, if $\mathcal T U = U^{-1}=U$, and $\mathcal{S}\mathcal T$-symmetric (where $\mathcal S$ is another local symmetry transformation), if $\mathcal{S} \circ U = U^{-1} \circ \mathcal{S}$. 

The aforementioned local symmetries of the rule $U$ then induce corresponding global symmetries of the many-body dynamical map $\mathcal V$.
For instance, if $\{s(t)\}_{t=-\infty}^\infty$ is an arbitrary orbit (\ref{eq:orbit}), then $\mathcal T-$symmetry implies that 
$\tilde{s}(t)=s(-t+1)$ is an orbit, 
$\mathcal C\mathcal T$-symmetry implies that
$\tilde{s}(t)=(s^*_1(-t+1),s^*_2(-t+1)\ldots,s^*_L(-t+1))$ is an orbit, $\mathcal P\mathcal T$-symmetry implies that
$\tilde{s}(t)=(s_L(-t+1),s_{L-1}(-t+1)\ldots,s_1(-t+1))$ is an orbit, and $\mathcal C\mathcal P\mathcal T$-symmetry implies that
$\tilde{s}(t)=(s^*_L(-t+1),s^*_{L-1}(-t+1)\ldots,s^*_1(-t+1))$ is an orbit.

We determine how many local update rules satisfy the symmetry classes defined above by exhaustively enumerating all admissible rules and list the total number of such rules for each symmetry class in Table~\ref{tab:sample_table}. The $11$ groups of rules (out of possible $15$ groups of transformation patterns) we analyze in detail are those invariant under $\mathcal{P}$, $\mathcal{P+T}$, $\mathcal{C+T}$, $\mathcal{C+P+T}$,  $\mathcal{CP}$, $\mathcal{CP+T}$, $\mathcal{CPT}$, $\mathcal{CT+P}$,  $\mathcal{PT+C}$ and $\mathcal{CP+PT}$. We note that many rules within these groups can be related to each other. Namely, given the three observables we use in the classification of ergodic behavior (they are discussed in detail below in Section~\ref{sec:dynobs}), many of the rules within each symmetry class are equivalent to each other in that it can easily be seen that all three observables give the same results. Our classification in this work will omit such equivalent `duplicates'.
\vspace{-5pt}
\begin{table}[h!]

    \centering
    \begin{tabular}{|c|c|c|c|c|c|}
        \hline
        Symmetry&$\mathcal{C}$ & $\mathcal{P}$ & $\mathcal{T}$ &$\mathcal{C}+\mathcal{P}$& $\mathcal{C}+\mathcal{T}$\\ 
        \hline
        Number of rules&764 & 96 & 764 & 16& 76 \\ 
        \hline
        Symmetry&$\mathcal{P}+\mathcal{T}$& $\mathcal{C}+\mathcal{P}+\mathcal{T}$& $\mathcal{C}\mathcal{P}\mathcal{T}$&$\mathcal{C}\mathcal{P}+\mathcal{T}$& $\mathcal{C}\mathcal{T}+\mathcal{P}$\\ 
        \hline
        Number of rules&40& 16& 764& 40& 40 \\ 
        \hline
        Symmetry&$\mathcal{P}\mathcal{T}+\mathcal{C}$& $\mathcal{C}\mathcal{P}$& $\mathcal{P} \mathcal{T}$&$\mathcal{C}\mathcal{T}$& $\mathcal{C}\mathcal{P}+\mathcal{P}\mathcal{T}$\\ 
        \hline
        Number of rules&76& 96& 764& 764& 96\\ 
        \hline
    \end{tabular}
    \caption{The number of different cellular automata rules with a given symmetry in each of the $15$ groups of distinct transformation properties. It is important to keep in mind that many rules within these groups can be related to each other.}
    \label{tab:sample_table}
\end{table}

\subsection{Additional dynamical constraints}
In addition to the discrete symmetries discussed above, it will also prove useful to distinguish two additional properties of rules, namely, the `dual reversible' and `chiral' rules. Each of these two dynamical structures imposes interesting constraints on the dynamics of a rule.   

\begin{itemize}
\item \textbf{Dual reversibility}:

For some $U$, there exist a dual rule $U^{D}:X^2 \rightarrow X^2$, such that 
\begin{equation}
    \forall s_1,s_2,s_3,s_4\in X:~~ U(s_1,s_2)=(s_3,s_4) \quad \text{and} \quad U_{D}(s_1,s_3)=(s_2,s_4) .
\end{equation}
\begin{figure}[h!]
    \centering{
        \includegraphics[width=0.4\linewidth]{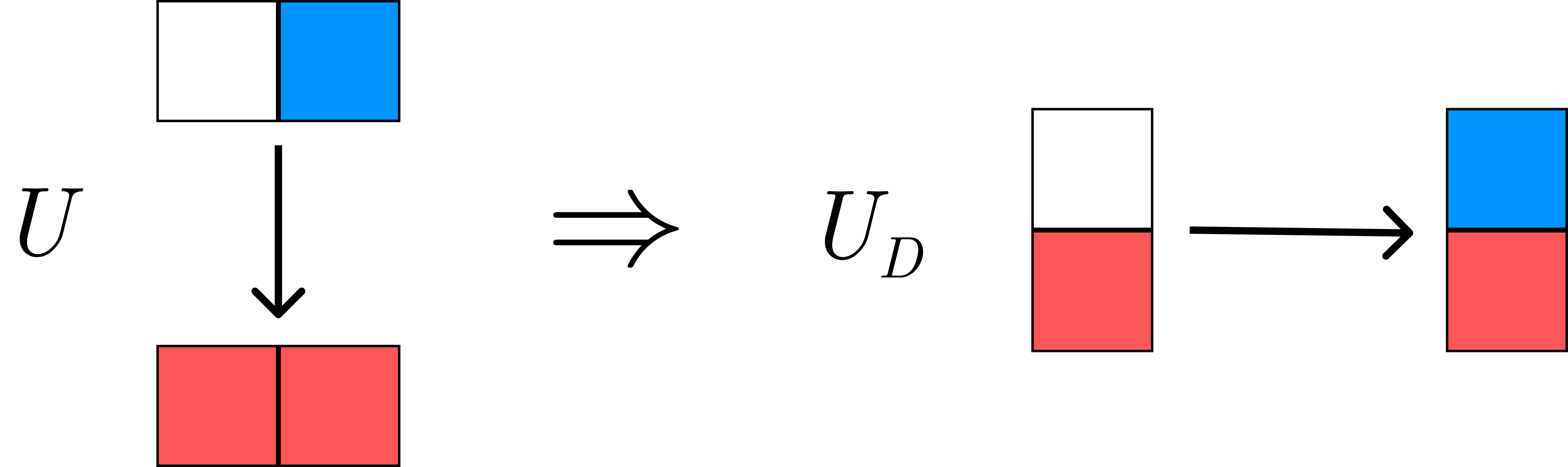}
    }
\end{figure}

Notice that, in general, $U^D$ may not be reversible even for a reversible $U$. The local reversible rule $U$ is therefore called {\it dual reversible} if there exist both $U^D$ and $(U^D)^{-1}$.

\item \textbf{Chirality}:

Chiral models are said to be those for which the local rule $U$ preserves the configuration of either the left or right variable. We distinguish the so-called \textit{East rules} also preserving the left variable, i.e.,
\begin{equation}
  \text{East}:\quad \forall s_1,s_2,s_3,s_4\in X:~~ U(s_1,s_2)=(s_3 = s_1,s_4) .
\end{equation}
Analogously, the \textit{West rules} preserve the right variable, i.e., 
\begin{equation}
  \text{West}:\quad \forall s_1,s_2,s_3,s_4\in X:~~ U(s_1,s_2)=(s_3,s_4 = s_2) .
\end{equation}
\noindent
Notice that the sets of all East and West local rules are isomorphic to each other since $U_{\rm East}=\mathcal{P}^{-1}\circ U_{\rm West} \circ \mathcal{P}$, and the only rule which is both East and West is the trivial rule 12345678. Since the rules are dynamically equivalent up to the parity transformation, we will only consider the East rules.
\end{itemize}
 
\section{Dynamical observables} \label{sec:dynobs}

In this section, we define the three dynamical observables used to classify the cellular automata models summarized in Section~\ref{sec:Summary}. In particular, we discuss how we compute the return times, find and analyze various conserved charges (as well as related Ruelle-Pollicott resonances) and compute two-point correlation functions of local observables. 

\subsection{Return time}

In order to discuss return times of a cellular automaton, we begin by defining the orbit of the configuration $s\in X^L$ as
\begin{equation}
    \mathcal{O}_{s} \defeq  \{s' \in X^L \,|\, \exists t:~ s'=\mathcal{V}^t s\},
\end{equation}
meaning that $\CO_s$ is a set of all configurations generated by the time-evolution from $s$, $\CV^t s$, at all discrete (double) time-steps $t$ until the configuration periodically returns to its original configuration $s$. Once we define the update rule, the whole configuration space splits into a dynamically disconnected union of orbits 
\begin{equation}\label{frag}
  X^L=   \bigcup_{\alpha=1}^{N_O} \mathcal{O}_{s_\alpha},~\text{where} ~s_{\beta} \notin  \mathcal{O}_{s_{\alpha} } ~\text{for}~ \alpha \neq \beta,
\end{equation}
where the number of orbits $N_O$ is bounded: $1\leq N_O \leq 3^L$. Here, the upper bound is given by the trivial local update rule and lower bound is given  by one orbit with maximum length $|X^L|=3^L$. Notice that, due to the reversibility of the local update rules $U$, every trajectory in configuration space must be a loop, i.e., any initial condition will return back to itself after a certain number of iterations.

We are interested in when an initial configuration comes back to itself, for the first time, under evolution with the local update rule. This, so called, return time for the configuration $s$ is then given by the length (or the period) of the orbit $\mathcal{O}_s$ and is the same for all 
$|\mathcal{O}_s|$ elements of $\mathcal{O}_s$. In other words, dynamical map $\mathcal V$ acts on $\mathcal{O}_s$ as a cyclic permutation.

As a quantitative indicator of an automaton's ergodic behavior, we will calculate the average return time $T$ for an update rules $U$ as follows:
\begin{equation}\label{T_frag}
    T \defeq \frac{1}{3^L} \sum\limits_{s\in X^L}|\mathcal{O}_s| =\frac{1}{3^L} \sum\limits_{\alpha=1}^{N_O} |\mathcal{O}_{s_\alpha}|^2,
\end{equation}
where we used Eq.~\eqref{frag} for splitting the configuration space.

\begin{figure}[h!]
    \hspace*{-0.55cm}
    \centering{
    \includegraphics[width=0.9\linewidth]{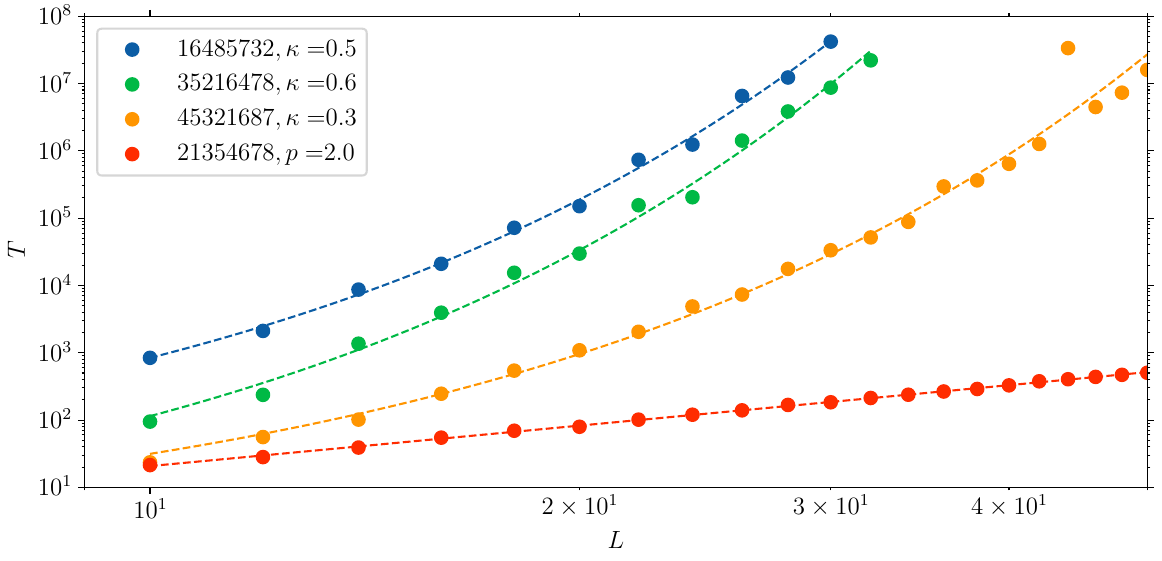}
        \caption{Mean return times for cellular automata with rules 1648573, 35216478, 45321687, 21354678. In their respective order, they represent examples from Class I to Class IV. For the first three examples, the mean return time exhibits exponential growth $T(L) \sim e^{\kappa L}$, while for the last one, it grows as a power law $T(L) \sim L^{p}$.
        }
        \label{fig:TL_reps}
    }
\end{figure}

For (the most) ergodic cellular automaton we expect that 
the dominant (largest) orbit spans almost entire 
phase space $X^L$, i.e. $T(L) \simeq 3^L$.
In other cases, say cases with symmetries or other more hidden constraints, we find that $T(L) \ll 3^L$ meaning that (exponentially) many distinct orbits of comparable size are needed in 
Eg.~(\ref{frag}) to span $X^L$. This can be viewed as a manifestation of ergodicity breaking, and in some sense, a deterministic analogue of Hilbert space fragmentation which is recently intensively studied in quantum lattice systems \cite{moudgalya2022quantum,sala2020ergodicity,khemaniPRB,zadnik2021folded}.

In our deterministic setup, we can place a bound on the number of orbits and the maximal orbit length by knowing the average return time. From Eq.~\eqref{T_frag}, it follows that $T(L) \geq  \mathcal{O}_{max}^2/3^L$. We can then write the upper bound on the maximal orbit length as follows:
\begin{equation}\label{O_bound}
   \frac{{\mathcal{O}}_{max}}{3^L}\leq \left( \frac{T_U}{3^L}\right)^{1/2},~\text{where}~~{\mathcal{O}}_{max}\defeq \max\limits_\alpha|\mathcal{O}_{s_\alpha}|.
\end{equation}
Then, using the fact that the total number of configurations is given by $|X^L|=3^L=\sum\limits_{\alpha=1}^{N_O}|\mathcal{O}_{s_\alpha}|$, the maximal orbit length can be bounded from below by $\mathcal{O}_{max} \geq 3^L/N_O$. Hence, $T \geq \mathcal{O}_{max}^2/3^L \geq 3^L/N_O^2$, which implies a lower bound on number of orbits:
\begin{equation}\label{N_bound}
    N_O\geq \left(\frac{3^L}{T_U}\right)^{ 1/2}.
\end{equation}

\begin{figure}[h!]
    \hspace*{-0.55cm}
    \centering{
    \includegraphics[width=1.0\linewidth]{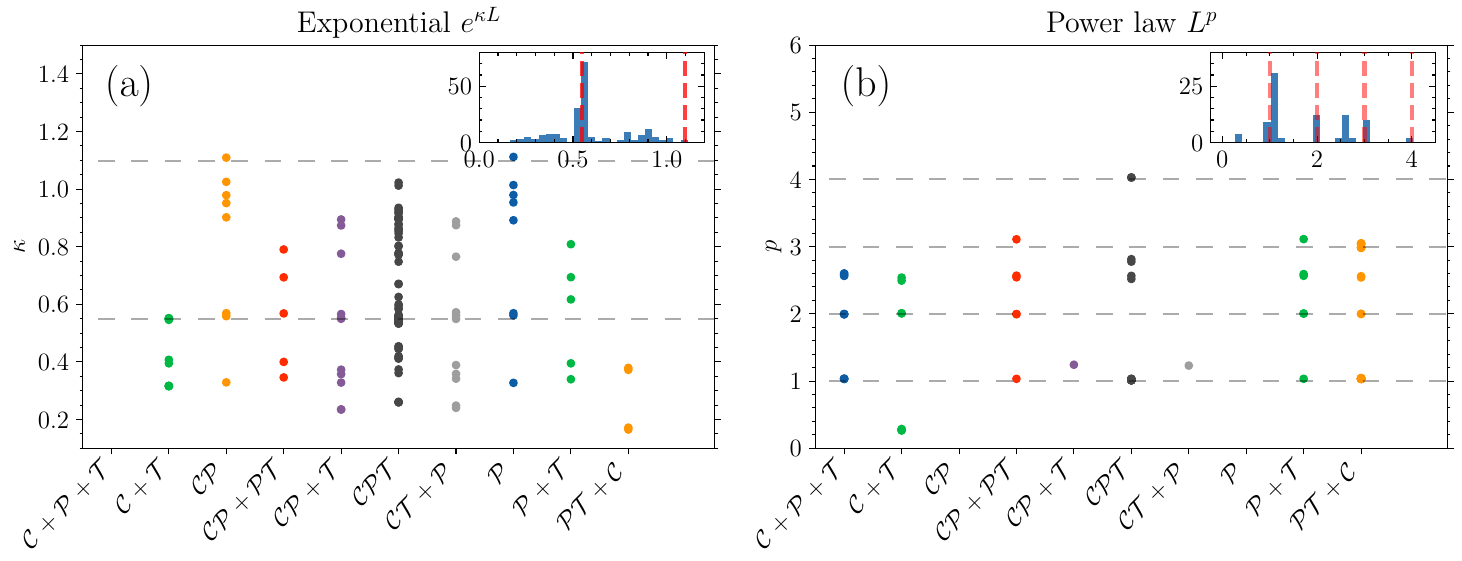}
        \caption{The plots show the $p$ parameter for rules in each symmetry group with \textbf{(a)} exponential ($T(L) \sim e^{\kappa L}$) or \textbf{(b)} power-law ($T(L) \sim L^p$) dependence of mean return times with system size. The horizontal dashed lines on the left plot show the values $\kappa=\ln 3$ and $\kappa=\ln 3/2$ while on the right plot they show integer values of $p$. The insets show the histograms of $\kappa$ and $p$ of the entire dataset with the same (now vertical) dashed lines.
        }
        \label{fig:all_times}
    }
\end{figure}

We calculate average return times as a function of the number of sites $L$ for the RCA models given by the local update rules discussed above, see Figure \ref{fig:all_times}. These return times are obtained by sampling random initial configurations and explicitly iterating the deterministic dynamics until the initial configuration is reached again. For generic models with the absence of conserved quantities, the average return time grows exponentially in system size $T=e^{\kappa L}$. In the majority of cases, the exponent parameter is submaximal  $\kappa < \ln{3}$. It follows from Eqs.~\eqref{N_bound} and \eqref{O_bound} that in such cases,
\begin{equation}
    N_O\geq e^{\frac{\ln(3)-\kappa}{2} L},\qquad\frac{{\mathcal{O}}_{max}}{3^L}\leq e^{-\frac{\ln(3)-\kappa}{2} L}.
\end{equation}
This means that the number of orbits is exponentially increasing while the longest orbit is exponentially smaller than total number of configurations.

\subsection{Conserved charges}
\label{sec:charges_def}
\textbf{Observables.} A global observable $\mathcal{O}:X^L\rightarrow \mathbb{C}$ is a complex-valued function in the space of configurations. The global observables form a multiplicative algebra $\mathcal{A}$ of functions over the configuration space. Multiplication is given by $\left(\mathcal{O}_1 \boldsymbol{\cdot} \mathcal{O}_2 \right)(s)= \mathcal{O}_1(s)  \mathcal{O}_2 (s)$, so the algebra is commutative. The algebra $\mathcal{A}$ is also a vector space, allowing us to define the inner product $\left<~,~\right> :\mathcal{A}\times \mathcal{A}\rightarrow \mathbb{C}$:
\begin{equation}
\left<\mathcal{O}_1,\mathcal{O}_2\right> \defeq \frac{1}{|X^L|}\sum_{s}\mathcal{O}^*_1(s)\mathcal{O}_2(s).
\end{equation} 
We start by defining a subalgebra of local observables.
First, we introduce local generators with respect to a point $x$ on the lattice:
\begin{equation}
    [\alpha]_x (s)  \defeq  \delta_{\alpha,s_x},~~~\alpha\in X.
\end{equation}
\noindent
Note that all local generators add up to a unit element of the algebra (local decomposition of identity): $[+]_x+[-]_x+[\varnothing]_x\equiv \mathds{1}_x\equiv \mathds{1} $. Using the multiplicative property of the generators we then define the following generators
\begin{equation}
    [\alpha_1 \alpha_2\ldots\alpha_r]_x\defeq  [\alpha_1]_x \boldsymbol{\cdot} [\alpha_2]_{x+1}...\boldsymbol{\cdot} [\alpha_r]_{x+r-1},
\end{equation}
which spans the algebra $\mathcal A^{(r)}_x$ of all observables supported on the interval $[x,x+r-1]$ of length $r$:
\begin{equation}\label{obs_space}
    \mathcal{A}_x^{(r)}\defeq \text{span}\{ [\alpha_1 \alpha_2\ldots\alpha_r]_x \,|\,\alpha_i\in X \},~x=1,2,\ldots,L.
\end{equation}
\noindent
See Appendix \ref{sec:numerical_charges} for more explicit definitions of the concepts introduced in this subsection. The minimal support (or simply, support) of an observable $\mathcal{O} \in \mathcal{A}$ is defined by
$\text{supp}(\mathcal{\mathcal{O}})\equiv \text{min}\{r\,|\,\exists x: \mathcal{O}\in \mathcal{A}_x^{(r)} \}$.
Different algebras of local observables can be mapped to each other by the translation operator
or $\mathbb{T}: \mathcal{A} \rightarrow\mathcal{A}$, giving
\begin{equation}
    \mathbb{T} \mathcal{O}(s_1,s_2,...,s_L) \defeq  \mathcal{O}(s_2,...,s_L,s_1).
\end{equation}
The local operator algebras are connected by $\mathcal{A}_x^{(r)}=\mathbb{T}^y \mathcal{A}_{x-y}^{(r)}$. One step of evolution of an observable is then given by the evolution operator $\mathbb{U}:\mathcal{A} \rightarrow\mathcal{A}$: 
\begin{equation} \label{eq:evolution}
   \mathbb{U} \mathcal{O}(s)\defeq \mathcal{O}\left(\mathcal{V} s\right),\quad\mathcal{O}^t\defeq \mathbb{U}^t \mathcal{O}.
\end{equation}
\noindent
Notice that, due to brickwork geometry of our cellular automata setup, the map $\mathbb{U}$ is invariant up to translation with respect to two sites\footnote{In fact, in the special case where $\mathcal{V}_o$ commutes with $\mathcal{V}_e$, the map $\mathcal{V}$ is invariant up to a translation of one site.}, i.e., $[\mathbb{U},\mathbb{T}^2]=0$. Thus we may consider dynamics confined to \textit{extensive} observables invariant under a two-site translation:
\begin{equation}
   A=\sum_{j=1}^{L/2-1} \mathbb{T}^{2j} a,\quad a\in \mathcal{A},
\end{equation}
where $a$ is the \textit{density} of an extensive observable. The set of all extensive observables forms a vector space $\mathcal{A}_\mathbb{T}$. In the space $\mathcal{A}_\mathbb{T}$, we consider the subspace given by the \textit{extensive local} observables with density of finite support, 
\begin{equation}    \mathcal{A}_{\mathbb{T}}^{(r)}=
\left\{\sum\limits_{j=1}^{L/2-1}\mathbb{T}^{2j}a\,|\,~a\in \mathcal{A}_1^{(r)}\right\}.
\end{equation}
We will say that the extensive observable $A$ has \textit{range} $r$ if $A \in \mathcal{A}_{\mathbb{T}}^{(r)},$ and $ A\notin \mathcal{A}_{\mathbb{T}}^{(r-1)} $. In order to discuss locality we will use the decomposition of an extensive observable to components of different ranges. Namely, observable $A$ can be written as:
\begin{equation}
    A=\sum_{r} A^{(r)},~~~A^{(r)}=P_{\mathcal{A}^{(r)}_{\mathbb{T}}/\mathcal{A}^{(r-1)}_{\mathbb{T}}}A,
\end{equation}
where $P_{\mathcal{A}^{(r)}_{\mathbb{T}}/\mathcal{A}^{(r-1)}_{\mathbb{T}}} $ is projector onto the vector subspace $\mathcal{A}^{(r)}_{\mathbb{T}}/\mathcal{A}^{(r-1)}_{\mathbb{T}}$. In this decomposition operators $a^{(r)}$ does not have components with lower than $r$ range.
\\

\textbf{Charges.} An observable $Q \in \mathcal{A}$ is a conserved charge, or simply a \textit{charge}, if it is conserved under time evolution
\begin{equation}
\mathbb{U} Q = Q \quad \Leftrightarrow \quad Q(\mathcal{V}s)=Q(s),~\forall s \in X^L.
\end{equation}
\noindent

To build a connection with the usual notion of integrability in quantum spin chains, we need to introduce the notion of locality of the charges. In analogy with quantum spin chains (see, e.g., \cite{ilievski2016quasilocal}), we introduce the following possible \textit{locality properties}. 
\begin{itemize}
    \item \textbf{Locality.} An extensive charge $Q^{(r)}$ is \textit{local} with range $r$ if it is an extensive local observable conserved under the evolution
    \begin{equation}
        Q^{(r)}=\sum_j \mathbb{T}^{2j} q^{(r)},~ q^{(r)}\in \mathcal{A}^{(r)}_1.
    \end{equation}
    \item \textbf{Quasilocality.} An extensive charge $Q$ is \textit{quasilocal} if its density $q$ can be written as a sum of mutually orthogonal local terms $q^{(r)}$ of range $r$  with the following properties in the $L\rightarrow \infty$ limit: 
    \begin{equation}\label{qlocdef}
        q=\sum_{r=1}^\infty q^{(r)},~\|q^{(r)}\|<C e^{-\xi r},
    \end{equation}
    where $\| a \| = \lim_{L\to\infty} \sqrt{\langle a,a \rangle}$, and $\xi \in \mathbb{R}_+$.
    \item \textbf{Pseudolocality.} In the definition of locality and quasilocality, we consider an infinite size limit $L\rightarrow \infty $. For a finite number of sites, the charge cannot be written as a sum of an infinite number of terms with increasing support as in Eq.~\eqref{qlocdef}. Thus, we have to consider a sequence of observables truncated by the maximal range $L$,
    \begin{equation}
        Q_L=\sum_r Q^{(r)}_L,\quad Q^{(r)}_L=\sum\limits_{j=1}^{L/2-1}\mathbb{T}^{2j} q^{(r)}_{L},
    \end{equation}
    where $Q^{(r)}=P_{\mathcal{A}^{(r)}_{\mathbb{T}}/\mathcal{A}^{(r-1)}_{\mathbb{T}}} Q_L$.
An extensive charge $Q=\lim\limits_{L\rightarrow \infty}Q_L$ is \textit{pseudolocal} if the sequence of observables $Q_{L}$ scales extensively:
\begin{equation}
    0<\lim\limits_{L\rightarrow\infty}\frac{1}{L}\left< Q_L,Q_L \right> <\infty,
\end{equation}
and has a fixed ($L$-independent) nonzero overlap with any strictly local observable (to set the scale).
Notice that all quasilocal charges are pseudolocal, while, if $\|q^{(r)}\|\sim 1/r^{\alpha}, \text{where}~ \alpha>1$, the charge will be pseudolocal, but not quasilocal.
    \item \textbf{Dynamical charges.} We also consider \textit{dynamical charges} which are conserved only after $n$ steps of the evolution (and its integer multiplicities), such that 
\begin{equation}
    \mathbb{U}^nQ=Q ~ \text{and}~ \mathbb{U}^m Q\neq Q, ~\text{for}~1\le m<n.
\end{equation}
\end{itemize}
\vspace{-10pt}
\subsection{Transfer matrix of dynamics of observables}\label{sect:trans}
As a measure of ergodicity of each of the local evolution (update) rules, we will consider the dependence of their number of extensive local charges and dynamical charges $N_r$ for a given support size $r$. In order to find the extensive charges, we need to find the eigenspace of the evolution operator $\mathbb{U}$ acting on vector space $\mathcal{A}_{\mathbb{T}}$ of extensive observables. The evolution increases the range of local observables as $\mathbb{U}:\mathcal{A}^{(r)}_{\mathbb{T}}\rightarrow\mathcal{A}^{(r+3)}_{\mathbb{T}}$. This complicates the numerical diagonalization of the evolution operator acting on the space of extensive local observables. Following Refs.~\cite{Prosen_2004, Prosen_2007, Klobas_2022},  we introduce the truncated transfer matrix, which is a linear map $\mathcal{T}^{(r)}: \mathcal{A}_{\mathbb{T}}^{(r)} \rightarrow \mathcal{A}_{\mathbb{T}}^{(r)}$, such that
\begin{equation} \label{T_def}
    \mathcal{T}^{(r)} \defeq P_{\mathcal{A}_{\mathbb{T}}^{(r)}} \,\cdot\,  \mathbb{U},
\end{equation}
where $P_{\mathcal{A}_{\mathbb{T}}^{(r)}}$ is an orthogonal projection
to subspace $\mathcal A^{(r)}_{\mathbb T}$.
    \vspace{-10pt}
\begin{figure}[h!]
    \hspace*{0.4cm}
    \centering{
        \includegraphics[width=0.7\linewidth]{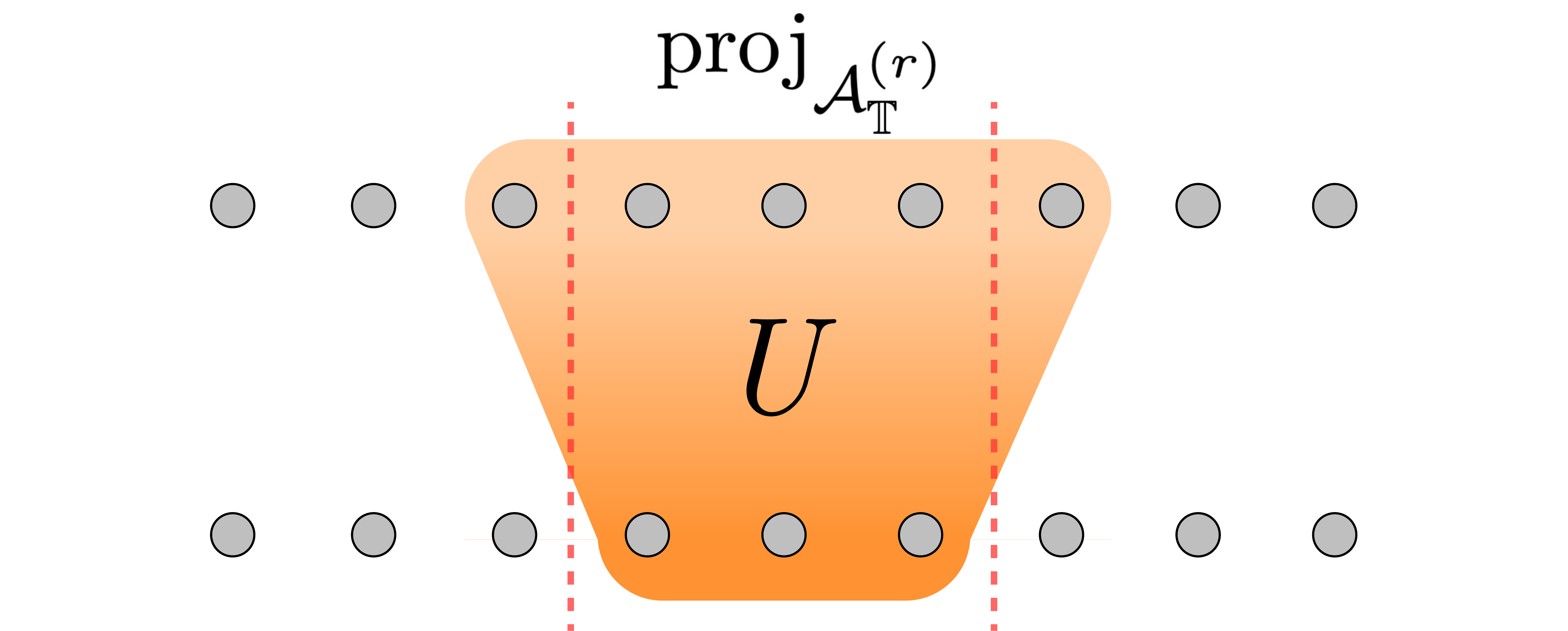}
        \caption{A sketch of the action of the transfer matrix $\mathcal{T}^{(r)}$ on the space of extensive local operators. We begin with an operator with support $r=3$, which is spread to support $5$ after $\mathbb U$, and is projected back to the $3$-local space as denoted by the red dashed lines.}
        \label{fig:algorithm}
    }
\end{figure}

Below, we will study the spectra of the transfer matrices $\mathcal T^{(r)}$ for increasing range $r$. Because we introduce the projection operator in the definition, the transfer matrix is not unitary, but is bounded in the operator norm  
$\| \mathcal{T}^{(r)} \|\equiv \sqrt{{\rm sup}_{A} 
\langle \mathcal T^{(r)}A, \mathcal T^{(r)}A\rangle/\langle A,A\rangle}
\leq 1$. The eigenvalues of the transfer matrix $\mathcal{T}^{(r)} \mathcal{O_\lambda}=\lambda~ \mathcal{O_\lambda}$ therefore lie on the unit disk $|\lambda|\leq 1$.  

Notice that the truncation preserves the local extensive charges, namely, that 
\begin{equation}
    \mathbb{U}Q^{(r')}=Q^{(r')}~\Rightarrow ~\mathcal{T}^{(r)}Q^{(r')}=Q^{(r')},~\text{for}~r' \leq r.
\end{equation} 
The local extensive charges correspond to the eigenvectors with eigenvalue $1$ of $\mathcal T^{(r)}$. By knowing the spectrum of the transfer matrices $\mathcal{T}^{(r)}$
for any $r$, we can find all extensive local charges. Note that any map $\mathbb{U}$ will have a trivial (non-dynamical) conserved quantity whose density is the identity element of the operator algebra, thus spectrum of $\mathcal{T}^{(r)}$ will always contain one trivial eigenvalue 1.

\begin{figure}[h!]
    \centering{
    \includegraphics[width=0.88\linewidth]{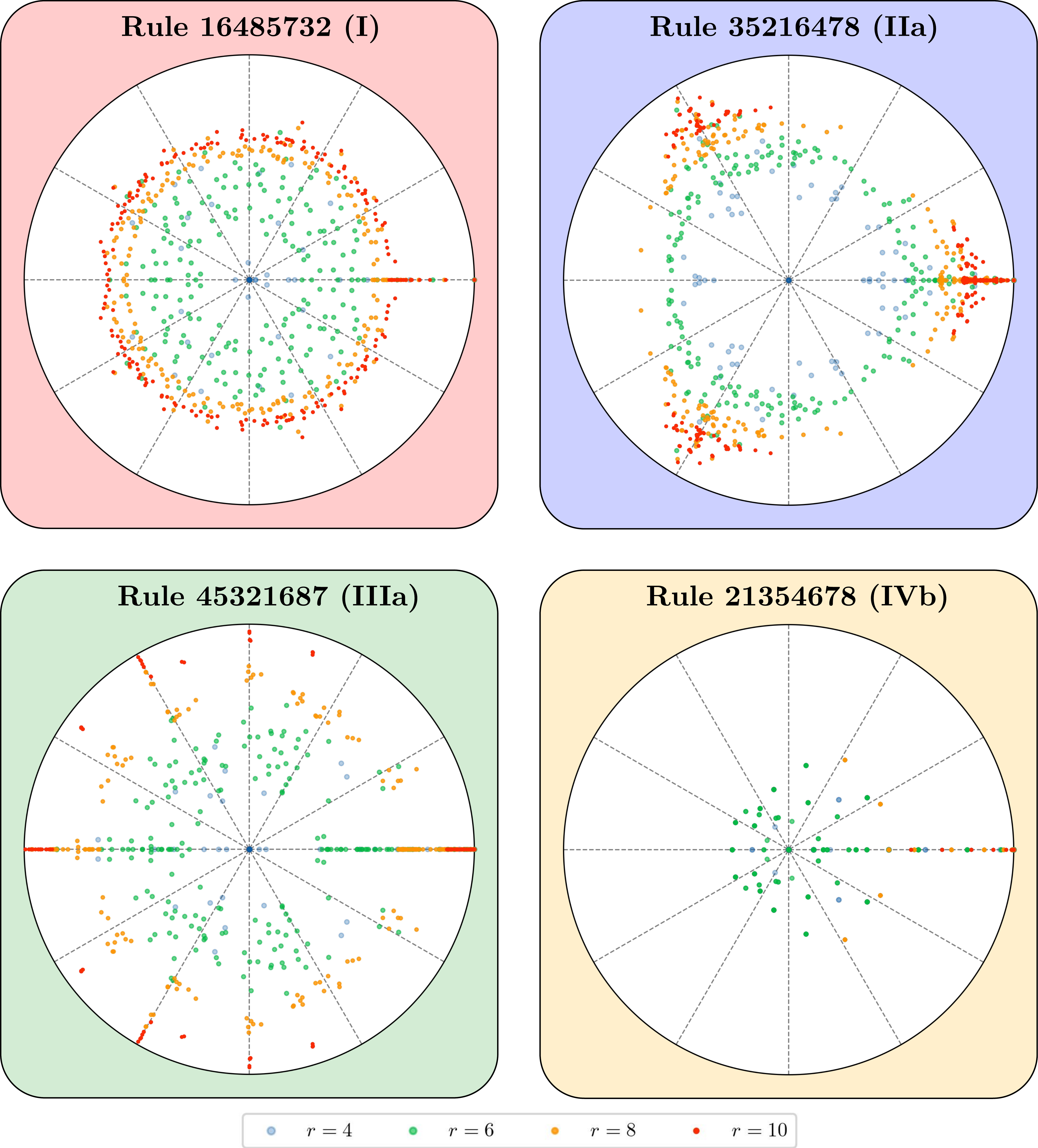}
    }
        \caption{Spectra of the 200 largest  eigenvalues of the transfer matrix $\mathcal{T}^{(r)}$ plotted for the four rules 16485732, 35216478, 45321687, 21354678 representative of each of the classes. Parameters $r=4,6,8,10$ correspond to red, orange, green, and red points, respectively. Note that every rule has an eigenvalue at $\lambda=1$ which corresponds to the identity of the operator algebra.
    }
    \label{fig:spectra_alt}
\end{figure}

We will call models with an increasing number of extensive local charges (with increasing the range $r$) \textit{integrable}\footnote{We stress that the notion of integrability in classical cellular automata is not understood, hence, we use this terminology in a more `colloquial' manner.}. In addition, we will also distinguish a subclass of \textit{superintegrable} models in which the number of charges grows faster than linearly, which is expected to constrain the dynamics more stringently. 

For dynamical charges that are conserved only after $n>1$ steps of the time evolution, we have to satisfy the condition 
\begin{equation}
    \mathbb{U}^n Q^{(r')}=Q^{(r')}~\Rightarrow~ \left[\mathcal{T}^{(r)}\right]^n Q^{(r)}=Q^{(r)},~ \textrm{for}~ r\geq r'+2 (n-1).
\end{equation} 

We note that due to a larger range of the transfer matrix that needs to be diagonalized, it is numerically harder to find dynamical charges of a given range, but they will be present in the spectrum for large enough $r$. Dynamical symmetries also influence transport \cite{dynamical_lenart}.

In addition to the eigenoperators with $\lambda=1$, we will be interested in eigenvalues close to the identity. Due to the presence of quasilocal charges in some of the models, the spectral gap $\Delta (\lambda)=|1-\lambda|$ will go to zero in large range $r$ limit. For each quasilocal charge present in the system we will find, in the $r\rightarrow \infty$ limit, an eigenvalue for which the spectral gap closes exponentially $\ln \Delta(r) \propto -r$. 
In Figure \ref{fig:spectra_alt}, we show the spectrum of transfer matrix of range $r$ up to 10 for representatives of the four ergodic classes discussed above.

\subsection{Correlation functions}\label{sec:CF}
To characterize the transport phenomena, we compute dynamical two-point of various observables. In particular, we compute correlators of all single-site observables. This space is constrained by the fact that $[\varnothing]_x + [+]_x + [-]_x \equiv \mathds{1}_x$, which means that one has to check correlators of only two independent observables. Hence, we consider connected correlators of the \textit{charge} $c_x: = [+]_x - [-]_x$ and \textit{density} $d_x \equiv c_x^2 =[+]_x + [-]_x$ operators sampled according to a maximum entropy (infinite temperature) distribution,
\begin{equation} \label{eq:correlation}
    \hat{C}(x,t) = \langle c_x(t) c_0(0) \rangle_{\infty}^C,
    \quad
    \hat{D}(x,t) = \langle d_x(t) d_0(0) \rangle_{\infty}^C.
\end{equation}
\noindent
We are interested in the temporal decay of the correlation functions that track the evolution of the maxima (peaks) of $\hat C$ and $\hat D$. In particular, we define 
\begin{equation}
    C(t)\defeq \max\limits_{x\in[1,L]} \hat{C}(x,t),\quad D(t)\defeq \max\limits_{x\in[1,L]} \hat{D}(x,t).
\end{equation}
\noindent
This is mainly because of the presence of models without the $\mathcal{P}$ symmetry, which can exhibit a drift of the correlator. Additionally, if there are multiple modes present in the correlator, we will extract the slowest one that dominates the late-time dynamics. The correlation function $\hat{D}(x,t)$ can be expressed as an autocorrelation of the `vacuum':
\begin{equation}
    \hat{D}(x,t) = \langle [\varnothing]_x(t) [\varnothing]_0(0)\rangle^C_{\infty}.
\end{equation}

\begin{figure}[h!]
    \centering{
    \includegraphics[width=0.95\linewidth]{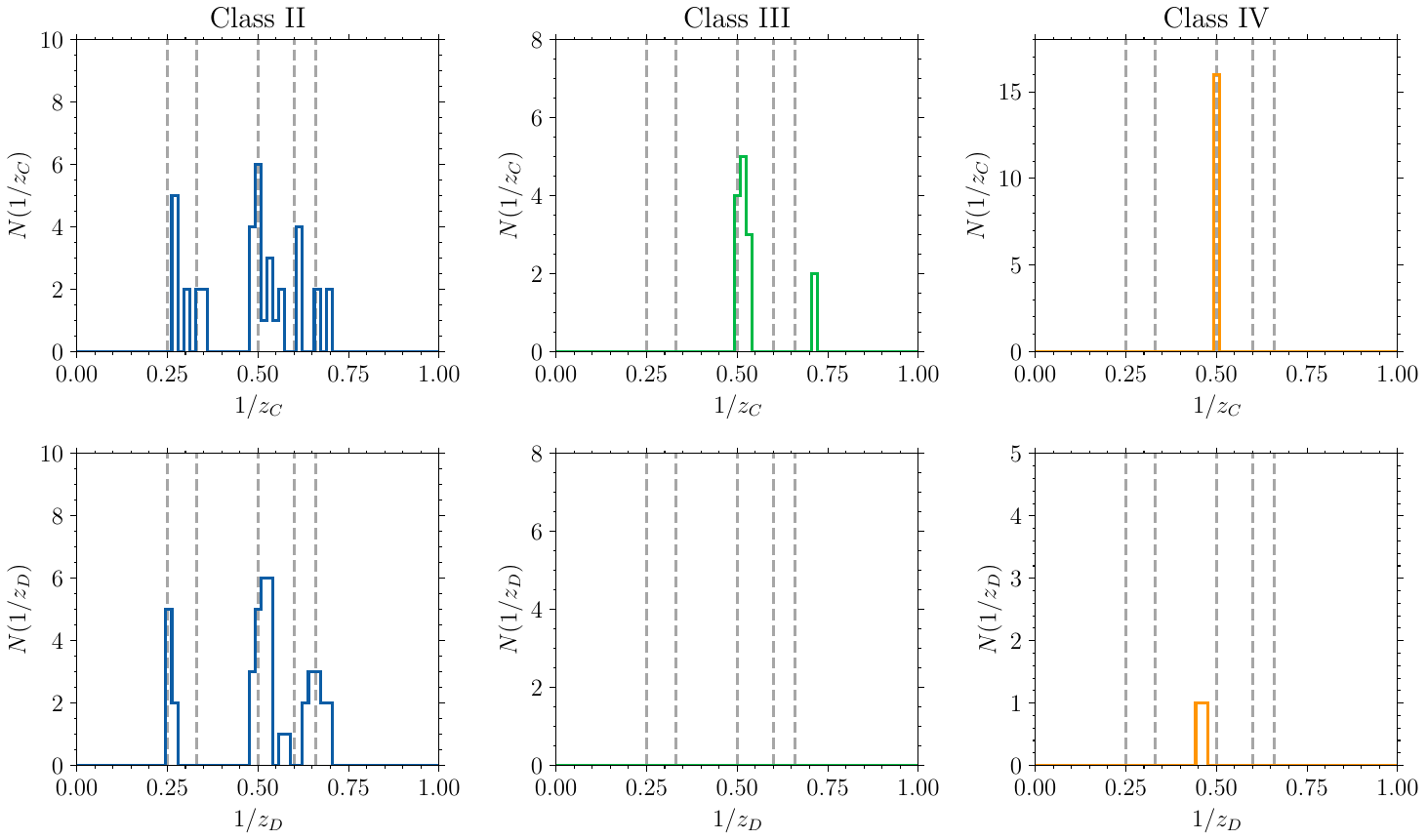}
        \caption{Histograms of all dynamical exponents $1/z_C$ and $1/z_D$ for Classes II, III and IV. The vertical dashed lines, from left to right, represent the values of $1/z \in \{\frac{1}{4},\frac{1}{3}, \frac{1}{2}, \frac{3}{5}, \frac{2}{3}\}$. The counting only includes the physically distinct RCA (with `duplicates' removed).
        }
    \label{fig:dynamical_exponents}
    }
\end{figure}

In chaotic models without any local conservation laws, it is expected that the correlation functions decay exponentially:
\begin{equation}
    C(t) \propto e^{- \alpha t}, \quad  D(t) \propto e^{- \beta t},
\end{equation}
and we will be able to distinguish these models in Class I by their decay rates $(\alpha, \beta)$. However, when a rule possesses additional conserved charges, the decay of either correlation function can become algebraic. To characterize such dynamics, we introduce the dynamical exponents $z_C$ and $z_D$:
\begin{equation}
    C(t) \sim t^{-1/z_C}, \quad  D(t) \sim t^{-1/z_D}.
\end{equation}
For example, dynamics exhibiting \textit{diffusion} with the diffusion constant $D$ has either of the two dynamical exponents equal to $1/z=1/2$, and the profile of the correlator is Gaussian:
\begin{equation}
         \lim_{t\to\infty} t^{1/2} \hat{C}(\xi t^{1/2},t) \propto \frac{1}{\sqrt{2 \pi \sigma^2}} e^{-\xi^2/2\sigma^2} ,\quad \sigma = 2\pi D,~~~\xi=x/\sqrt{t}
\end{equation}
The second common case is \textit{ballistic} transport $z=1$, which is typically associated with charge-carrying `free' quasiparticles. Another option is the so-called \textit{superdiffusive} transport with $1/2 < 1/z < 1$. A notable example is the Kardar-Parisi-Zhang-type (or KPZ-type) scaling  with $1/z=2/3$ which, in the Hamiltonian (non-noisy/non-dissipative) one-dimensional lattice systems, occurs in two different known contexts: (i) in anharmonic oscillator chains like the Fermi-Pasta-Ulam systems with exactly three coupled conserved quantities, where the mechanism is explained within the theory of non-linear fluctuating hydrodynamics\cite{mendl,spohn2014}, 
and (ii) in integrable systems with non-abelian global symmetries~\cite{krajnik2020kardar, ilievski2021superuniversality, krajnik2024dynamical, takeuchi2024partial, ljubotina2017spin, medenjak}, where
the precise mechanism of the KPZ scaling is still unclear.
Finally, the so-called \textit{subdiffusive} scaling has $0< 1/z < 1/2$ (see, e.g.,~Refs.~\cite{MBL,de2024absence,Brighi2023,Vasseur2021}). The distribution of all empirically determined dynamical exponents is shown in Figure~\ref{fig:dynamical_exponents}.
 
\newpage

\section{Class I}\label{sec:class_1}
The first category we consider includes the most generic  `chaotic' (ergodic or mixing) models, referred to as Class I. These models have no extensive local conserved charges and exhibit an exponentially decaying correlation function of local observables, in particular, for the observables of range 1. Additionally, the average return time for all models in Class I scales exponentially with the number of sites, 
$T(L)\sim e^{\kappa L}$.
In models with `least structure', we find that the exponent
saturates the upper bound, $\kappa = \kappa_{\rm max} = \ln 3$.
On the other hand, in models with no symmetry constraints other than a symmetry containing time reversal, say 
$\mathcal S\mathcal T $ (see Section~\ref{sec:Symmetries} for definitions), one may argue, following Ref.~\cite{Rannou}, that a typical orbit length will be
$T(L) \simeq \sqrt{|X^L|}$ and that $\kappa = \kappa_{\rm max}/2 = \frac{1}{2}\ln 3$.

To show the interesting physical features of RCA in this class, we focus on the cellular automaton with rule 16485732. We depict the rule along with a sample of its dynamics here. 
\begin{figure}[h!]
\centering
\includegraphics[width=0.55\linewidth]{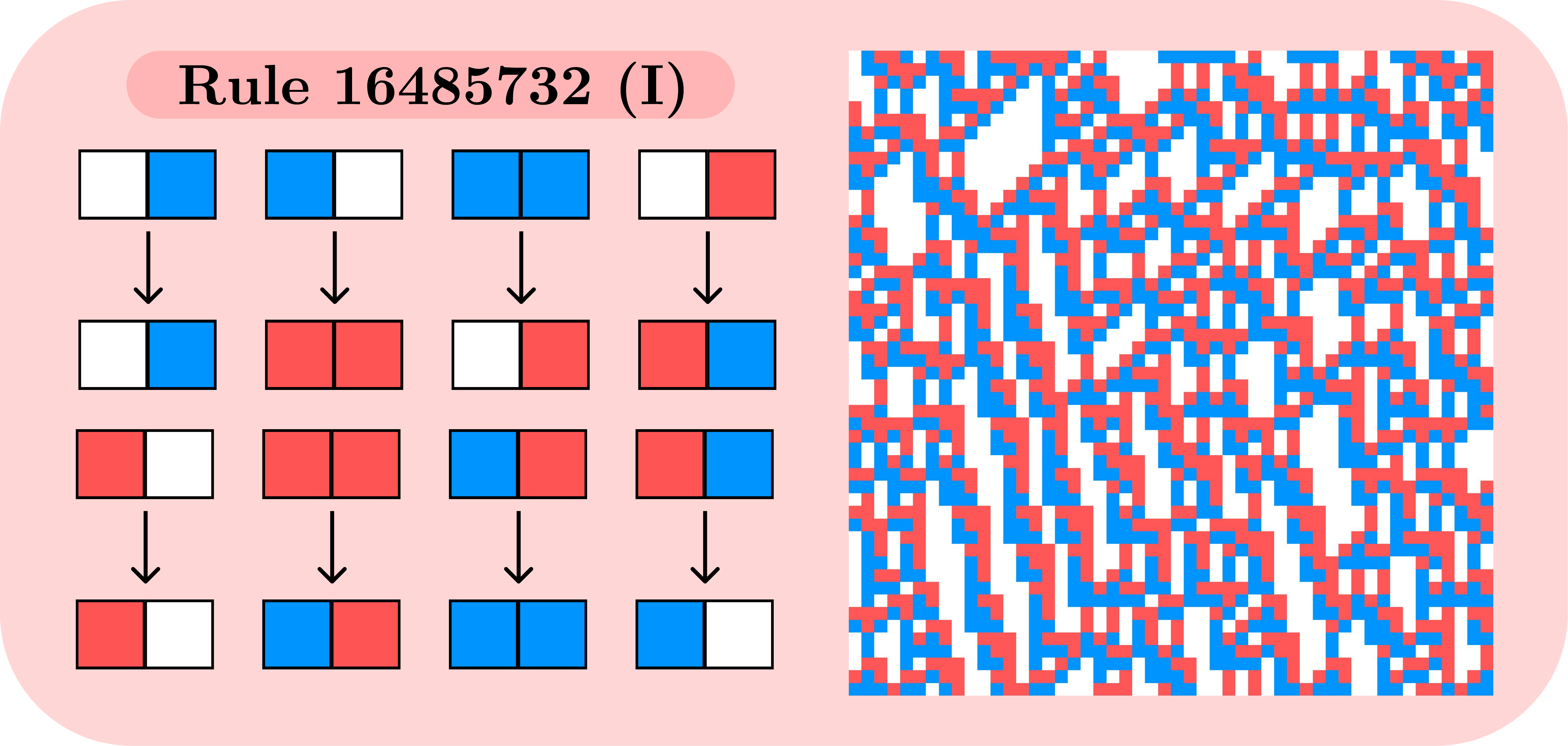}
  \label{fig:class_1a_dynamics}
  \label{Ia_rule}
\end{figure}

\begin{figure}[h!]  
\centering
  \includegraphics[width=0.9\linewidth]{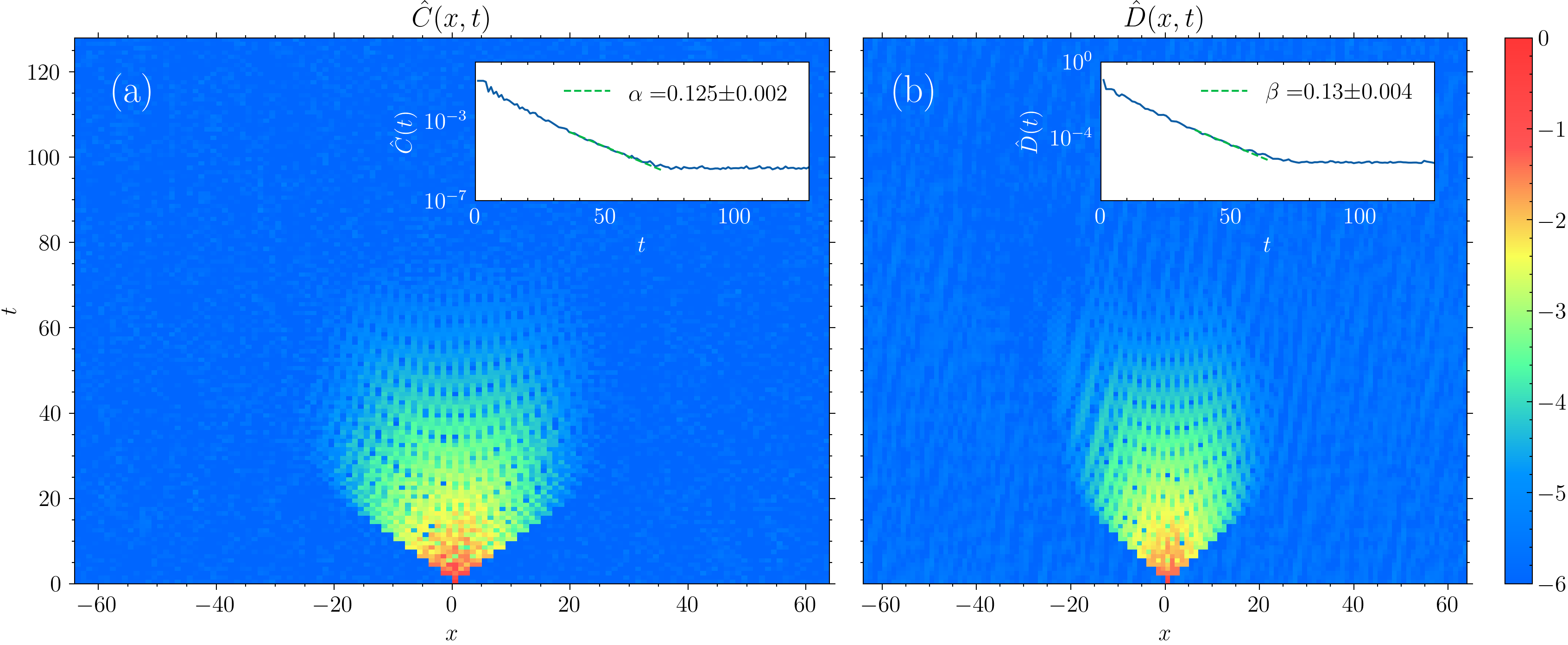}
  \caption{Correlation functions of the $\mathcal{CPT}$-symmetric cellular automaton with rule 16485732: \textbf{(a)} $\hat{C}(x,t)$ and \textbf{(b)} $\hat{D}(x,t)$. The colors are shown in the ${\rm log}_{10}$-scale. The insets show both correlators at maximal $x$ along with their respective exponential fits $e^{-\alpha t}$ and $e^{-\beta t}$ depicted by green dashed lines. We also state the values of $\alpha$ and $\beta$.
  }
  \label{fig:class_1_correlator}
\end{figure}

The model is $\mathcal{C}\mathcal{P}\mathcal{T}$ symmetric, but it does not possess independent $\mathcal{C}$, $\mathcal{P}$ or $\mathcal{T}$ symmetries. Due to the absence of $\mathcal{P}$ symmetry, its correlation functions are not symmetric under $x\rightarrow -x$. We present the correlation functions of range-1 local observables (see Section~\ref{sec:CF}) in Figure~\ref{fig:class_1_correlator}. Both correlators decay exponentially with similar rates.

\begin{figure}[h!]
\centering
\centerline{
    \hspace{-0.3cm}
  \includegraphics[width=0.5\linewidth]{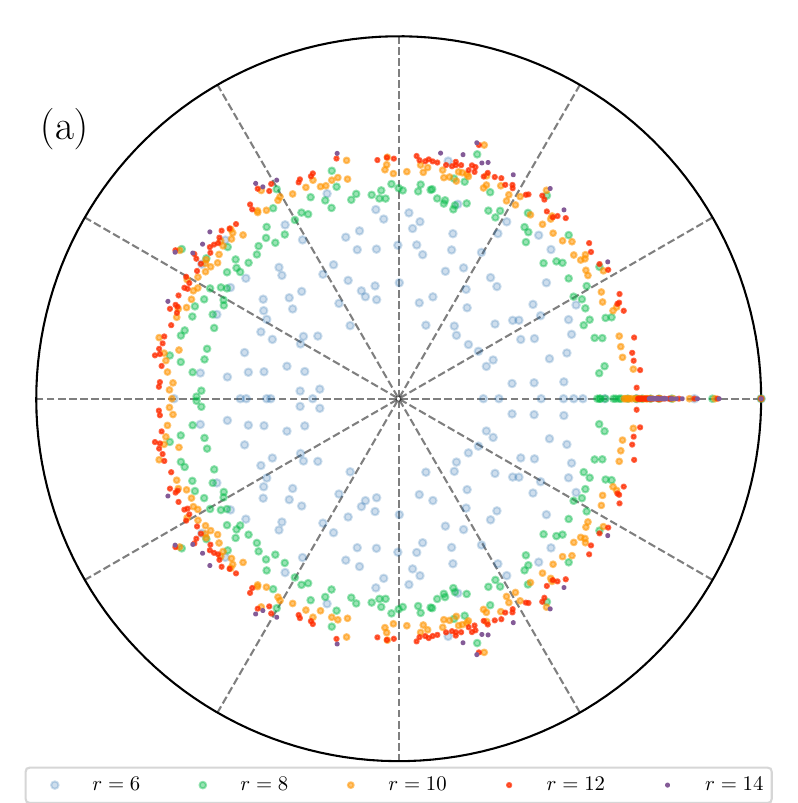}

  \hspace{-0.25cm}
   \includegraphics[width=0.5\linewidth]{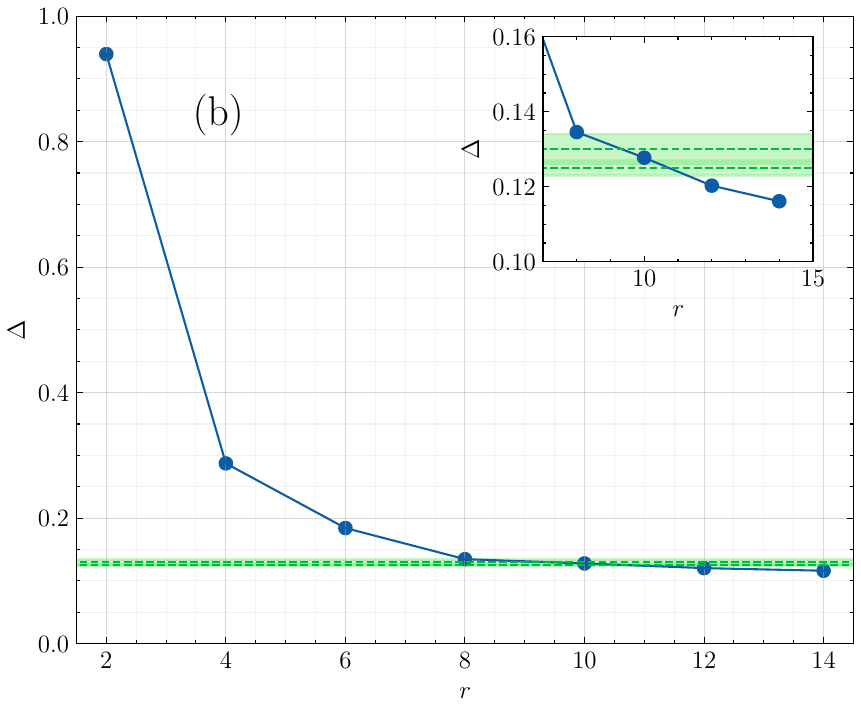}
}
  \caption{\textbf{(a)} Spectrum of the transfer matrix computed up to range $r=14$. \textbf{(b)} Decay of the gap $\Delta=1-\lambda^{(r)}$ with increasing support $r$. The horizontal lines show the fitted values of $\alpha$ and $\beta$ from the correlators $C(t)$ and  $D(t)$, with the shaded regions representing the error of the fitted parameters. The inset shows a closeup of the region $r \in [8, 14]$.
  }
  \label{fig:class_1_spectrum}
\end{figure}

Next, to confirm the absence of local charges, we study the spectrum of the transfer matrix of local extensive observables defined in Section~\ref{sect:trans}, which indeed shows the absence of any eigenvalues on the unit circle. To examine the absence of quasilocal charges, we then analyze the scaling of the gap in the transfer matrix spectrum, see Figure~\ref{fig:class_1_spectrum}.

Figure~\ref{fig:class_1_spectrum} shows that as we increase the range of the transfer matrix, we observe the `freezing' of the gap in the spectrum at some $\Delta_0$. Its value is approximately given by the decay rate of the correlation function $C(t) \sim e^{-\alpha t} $: $\Delta_0\approx e^{-\alpha}$. This behavior suggests that the freezing eigenvalue corresponds to the \textit{Ruelle-Pollicott (RP) resonance}. Conceptually, similar RP resonances have been proposed
in quantum spin lattice systems
\cite{TomazProsen_2002} and reinvestigated recently in Ref.~\cite{Marko_PR,zhang2024thermalizationratesquantumruellepollicott, jacoby2024spectralgapslocalquantum,TakashiMori}, where it was shown that the leading eigenvalue captures the decay of correlation functions in non-integrable systems.

We propose the following definition of the RP resonance for a lattice system, where, here, we focus on RCA. Let as assume that the truncated transfer operators $\mathcal T^{(r)}$ 
have a sequence of eigenvalues $\lambda^{(r)}$ and left/right eigenvectors
$\mathcal T^{(r)}u^{(r)} = \lambda^{(r)}u^{(r)}$,
$v^{(r)}\mathcal T^{(r)} = \lambda^{(r)}v^{(r)}$, where, by convention, $\langle{v^{(r)},u^{(r)}}\rangle=1$.
We make sure that $L>r$, which is equivalent to considering the limit $L\to\infty$.
If, for each $r$, $\lambda^{(r)}$ corresponds to the leading nontrivial eigenvalue, then the long-time asymptotics of the correlation functions
of translationally invariant observables $A,B\in\mathcal A_{\mathbb T}$ reads
\begin{equation}
    \langle A(t)B \rangle_\infty^C \simeq \lim_{r\to\infty}
    \langle A,v^{(r)}\rangle  \langle u^{(r)},B\rangle e^{-\lambda^{(r)} t}\,.
\end{equation}
We note that the inverse truncation range $1/r$ can be physically interpreted as the noise strength. The limit $r\to \infty$ thus corresponds to the noiseless limit of conservative dynamics. 

If it happens that the following limits exist: \begin{equation}
\lambda^*=\lim_{r\to\infty}\lambda^{(r)}\,,\quad 
u=
\lim_{r\to\infty} 
u^{(r)},\quad
v=
\lim_{r\to\infty} 
v^{(r)}, \end{equation}
then we define the triple $(\lambda^*,u,v)$ as the RP resonance. In practice, the convergence of these limits is rather fast and the essential information can be obtained from $r$ of around 10 (see Figure~\ref{fig:class_1_spectrum}). 

Furthermore, the convergence in the eigenvectors can be conveniently benchmarked with the convergence of the bi-norms
\begin{equation}
Z^{(r)}_\ell = 
\left< u^{(r,\ell)},
v^{(r,\ell)} \right>,\quad\ell=1,2\ldots r,
\end{equation}
where $u^{(r,n)}= P_{\mathcal{A}_{\mathbb{T}}^{(n)}\setminus\mathcal{A}_{\mathbb{T}}^{(n-1)}} u^{(r)}$  and $v^{(r,n)}= P_{\mathcal{A}_{\mathbb{T}}^{(n)}\setminus\mathcal{A}_{\mathbb{T}}^{(n-1)}} v^{(r)}$ are components of the observables with range $n$.

\begin{figure}[h!]
\centerline{
    \hspace{-0.3cm}
\includegraphics[width=0.6\linewidth]{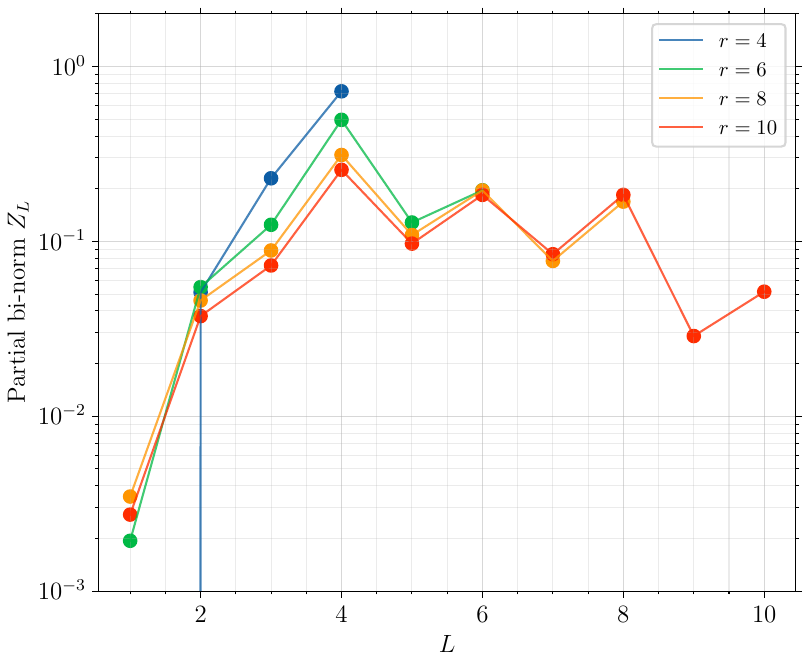}
    }
  \caption{Partial bi-norms for the  $\mathcal{CPT}$-invariant cellular automaton with rule 16485732. For increasing $r$, we observe the beginning of exponential decay of components with increased support.
  }
  \label{fig:binorms}
\end{figure}

Note that, by normalization convention, 
$\langle v^{(r)},u^{(r)}\rangle = \sum_{\ell=1}^r Z^{(r)}_\ell = 1$.
One also observes that $Z^{(r)}_\ell \ge 0$ in all cases that we have tried (although we are not able to prove it in general). Assuming that, we can define the distribution of bi-norms, and in particular,
if the limit $\mathcal P(\ell)=\lim_{r\to\infty} Z^{(r)}_\ell$ exists, the empirical tails $Z^{(r)}_{\ell \sim r}$ give excellent information on the convergence of the RP resonance. A typical example is shown in 
Fig.~\ref{fig:binorms}.

In cases when $\lambda^*=1$, but $\lambda^{(r)}\neq 1$ for finite $r$, then $Q=u=v$ corresponds to the pseudo-local or quasilocal charge (see definitions in Section \ref{sec:charges_def}). We can then benchmark the locality properties of the eigenvectors using the partial 2-norms, namely,
\begin{equation}
    Z^{(r)}_\ell = \sum_{\ell = r}
\left< u^{(r,\ell)},
u^{(r,\ell)} \right>,\quad\ell=1,2\ldots r\,,
\end{equation}
 which provides a useful criterion for convergence, complementing the scaling of the gap $1-\lambda^{(r)}$.

\newpage

\section{Class II} \label{sec:class_2}

The distinguishing feature of the rules that belong to what we refer to as Class II is the algebraic decay of the correlation functions for observables of support 1. Remarkably, the average return time grows exponentially for all of the rules in this class. We show that the algebraic decay of correlators is a consequence of the presence of local or quasilocal extensive charges. Motivated by the concept of integrability in quantum spin chains, we divide the local update rules into subclasses IIa, IIb, and IIc, corresponding to a constant number, and linear and exponential growth of the number of {\em strictly} local extensive charges.

\begin{figure}[h!]
    \centering{
    \includegraphics[width=1.0\linewidth]{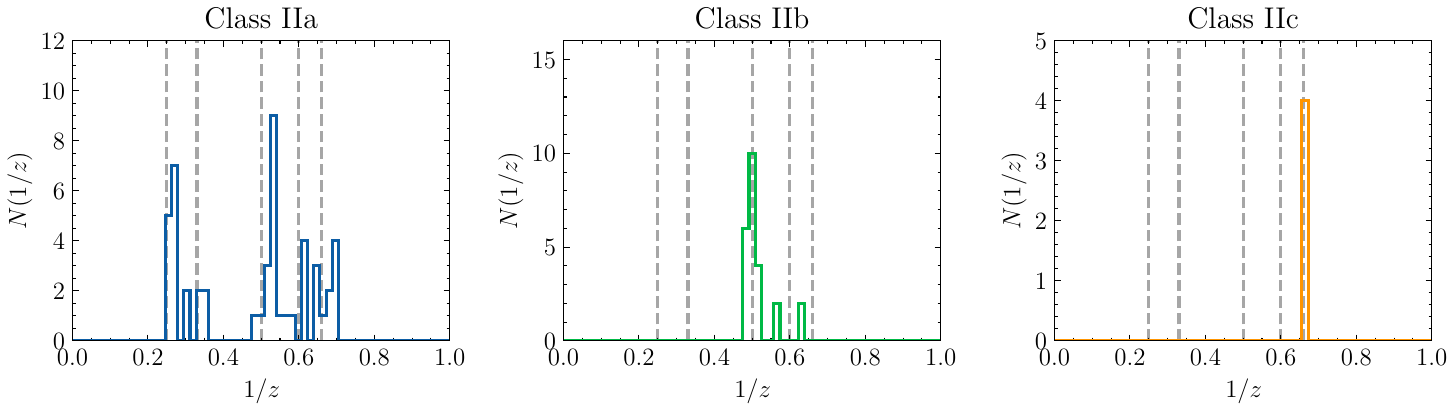}
        \caption{Histograms of all dynamical exponents $1/z_C$ and $1/z_D$ for sublcasses of Class II. The vertical dashed lines, from left to right, represent the values of  $1/z \in \{
        \frac{1}{4},\frac{1}{3}, \frac{1}{2}, \frac{3}{5}, \frac{2}{3}\}$. Note again that in all such histograms, the counting only includes the physically distinct RCA models with their `duplicates' removed. 
        }
\label{fig:dynamical_exponents_class_II}
    }
\end{figure}

Many rules in Class II exhibit `anomalous' correlations, i.e., they exhibit anomalous transport of the corresponding observable. In Figure~\ref{fig:dynamical_exponents_class_II}, we show the histogram of dynamical exponents for the models in Class II. While most of these models exhibit exponents close to $1/2$ (corresponding to diffusive transport), several models demonstrate subdiffusive and superdiffusive correlations. 

This plethora of `unconventional' dynamical exponents should, however, be interpreted with care. In some cases, we suspect the values of dynamical exponents to be the result of insufficient numerical data due to finite size effects. A correlation function usually exhibits a central peak (the so-called `heat' peak) which then spreads. In many models, this central peak is secondary to additional propagating sound-like peaks, which propagate away from the center $x=0$ with some drift velocity.

\subsection{Subclass IIa}
This subclass contains models with a constant number of charges as the range increases. Based on our definition of integrability, such models can be regarded as chaotic. In contrast to the chaotic rules from Class I, which have no conserved charges, models in Subclass IIa have at least one local or quasilocal charge. The presence of charges constrains the decay of correlation functions of local observables that overlap with the charge density. For the rules in Class IIa, where the correlators of range-1 observables decay exponentially, we find range-2 observables that exhibit algebraic decay of the corresponding correlators due to the overlap with charge density. Here, we discuss a number of interesting physical properties of these models, in particular, the existence of quasilocal charges, superdiffusive and subdiffusive transport (for some models with rather unusual dynamical exponent $z=3$), and exponential decay of range-1 observables. 
\\

\textbf{Quasilocal charges}. Despite exhibiting algebraic decay of correlators, some of the models do not possess any strictly local charges. Instead, it is the quasilocal charges that are responsible for slowly decaying correlations. By examining the spectrum of the transfer matrix $\mathcal{T}^{(r)}$ with increasing range $r$, we observe several eigenvalues that approach the eigenvalue 1 with the gap closing exponentially in $r$. An example is the $\mathcal{CP}$-symmetric model with rule 32546187.
    
\begin{figure}[h!]
\centering
\includegraphics[width=0.55\linewidth]{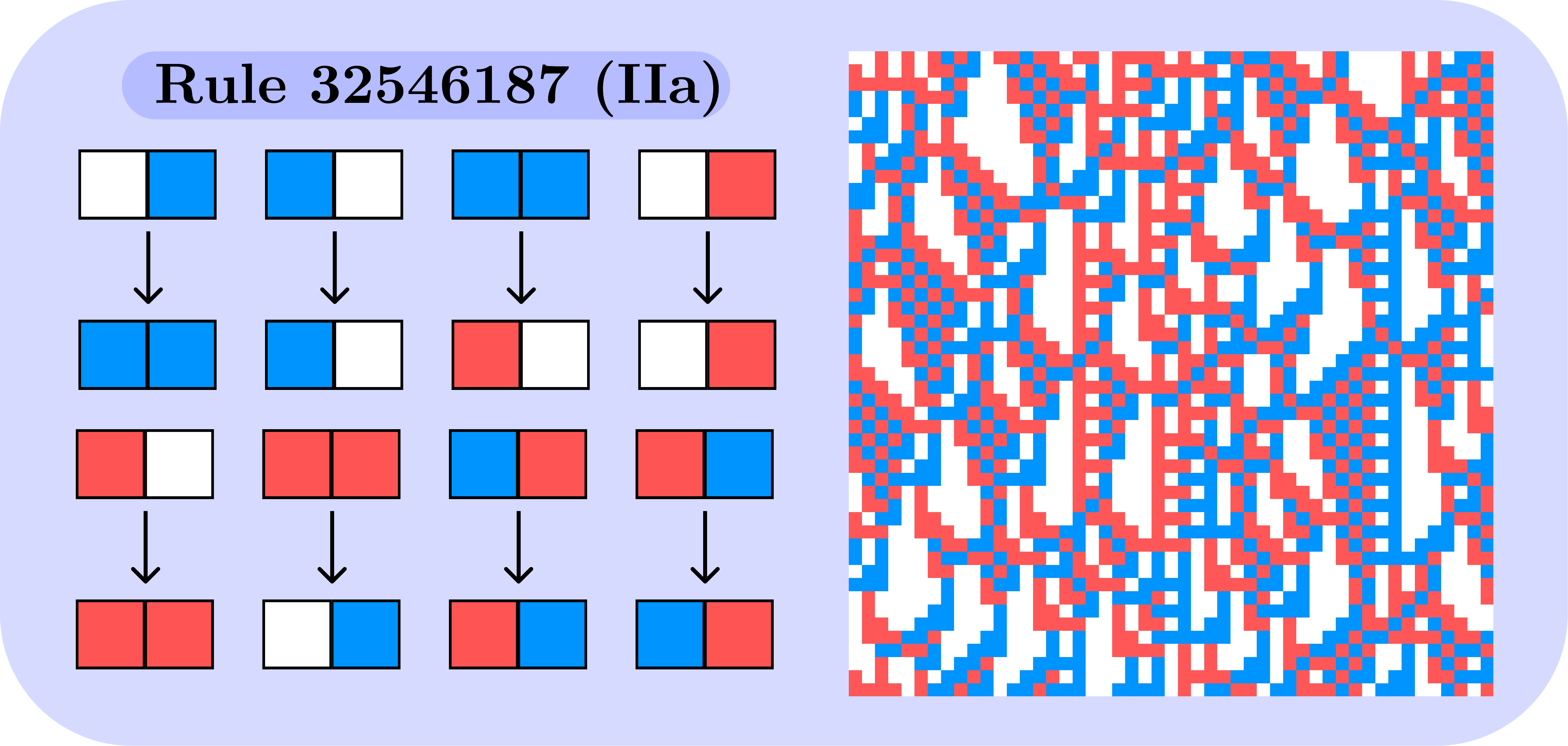}
  \label{fig:class_IIb_dynamics}
  \label{IIb_rule}
\end{figure}
    
\begin{figure}[h!]
\centerline{
    \hspace{-0.3cm}
  \includegraphics[width=0.5\linewidth]{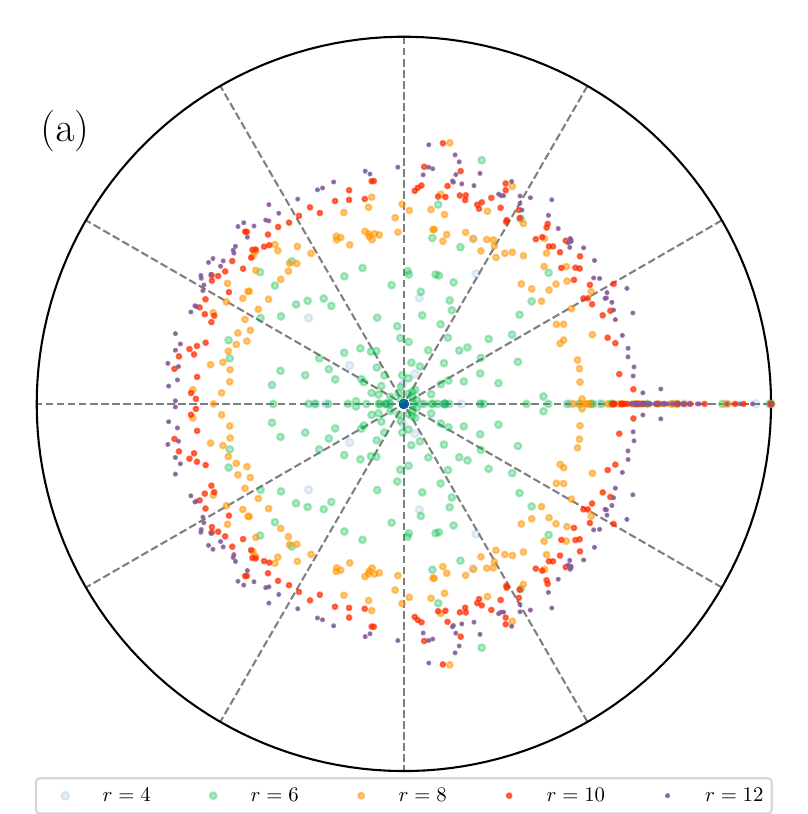}

  \hspace{-0.25cm}
   \includegraphics[width=0.5\linewidth]{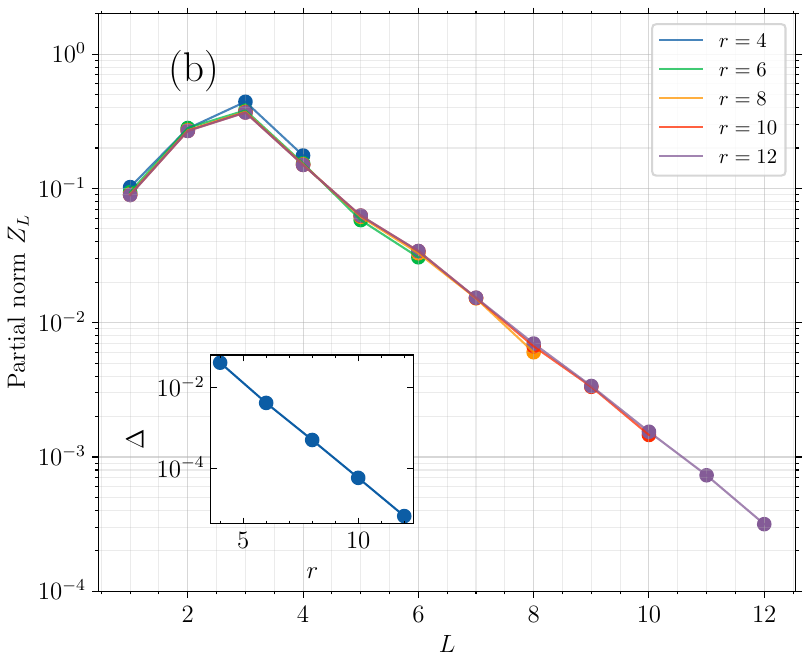}
}
  \caption{$\mathcal{CP}$-symmetric cellular automaton  with rule 32546187. \textbf{(a)} Spectrum of the transfer matrix for increasing $r$. \textbf{(b)} Partial norms of the eigenvector associated with the gap for increasing $L$. The inset shows the exponential decay of the gap with increasing support $r$.  
  }
  \label{fig:quasilocal}
\end{figure}

In Figure~\ref{fig:quasilocal}, we present the spectrum of the transfer matrix for this model, along with the exponential closing of the gap. In addition, we investigate the locality properties of the eigenvector corresponding to the eigenvalue closest to 1. We present the decay of partial norms of the eigenvector with the eigenvalue closest to 1 in Figure~\ref{fig:quasilocal}. This exponential decay of the partial norms affirms the quasilocal nature of the charge.
 
Similarly as in quantum spin chains, quasilocal charges significantly influence the transport properties of the system (see Refs.~\cite{ilievski2016quasilocal,TomazEnej,Enej2015}). However, unlike in quantum spin chains with quasilocal charges, such as the XXZ model, we observe quasilocal charges even in models lacking strictly local charges. We note that a model with an extensive number of quasilocal charges but without any strictly local charges actually exists in Class III. It will be discussed in Section~\ref{sec:class_3a}. 
\\

{\bf Superdiffusive transport.} Models in Subclass IIa can exhibit anomalous transport properties quantified by their correlation functions. An example is the $\mathcal{C}\mathcal{P}$- and $\mathcal{P}\mathcal{T}$-symmetric cellular automaton with rule 35216478. 

\begin{figure}[h!]
\centering
\includegraphics[width=0.55\linewidth]{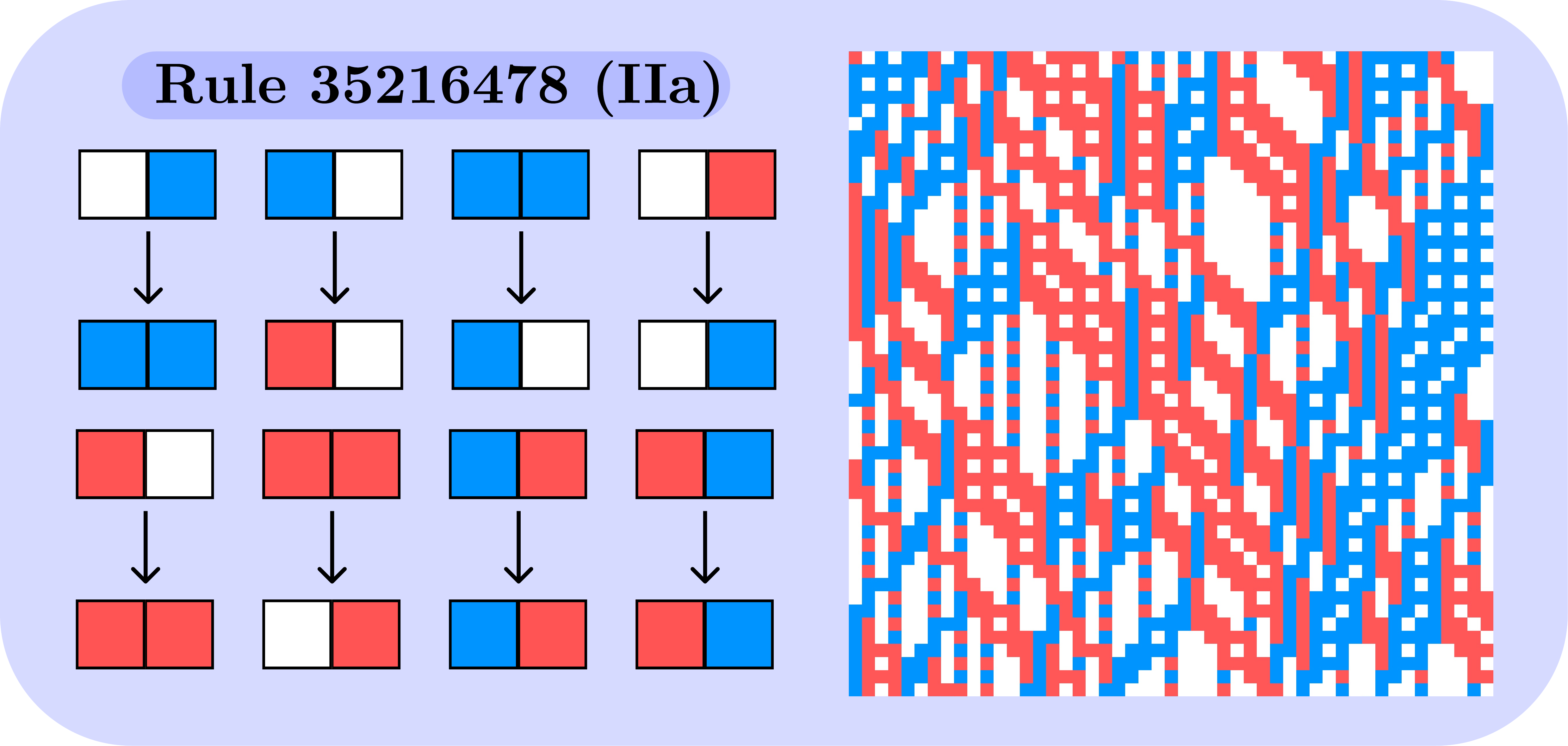}
  \label{fig:class_IIb_dynamics}
  \label{IIb_rule}
\end{figure}

\begin{figure}[h!]
\centering
\includegraphics[width=1.0\linewidth]{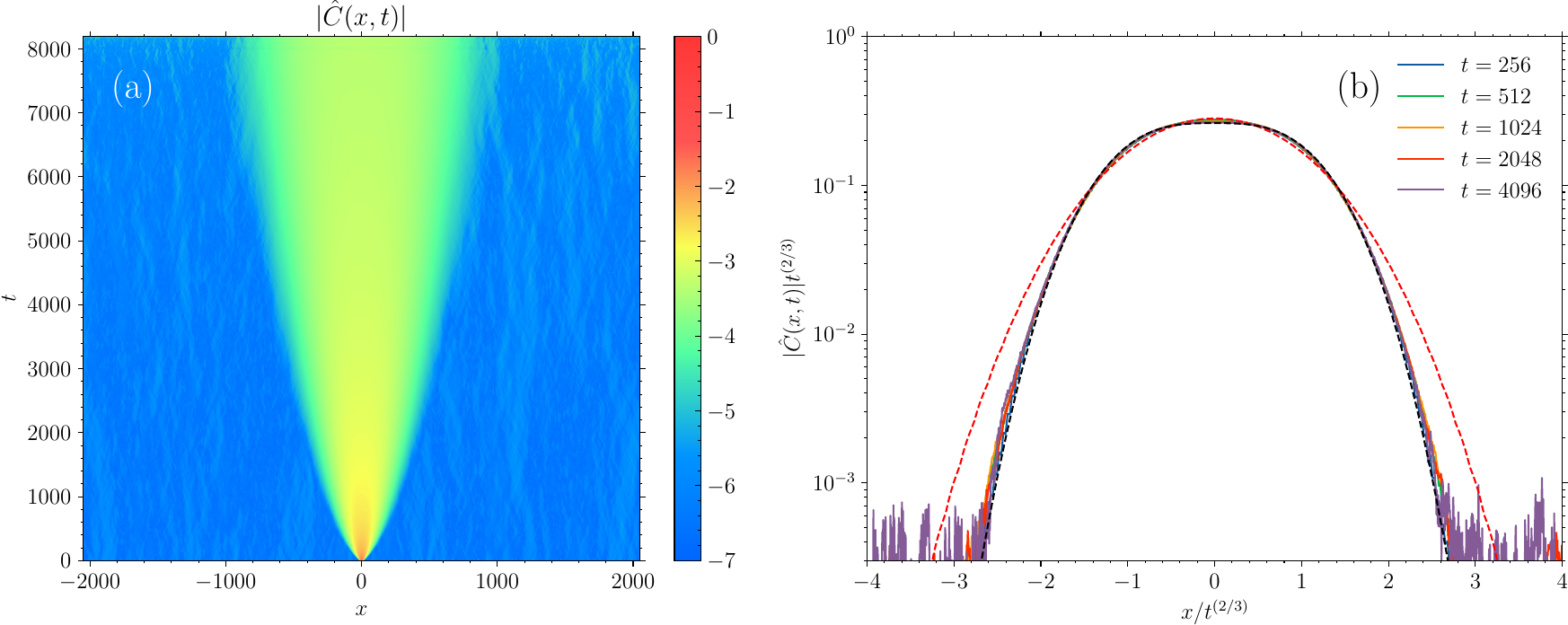}
  \caption{$\mathcal{C}\mathcal{P}$- and $\mathcal{P}\mathcal{T}$-symmetric rule 35216478.\textbf{(a)} $\hat{C}(x,t)$. \textbf{(b)} Cross-sections of $\hat{C}(x,t)$ at increasing fixed times $t$ denoted in the legend. The red dashed line shows the best fit of the Prahofer-Spohn scaling function $f_{PS}$, while the black dashed line shows a fit of $\mathrm{exp}(-c|x|^3)$ with $c\approx0.2621 \pm 10^{-5}.$
  }
  \label{fig:kpz}
\end{figure}

The spectrum of the transfer matrix for this model was plotted above in Figure~\ref{fig:spectra_alt}. It shows that the rule has three local conserved quantities and no quasilocal charges. The rule has an extensive charge with a density of support 1: $Q_1=\sum_{x}[+]_x-[-]_x$. Indeed, as one can simply infer from the elementary update rule, the difference between the number of red and blue cells is conserved. The correlation function $\hat{C}(x,t)$ then reveals that the transport of the charge $Q_1$ is superdiffusive with the KPZ-like dynamical exponent $1 / z_C \approx 2/3$ (see Figure~\ref{fig:kpz}).

In Figure~\ref{fig:kpz}, we compare $\hat C(x,t)$ at increasing times $t$ to the  Prähofer–Spohn scaling function $f_{PS}$ \cite{prahofer2004exact}, which is associated with the scaling of the correlation function in models belonging to the KPZ universality class. We find that this function does not fit well with our computed correlator. We observe a scaling function that is significantly flatter at $x = 0$ and decays more rapidly with increasing $x$ than either the Gaussian or KPZ profiles.

\begin{figure}[h!]
\centering
\includegraphics[width=1.0\linewidth]{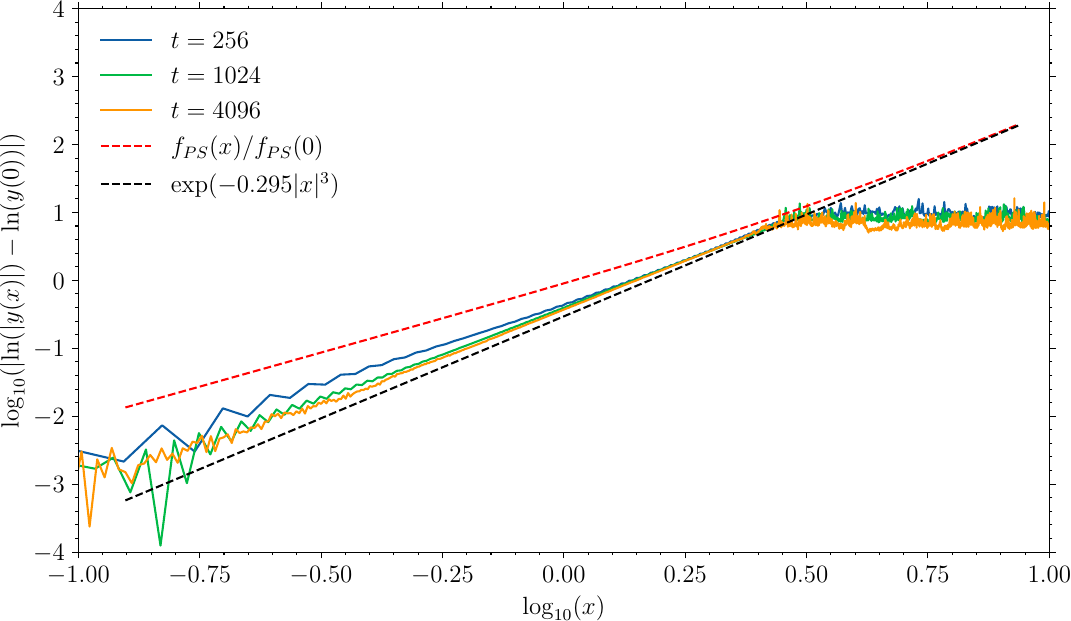}
  \caption{Asymptotic behavior of the scaling function. Linearized scaling functions for $x>0$, fitted to $y(x) = A e^{-x^n}$, for increasing times $t$. The slope of the scaling function is proportional to $n$. The red and black dashed lines represent the  Prahofer-Spohn scaling function $f_{PS}$ and its asymptotic value $n=3$ \cite{prahofer2004exact}, respectively. 
  }
  \label{fig:kpz_asymptotics}
\end{figure}

The asymptotic $x\to\infty$ behavior of the scaling function is of the form $y(x) = A e^{-x^n}$. To analyze it we linearize $y(x>0)$ at increasing times, as shown in Figure~\ref{fig:kpz_asymptotics}. We observe that, asymptotically, the scaling function decays like the Prahofer-Spohn scaling function with a slope $n=3$. We find that at small $x$ the scaling function also seems to approach $n=3$, suggesting the profile $\mathrm{exp}(-c|x|^3)$. The fit of this profile is in excellent agreement with our data, and it is shown in black in Figure~\ref{fig:kpz}.

\begin{figure}[h!]
\centering
\includegraphics[width=0.6\linewidth]{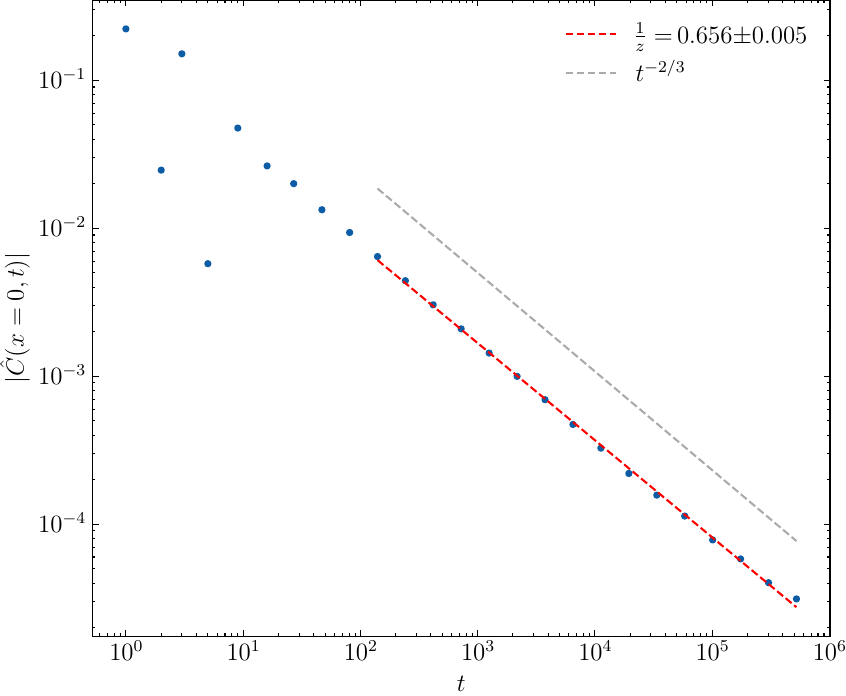}
  \caption{Extended simulation of the CA with rule 35216478. The decay of the correlator at $x=0$ is best fit with $1/z=0.656\pm0.005$. The dashed gray line indicates a decay of $t^{-2/3}$ to guide the eye.
  }
  \label{fig:kpz_long_times}
\end{figure}

Because our simulations have so far been done using a relatively small number of time steps (up to $t=4096$ for systems of similar size), here, we perform a more extensive simulation to confirm the existence of the anomalous dynamical exponent $z=3/2$. In particular, we compute the peak of the correlation function $\hat{C}(0,t)$ for times up to $t=2^{19}$ and for a system with size $2^{20}$. The results are shown in Figure~\ref{fig:kpz_long_times}, confirming that the model also exhibits the dynamical exponent $z=3/2$ at late times.

\textbf{Subdiffusive Transport with $z=3$}. Some chaotic models in Class IIa exhibit a slower decay of correlators than the diffusive $1/z=1/2$. An example of such a subdiffusive model is the cellular automaton with rule 45621378. This model is $\mathcal{P}$ and $\mathcal{C}\mathcal{T}$ symmetric.

\begin{figure}[h!]
\centering
\includegraphics[width=0.55\linewidth]{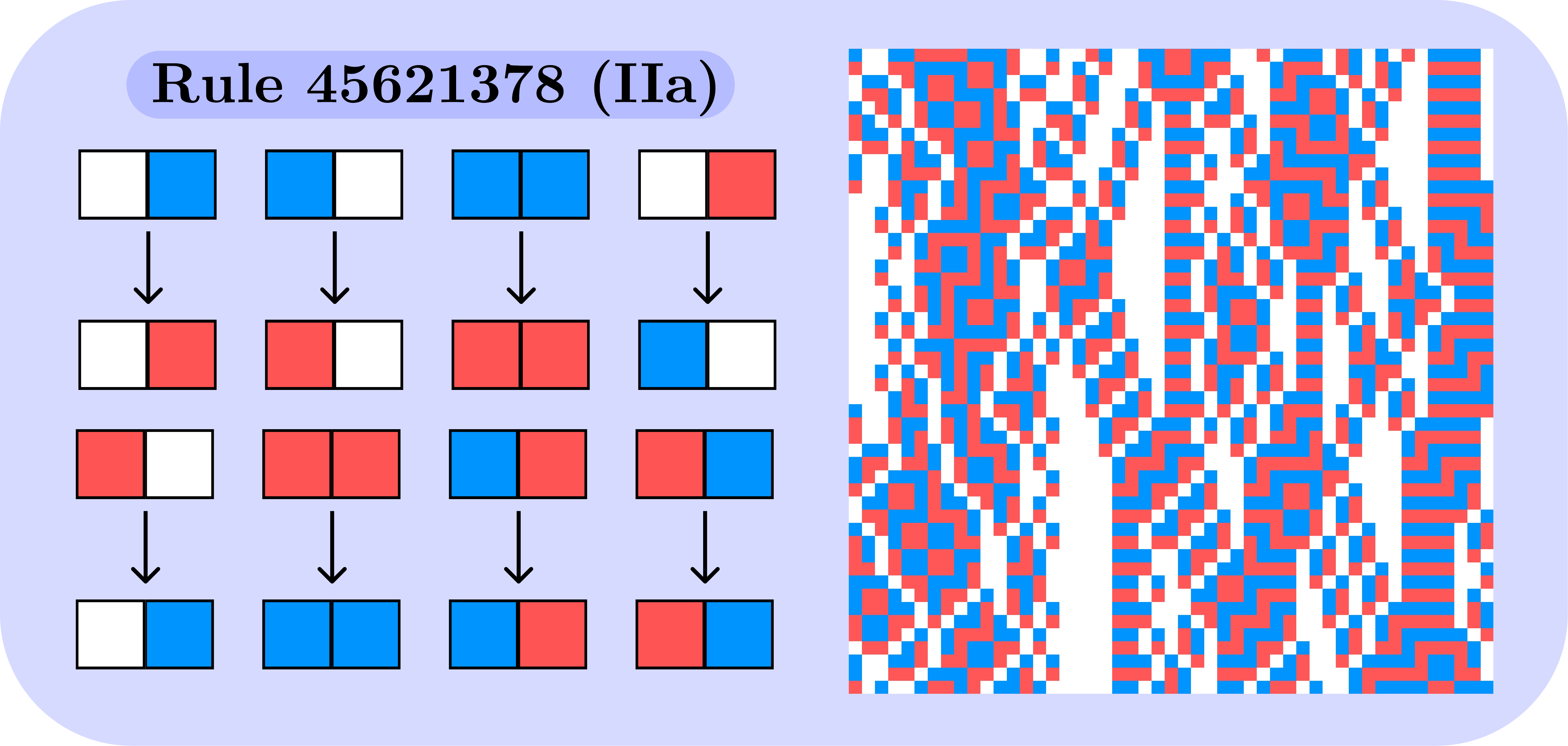}
  \label{fig:class_IIb_dynamics_2}
  \label{IIb_rule}
\end{figure}

\begin{figure}[h!]
\centering
\includegraphics[width=1.0\linewidth]{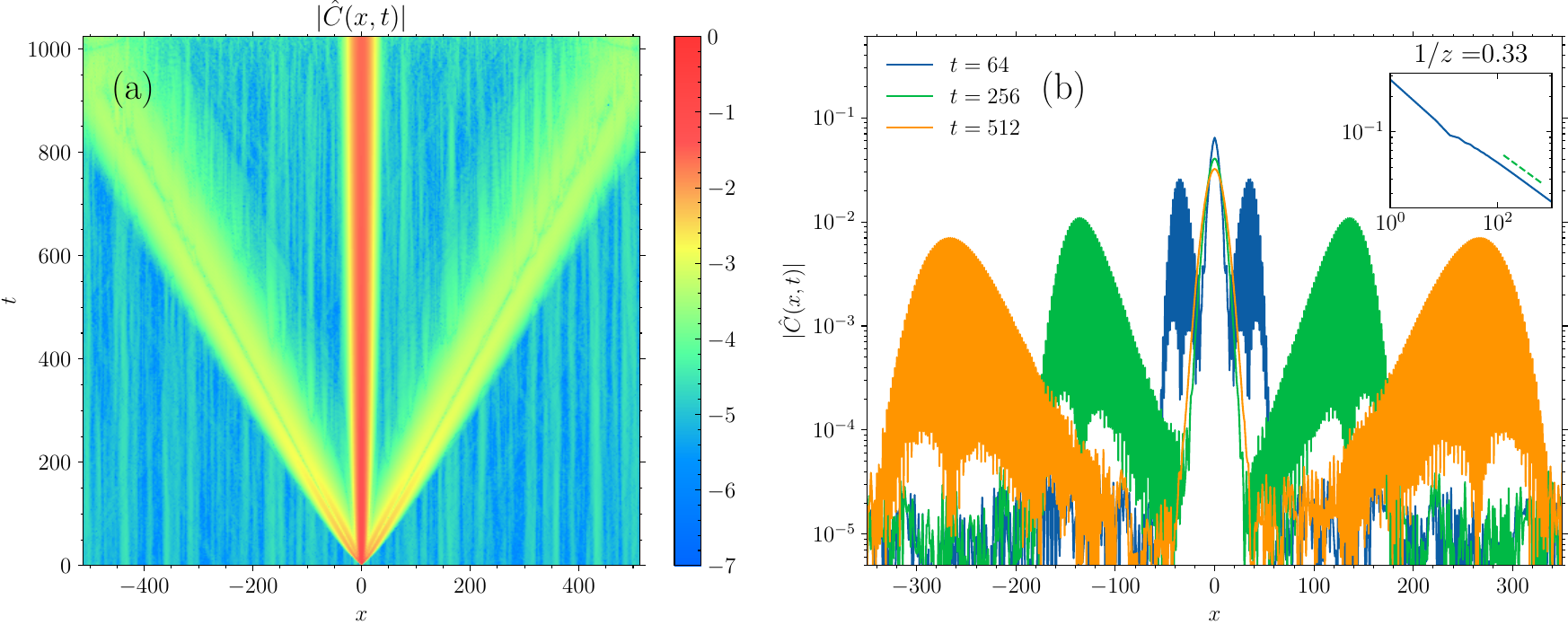}
\includegraphics[width=1.0\linewidth]{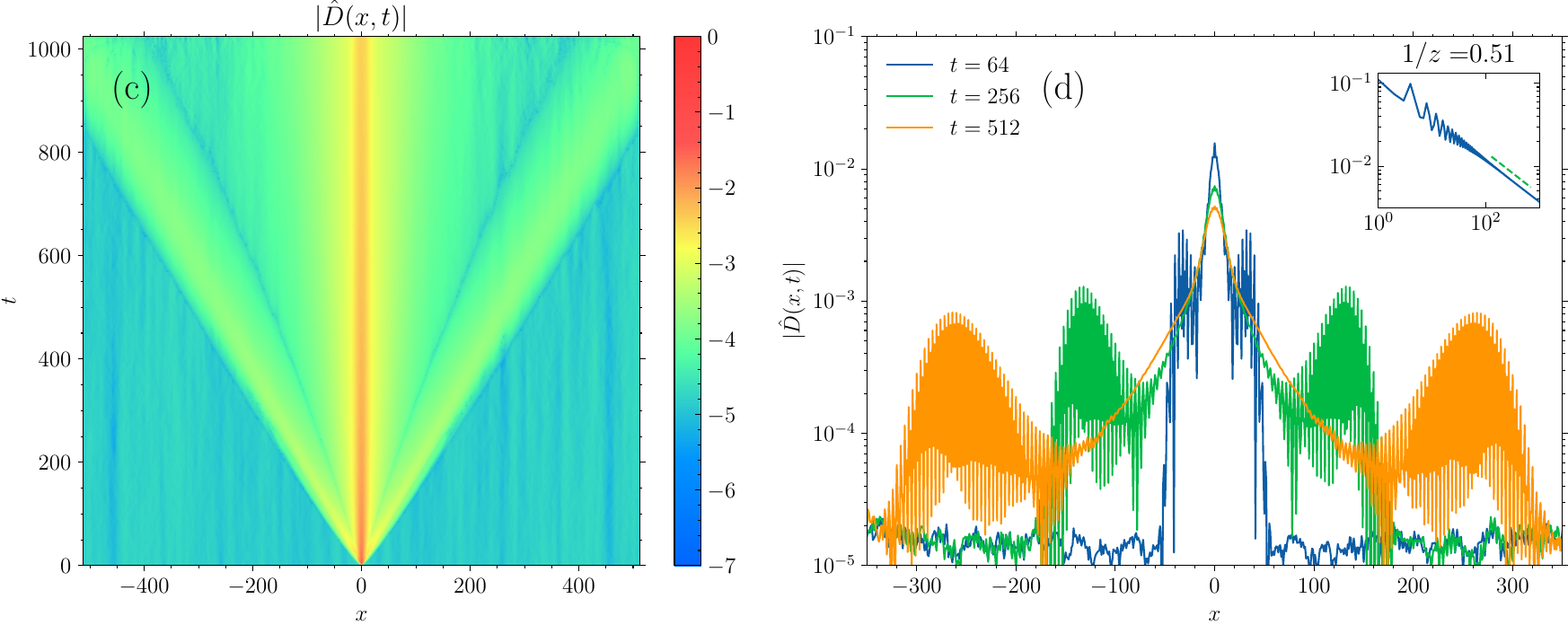}
\vspace{-15pt}
  \caption{Rule 45621378. \textbf{(a)} $\hat{C}(x,t)$ of the observable $[+]-[-]$, \textbf{(b)} Cross-sections of $\hat{C}(x,t)$ at increasing $t$. The inset shows the $\hat{C}(0,t) \sim t^{-1/z}$, with the dashed green line showing the best fit $1/z_C=0.33$. \textbf{(c)} $\hat{D}(x,t)$ of the observable $[\varnothing]$. \textbf{(d)} Cross-sections of $\hat{D}(x,t)$ at increasing $t$, with the inset showing the decay of the maximum over $x$.
  }
  \vspace{-10pt}
  \label{fig:corr_45621378}
\end{figure}

The model has three local extensive charges: two with a density of support 1 and one with a support of 4. From the local update rule, two independent charge densities of range 1 can be identified, namely, $q_1^{(1)}= [+]-[-]$ and $q_2^{(1)}= [\varnothing]$. These conserved quantities correspond to the conservation of the total number of each `particle' species. In Figure~\ref{fig:corr_45621378} we demonstrate the transport of the charges.

Due to $\mathcal{P}$ symmetry, the correlation function is symmetric with respect to $x \rightarrow -x$. For the density of the first charge, the correlator has a central peak, which corresponds to subdiffusive transport $\hat{C}(0,t) \sim t^{-1/3}$ and additional propagating sound modes. The correlator of the density of the second charge $D(x,t)$ decays with the dynamical exponent $z=2$. It also exhibits a pair of sound modes.

\begin{figure}[h!]
\centering
\includegraphics[width=0.6\linewidth]{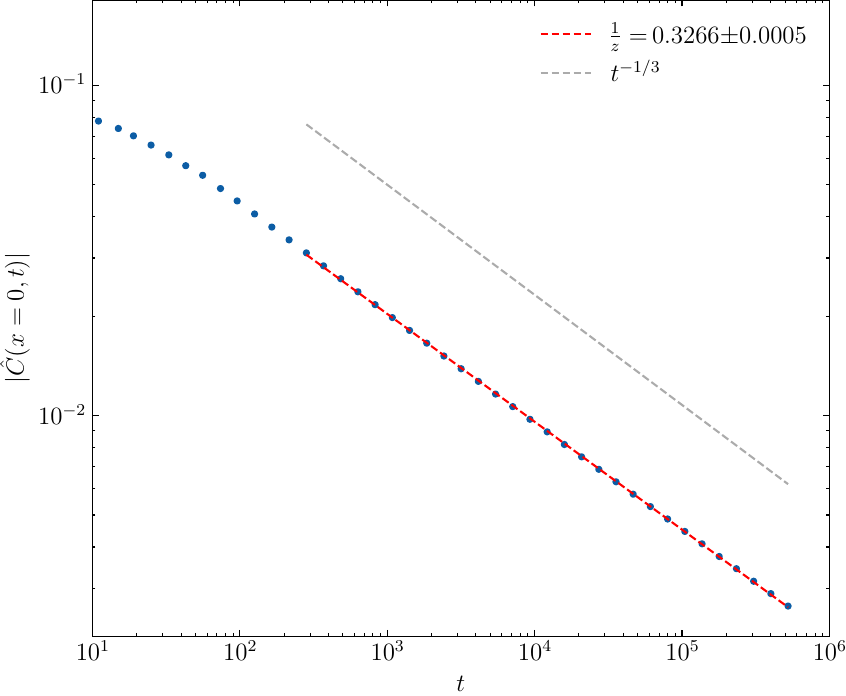}
\vspace{-5pt}
  \caption{Decay of the correlator $\hat{C}(x,t)$ of the CA with rule 45621378 at $x=0$ with an algebraic best fit $1/z =0.3266\pm0.0005$. The dashed gray line denotes a decay of $t^{-1/3}$ to guide the eye.}
  \vspace{-5pt}
  \label{fig:subdiff_long_times}
\end{figure}

As in the previous example, we examine whether this type of transport persists for asymptotically long times. We perform an extended simulation for a system of size $2^{20}$, up to times of $2^{19}$. The results are shown in Figure~\ref{fig:subdiff_long_times} confirming that the dynamical exponent $z = 3$ remains valid on much longer timescales.

Subdiffusion is often associated with the conservation of dipole moments of some conserved charge density or higher multipole moments \cite{PollmannSub,SubStrong,Nandy2024}. However, the conservation of a dipole moment typically results in a larger dynamical exponent $z=4$. Subdiffusive transport has also been found in different spin chain models \cite{DeRoeck2020,Singh2021,DeNardis2022}, where the interplay between interactions, disorder, and boundary effects plays a crucial role in the observed dynamics. We also find a number of models in this class for which the dynamical exponents converge to $z=4$ (see Appendices \ref{app:uc} and \ref{AppD}).
Understanding the various microscopic mechanisms responsible for the distinct subdiffusive transport remains an open question.
\\

\textbf{Exponential decay of range-1 observables}. Certain models in Class IIa exhibit exponential decay of the correlation function for both independent range-1 observables. These rules have a constant number of charges, which would typically be expected to result in slower decay. An example is the $\CC\CP\CT$-symmetric cellular automaton with rule 13825476. 
\begin{figure}[h!]
\centering
\includegraphics[width=0.55\linewidth]{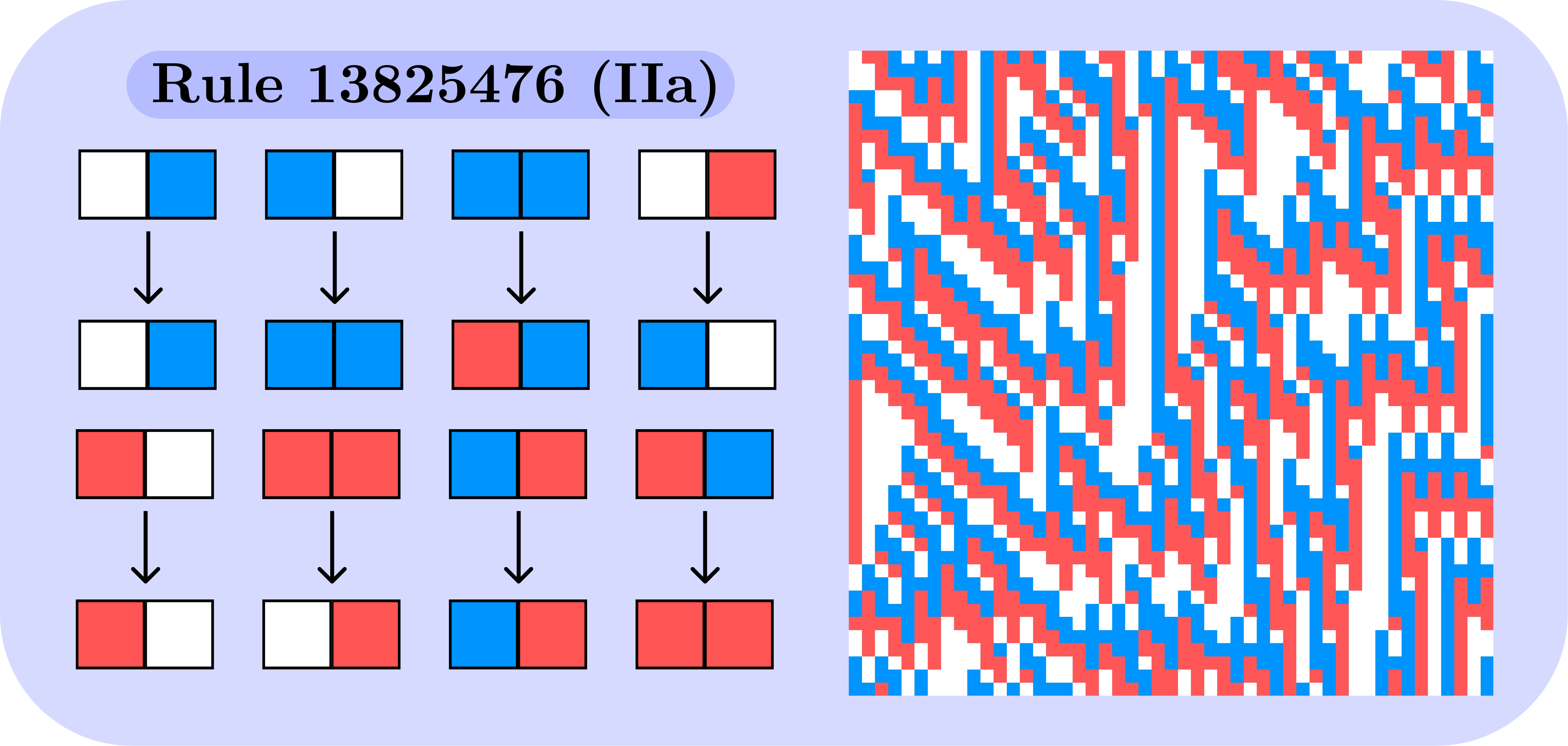}
  \label{fig:class_IIa_exp}
\end{figure}

\begin{figure}[h!]
\centering
\includegraphics[width=1.0\linewidth]{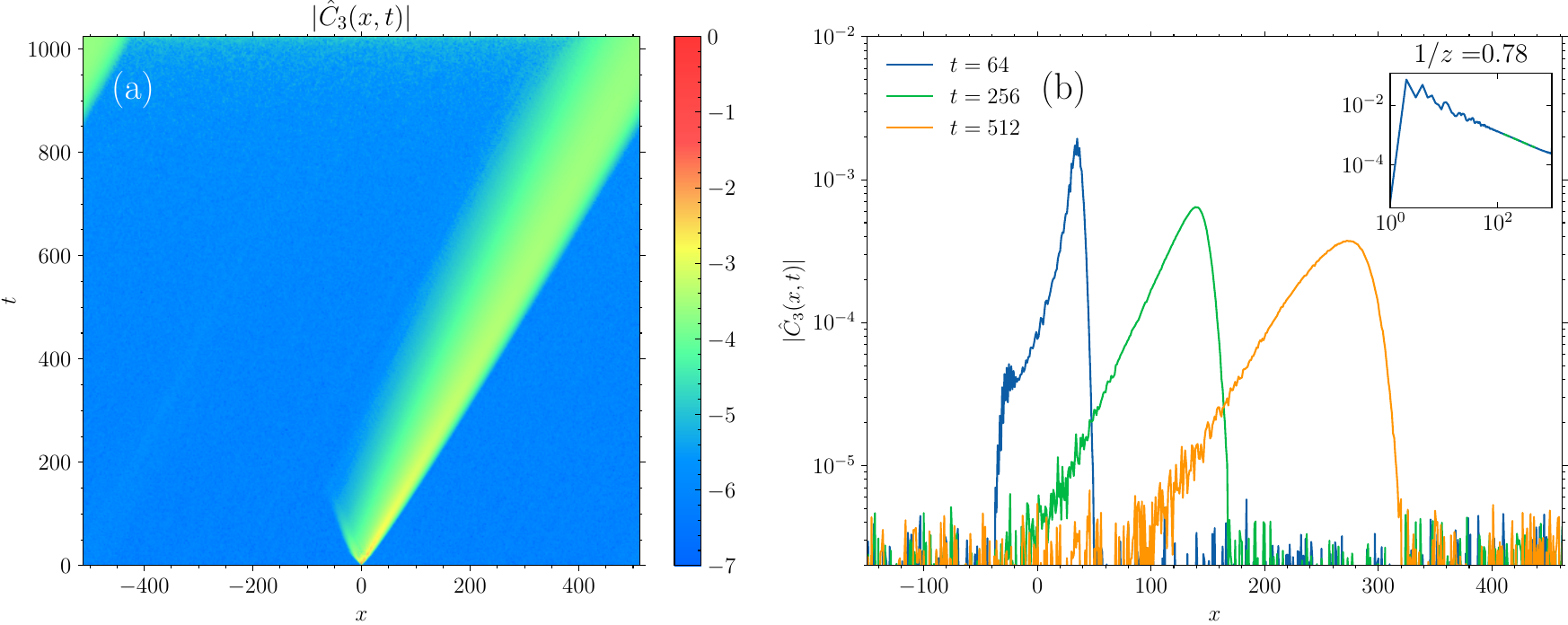}
  \caption{$\mathcal{CPT}$-symmetric cellular automaton with rule 13825476. \textbf{(a)} $\hat{C}_3(x,t)= \langle [11]_x(t) [11]_0(0) \rangle_\infty^C$ of an observable with support 2. \textbf{(b)} Cross-sections of $\hat{C}_3(x,t)$ at increasing times. The inset shows $C_3(t) \sim t^{1/z}$ with the dashed green line showing the best fit of $1/z=0.78$.
  }
  \label{fig:corr_13825476}
\end{figure}

This model has only one conserved charge density, which is of range $4$:
\begin{align}\label{eq:charge}
    q_1^{(4)}=&\,\, 3 [1 1]+3 [2 1]-3 [1 2] + [2 2]+3 [0 1 1 ]+[0 2 2]-3 [0 1 2]+3 [0 1 0 1] -3[0 1 0 2] \nonumber\\
    &-3[0 1 1 2]-3 [0 1 2 1]+3[0 2 0 1]+[0 2 0 2]+3[0 2 1]-3[0 2 1 1]+[0 2 2 2],
\end{align}

where we use an orthogonal basis of observables 
\begin{equation}    
[0]=[\varnothing]+[+]+[-] \equiv \mathds{1},\quad [1]=[+]-[-], \quad [2]=-2[\varnothing]+[+]+[-].
\end{equation}
Notice that the charge does not have any overlap with the range-1 observables since all terms in Eq.~\eqref{eq:charge} have higher support. Due to the absence of additional conserved quantities and a zero overlap of range-1 observables and charge, the correlation function decays exponentially. To show the slower decay of higher range observables, in Figure~\ref{fig:corr_13825476}, we plot the correlation function of the observable $\tilde{\mathcal{O}}=[11]_0$, which non-trivially overlaps with the charge, so that $ \langle Q_4,\tilde{\mathcal{O}} \rangle \neq 0 $.

\vspace{-10pt}
\subsection{Subclass IIb}
This subclass contains models with a linearly growing number of local charges, $N_Q \sim r$.\footnote{We remark that, due to a finite accessible support $r$, such behavior can only be conjectured.} Due to the presence of an extensive number of local conserved quantities, we will refer to these models as integrable, however, their local rules $U$ do not satisfy the set-theoretic Yang-Baxter equation and a possible theoretical framework for studying their integrability remains to be understood. Regardless of the presence of conserved charges, the average return time for all integrable rules increases exponentially with the system size.

Most models in Class IIb exhibit correlation functions characterized by a dynamical exponent close to $z=2$; see Figure~\ref{fig:dynamical_exponents_class_II}. However, certain models also exhibit anomalous correlations with dynamical exponents that do not match diffusive, ballistic or KPZ-like behavior. Here, we show an example of a diffusive and a superdiffusive model. 

\newpage
\textbf{Diffusive transport.} An example of a diffusive model is the cellular automaton with rule 26534178, which possesses the $\mathcal{CP}+\mathcal{PT}$ symmetry.

\begin{figure}[h!]
\centering
\includegraphics[width=0.55\linewidth]{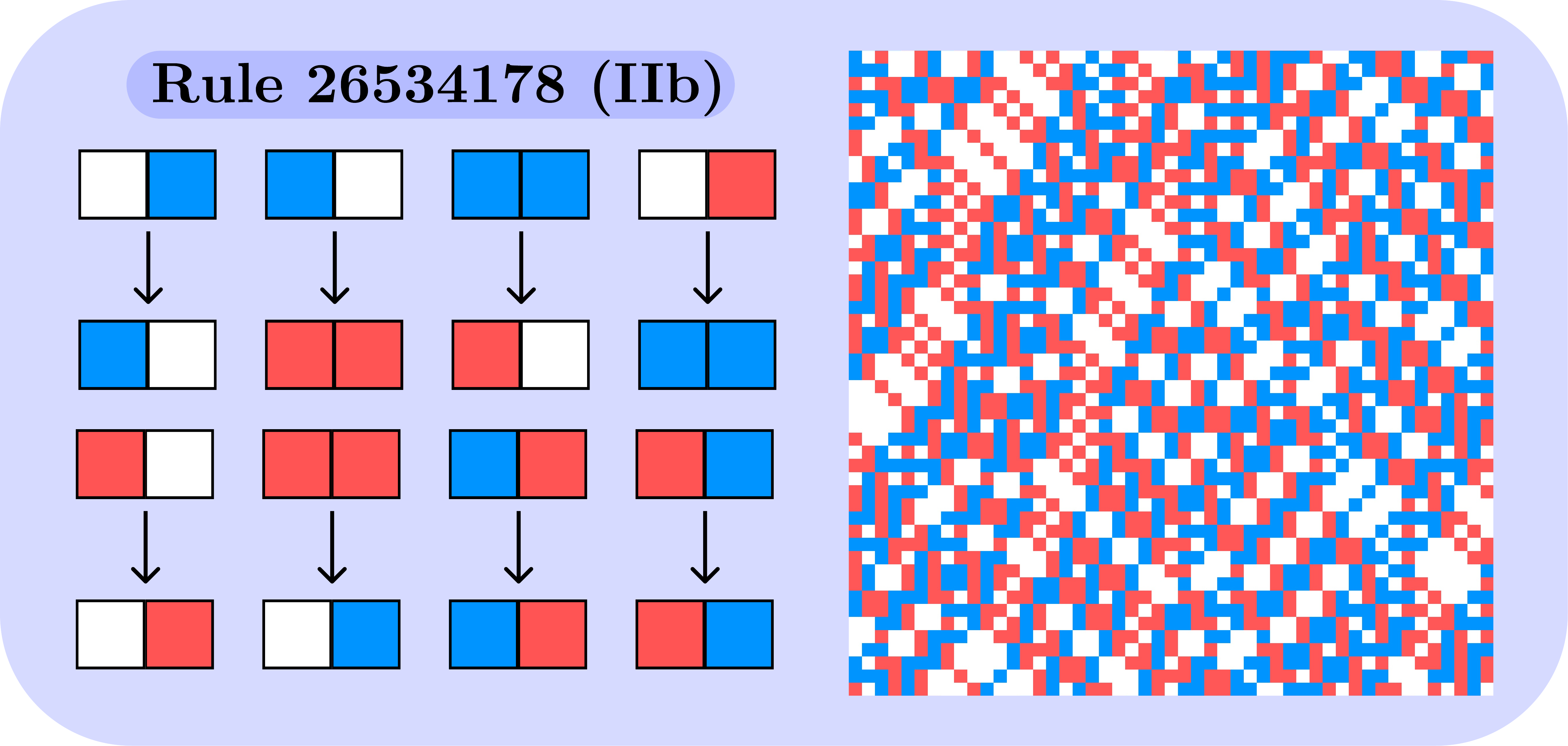}
  \label{fig:class_IIb_dynamics}
  \label{IIb_rule}
\end{figure}

The number of conserved charges indeed grows linearly with range $r$.
\begin{table}[h!]
\centering
\begin{tabular}{|p{3.5cm}|p{0.8cm}|p{0.8cm}|p{0.8cm}|p{0.8cm}|p{0.8cm}|p{0.8cm}|p{0.8cm}|}
\hline
Range $r$& 2 & 4 & 6 & 8 & 10 & 12 & 14 \\ \hline
Number of charges & 2 & 4 & 4 & 4 & 6 & 6 & 6 \\ \hline
\end{tabular}
\end{table}

\begin{figure}[h!]
\centering
\includegraphics[width=1.0\linewidth]{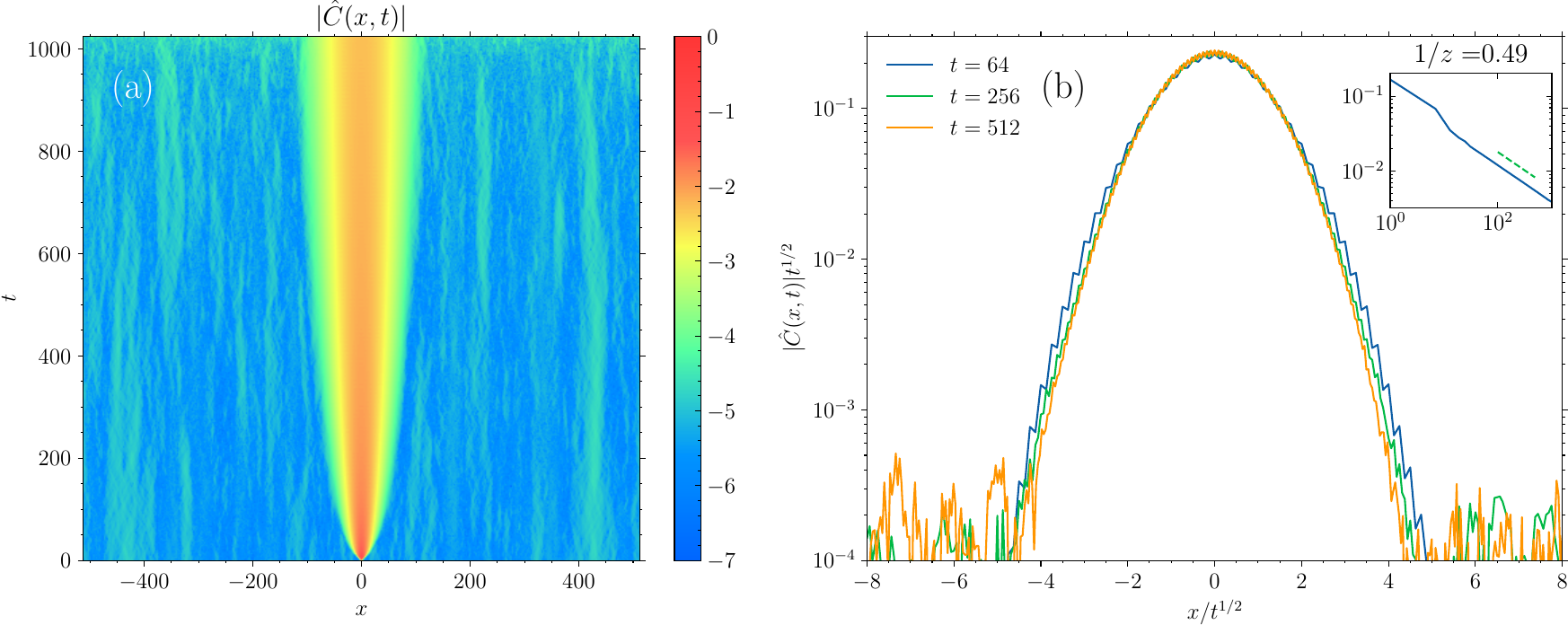}
  \caption{Integrable $\mathcal{CP+PT}$-symmetric cellular automaton with rule 26534178 with diffusive transport. \textbf{(a)} $\hat{C}(x,t)$ of the density of charge $[+]-[-]$. \textbf{(b)} Cross-sections of $\hat{C}(x,t)$ at increasing times. Note the rescaling of the axes by $t^{1/2}$. The insets show $\hat{C}(x=0,t)$ with the best fitting dynamical exponent $1/z=0.49$.
  }
  \label{fig:integrable_diffusion}
\end{figure}

One of the simplest charges is given by translations of the local density $ [+] - [-]$. The transport of this charge is depicted in Figure~\ref{fig:integrable_diffusion}, where a spreading Gaussian peak associated with diffusion is observed.

\textbf{Superdiffusive transport.} One example of a model with anomalous correlations and superdiffusive transport is rule 25314678 from the $\mathcal{CP}+\mathcal{PT}$ symmetry class.

\begin{figure}[h!]
\centering
\includegraphics[width=0.55\linewidth]{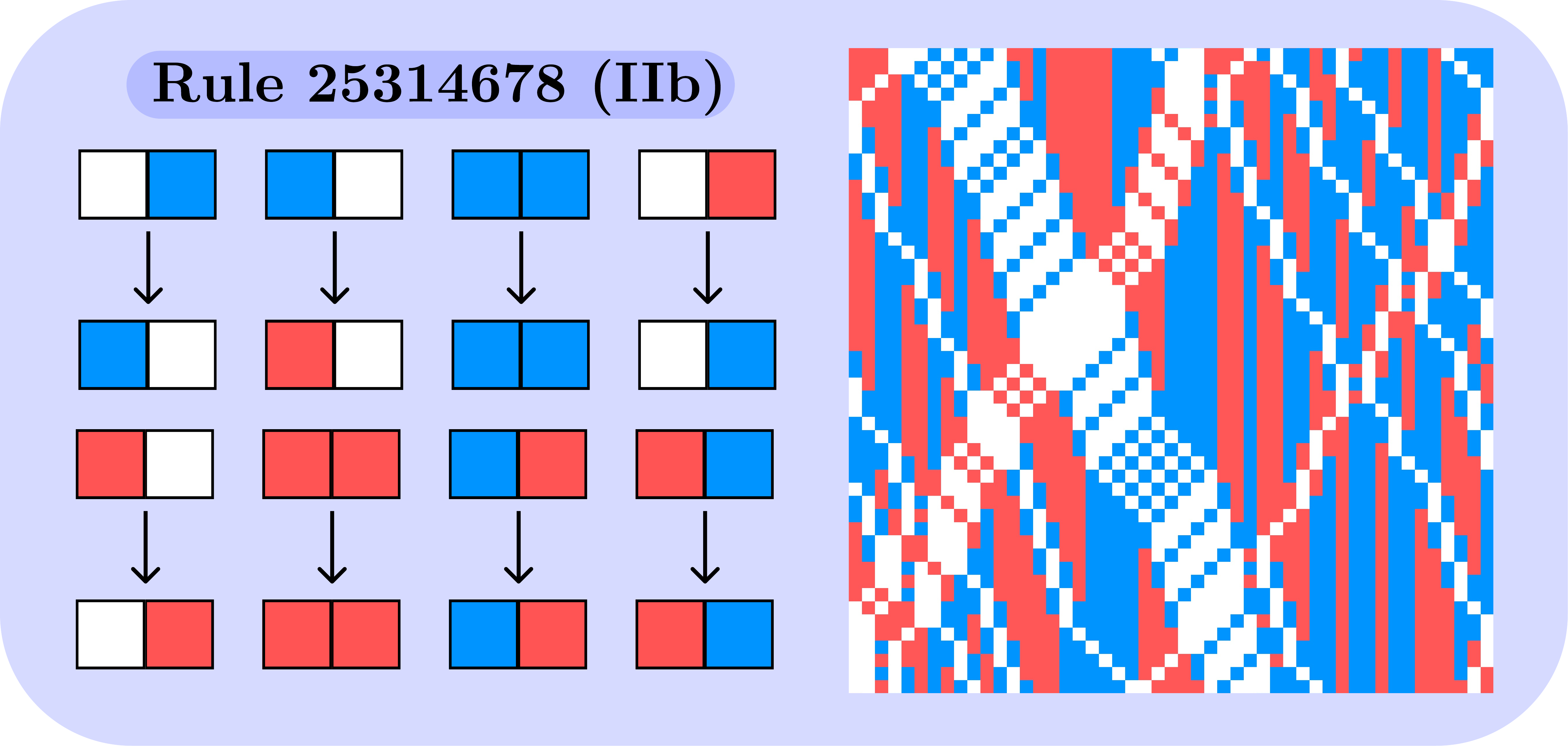}
  \label{fig:class_IIb_dynamics}
  \label{IIb_rule}
\end{figure}

Similarly to the diffusive example above, the number of strictly local conserved quantities appears to increase linearly. 

\begin{table}[h!]
\centering
\begin{tabular}{|p{3.5cm}|p{0.8cm}|p{0.8cm}|p{0.8cm}|p{0.8cm}|p{0.8cm}|p{0.8cm}|p{0.8cm}|}
\hline
Range $r$& 2 & 4 & 6 & 8 & 10 & 12 & 14 \\ \hline
Number of charges & 2 & 2 & 4 & 4 & 6 & 6 & 8 \\ \hline
\end{tabular}
\end{table}

\begin{figure}[h!]
\centering
\includegraphics[width=1.0\linewidth]{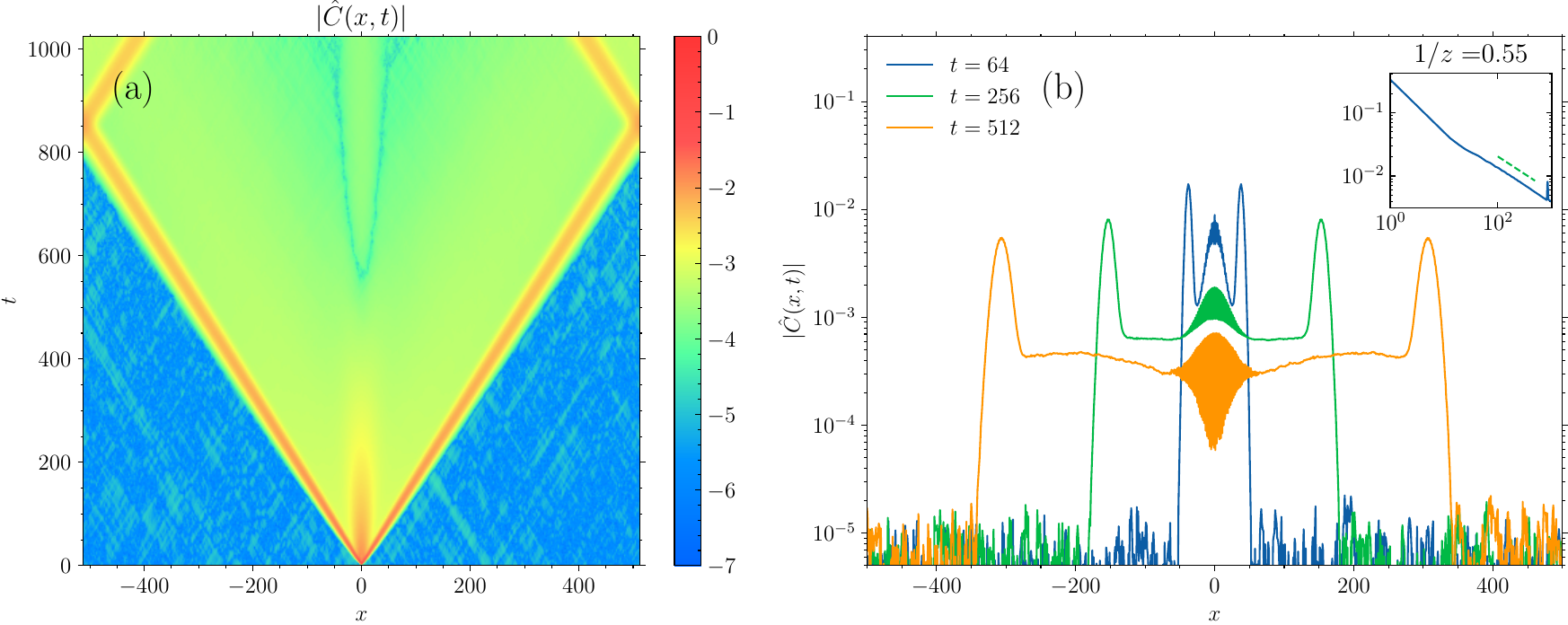}
\includegraphics[width=1.0\linewidth]{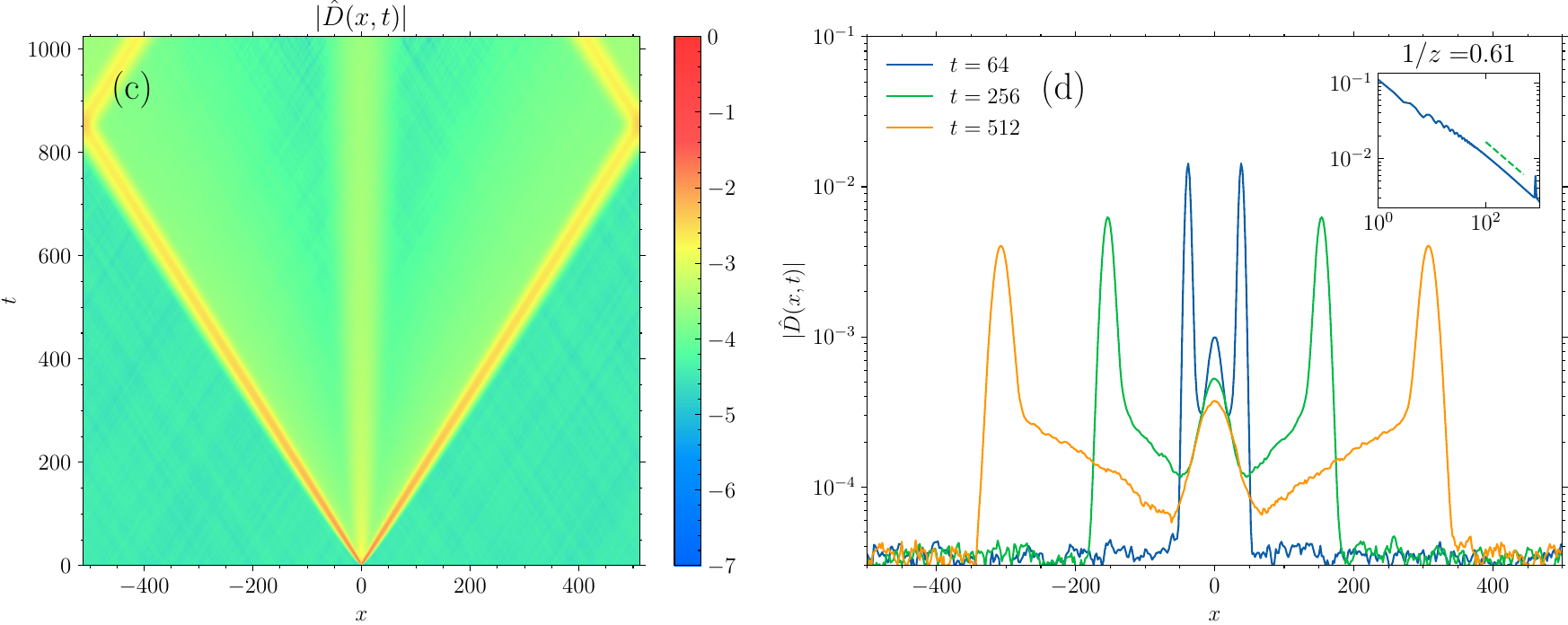}
  \caption{Correlation functions of the integrable $\mathcal{CP+PT}$-symmetric cellular automaton with rule 25314678. \textbf{(a)} $\hat{C}(x,t)$ of the observable $[+]-[-]$, \textbf{(b)} Cross-sections of $\hat{C}(x,t)$ at increasing $t$. The inset shows the $C(t) \sim t^{-1/z}$, with the dashed green line showing the best fit $1/z_C=0.55$. \textbf{(c)} $\hat{D}(x,t)$ of the observable $[\varnothing]$, which is a density of a conserved charge of this model. \textbf{(d)} Cross-sections of $\hat{D}(x,t)$ at increasing $t$, with the inset showing the decay of the maximum over $x$ with dynamical exponent $1/z_D = 0.61$.}
  \label{fig:integrable}
\end{figure}

The simplest charge of range 1 is a consequence of the conservation of the total number sites in the `vacuum' state: $\hat{Q}_1=\sum_x [\varnothing]_x$. Within our algorithm, this charge is identified as a range-2 charge density $q_1^{(2)} = [20]+[02]$, while the second charge of support 2 has density $q_2^{(2)} = [10]+[01]-\frac{2}{3}[02]$. Here, we demonstrate the transport of $q_1^{(2)}$ in Figure~\ref{fig:integrable}. It can be seen that the correlation function decays as $D(t) \sim t^{-0.62}$, and that there also exists an additional central peak that decays as $\sim t^{-1/2}$.

\clearpage
\subsection{Subclass IIc}

In this subclass, we have found two superintegrable rules 16435287 and 45378621. 

\begin{figure}[h!]
\centering
\includegraphics[width=1.0\linewidth]{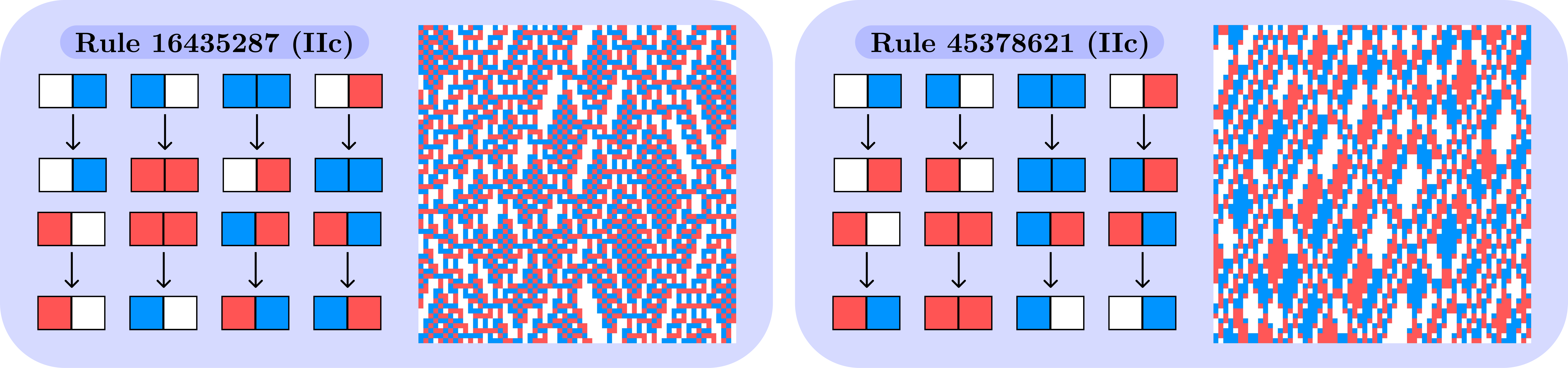}
  \label{fig:dynamics_class_IIc}
\end{figure}

\begin{figure}[h!]
\centerline{
    \raisebox{0.75cm}{
      \hspace{-0.3cm}
  \includegraphics[width=0.5\linewidth]{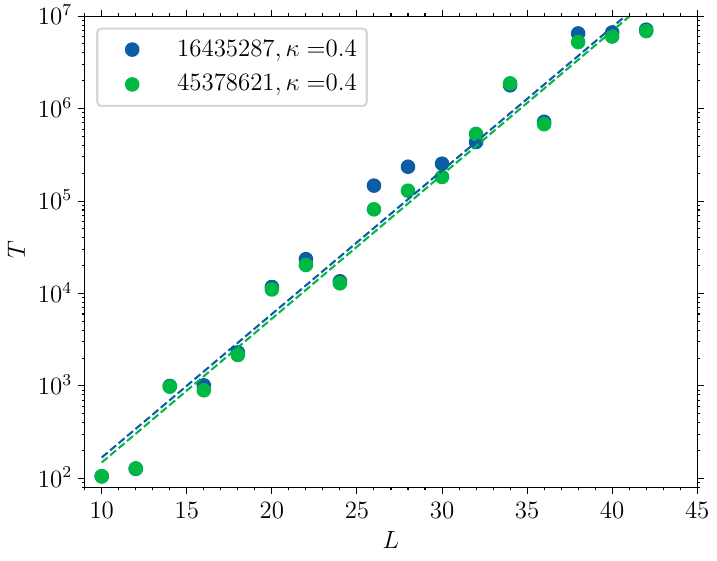}  
    }
  \hspace{-0.25cm}
   \includegraphics[width=0.5\linewidth]{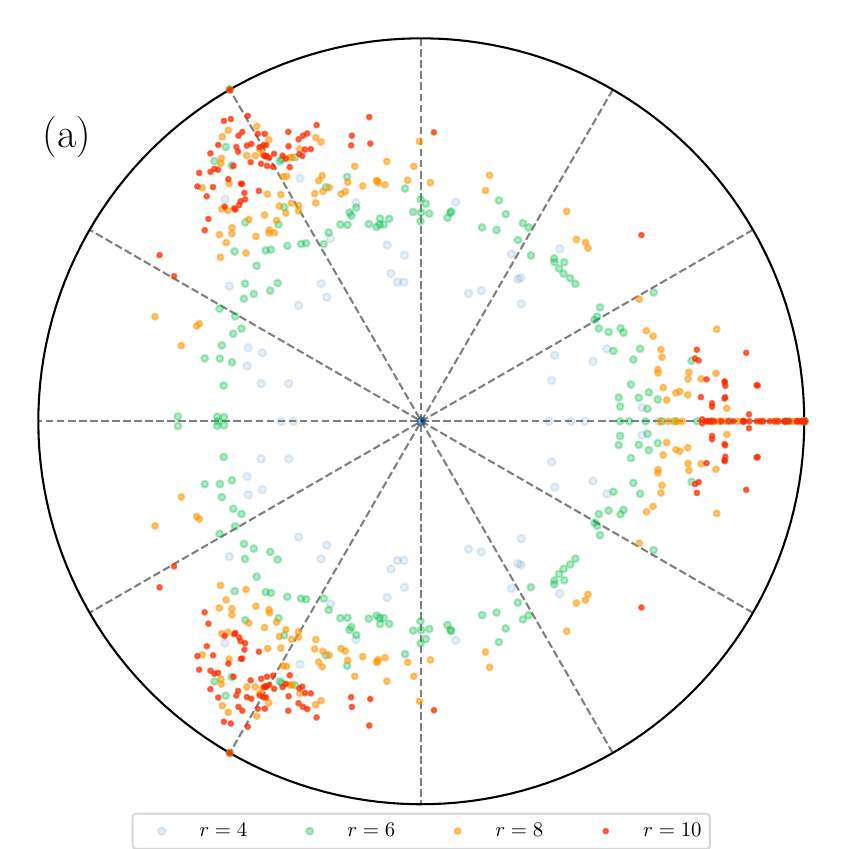}
}
  \caption{ \textbf{(a)} Exponential scaling of the mean return time with system size $T(L) \sim e^{\kappa L}$ of rules 16435287 and 45378621 in semilog scale. \textbf{(b)} The spectrum of the transfer matrix for the cellular automaton with rule 45378621.
  }
  \label{fig:spectrum_class_IIc}
\end{figure}

Both models have an exponentially increasing mean return time shown in the left panel of Figure~\ref{fig:spectrum_class_IIc}. Analogous behavior was previously observed in four-color cellular automata satisfying a set-theoretic Yang-Baxter equation \cite{Gambor}. In our case, the two rules do not possess this property; all models that are set-theoretically Yang-Baxter integrable appear to belong to Class IV. Notably, the physical quantities computed for both models are equivalent, even though the dynamics of the two models appears distinct. For the rest of the section, we focus on the model with rule 45378621. 

This rule appears to possess an exponentially growing number of strictly local conserved charges, suggesting a highly intricate underlying structure and the reason for calling it superintegrable.

\begin{table}[h!]
\centering
\begin{tabular}{|p{3.5cm}|p{0.8cm}|p{0.8cm}|p{0.8cm}|p{0.8cm}|p{0.8cm}|p{0.8cm}|p{0.8cm}|p{0.8cm}|}
\hline
Range $r$& 2 & 4 & 6 & 8 & 10 & 12 & 14  \\ \hline
Number of charges & 0 & 2 & 2 & 4 & 4 & 8 & 8  \\ \hline
\end{tabular}
\end{table}

Furthermore, as illustrated in the right panel of  Figure~\ref{fig:spectrum_class_IIc}, the transfer matrix spectrum shows that in addition to an exponential number of local conserved quantities, the spectrum features dynamical symmetries characterized by eigenvalues on the unit circle ($\lambda \neq 1$). Specifically, these eigenvalues are $\lambda_1 = e^{i \frac{2\pi}{3}}$ and $\lambda_2 = e^{i \frac{4\pi}{3}}$, indicating a periodic structure in the dynamics of the system. The extensive dynamical charges for rule 45378621,\footnote{For rule 1643528, the dynamical symmetries are the same, except that there is no $(-1)^j$ in $Q_1$.} which are conserved after 3 steps are given by the following densities:
\begin{subequations}\label{dyn_sym}
\begin{equation}
q_3^{(1)}= (-1)^j \left( [+]_j-[-]_j \right),\quad Q_1 = \sum_j \mathbb{T}^jq_3^{(1)}, 
\quad\mathbb{U}Q_1=e^{i \frac{2 \pi}{3}}Q_1,
\end{equation}
\begin{equation}
 q_4^{(1)}= (-1)^j  [\varnothing]_j,\quad Q_2 =\sum \mathbb{T}^j q_4^{(1)}, \quad \mathbb{U}Q_2=e^{i \frac{4 \pi}{3}}Q_2.
\end{equation}
\end{subequations}

\begin{figure}[h!]
\centering
\includegraphics[width=1\linewidth]{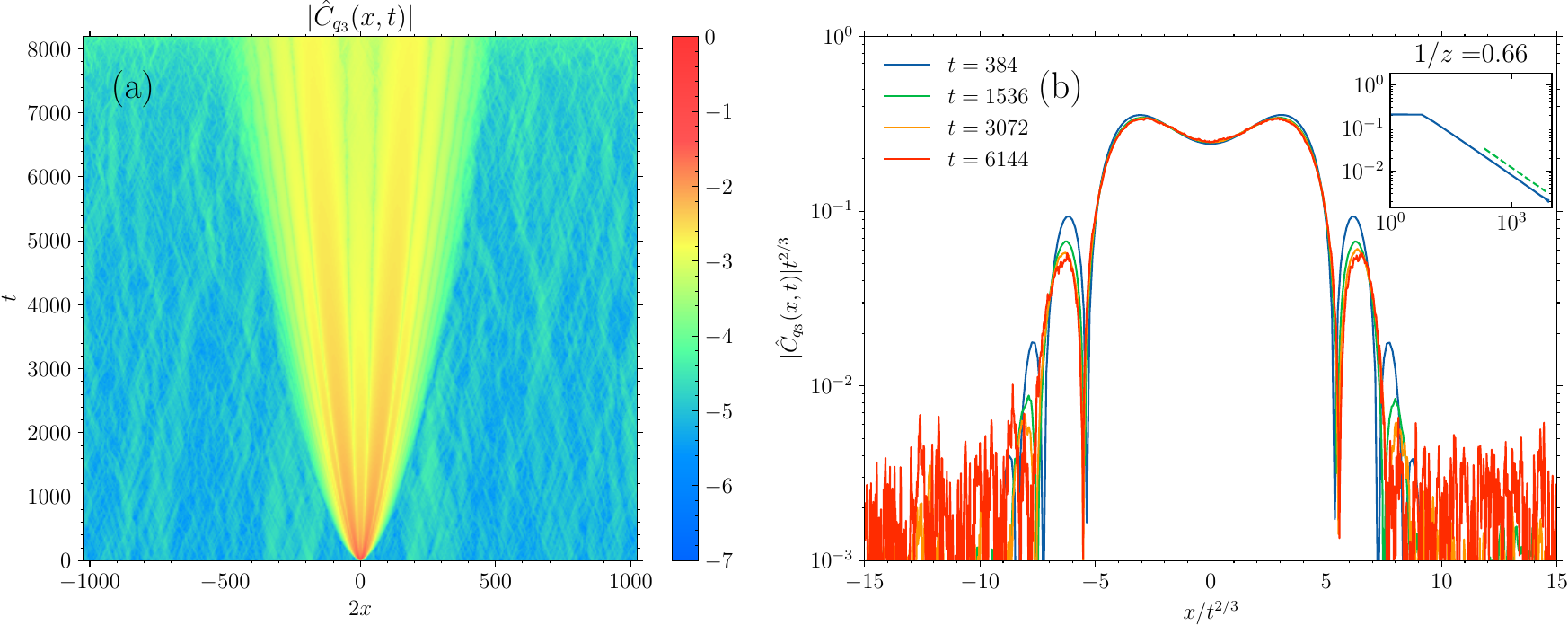}
\includegraphics[width=1\linewidth]{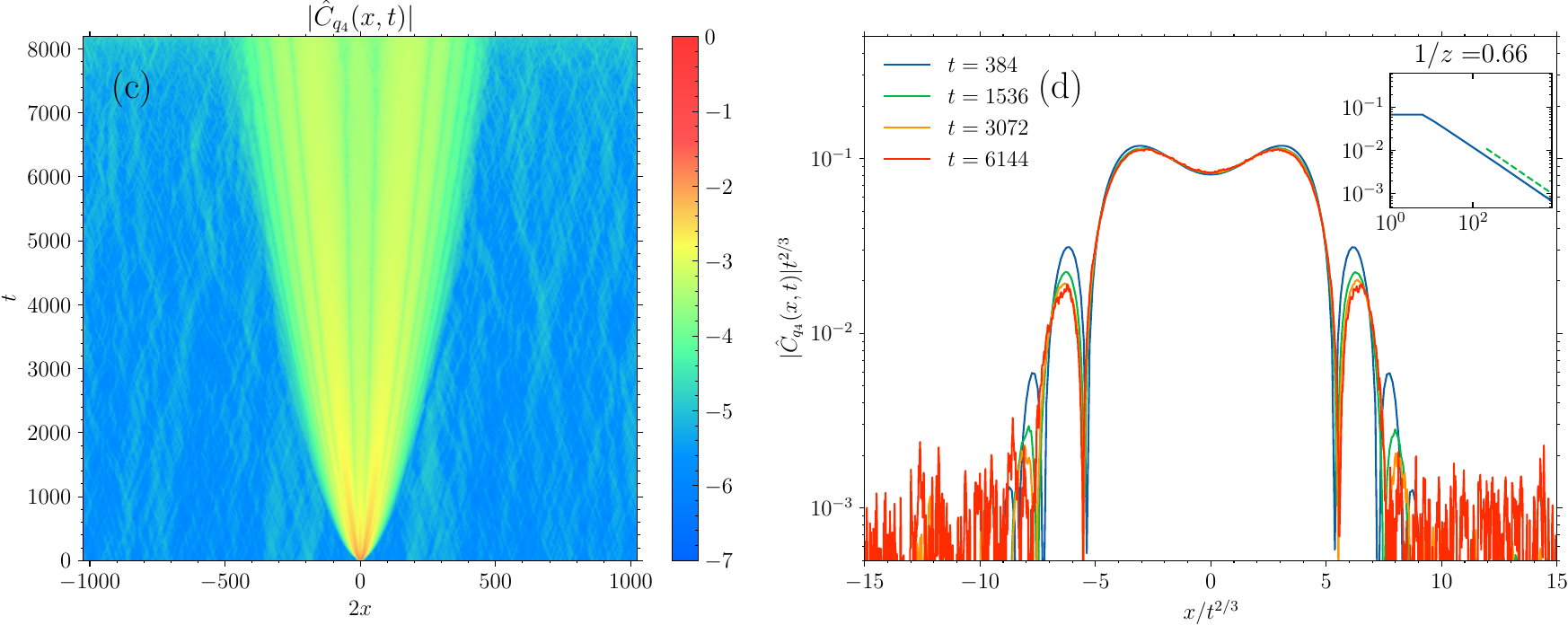}
  \caption{Correlators of dynamical charges of the $\mathcal{CT}+\mathcal{P}$-symmetric cellular automaton with rule 45378621. \textbf{(a)} $\hat{C}_{q_3}(x,t)$, \textbf{(b)} Cross-sections of $\hat{C}_{q_3}(x,t)$, \textbf{(c)} $\hat{C}_{q_4}(x,t)$, \textbf{(d)} Cross-sections of $\hat{C}_{q_4}(x,t)$.
  All cross-sections and insets are plotted for times $t~{\rm mod}~6 = 0$ as the correlator oscillates with period 3 due to the period of the dynamical charge. The insets on the plots in the right panel show $\hat{C}_{q_3}(0,t)$ and $\hat{C}_{q_4}(0,t)$ which decay algebraically $\sim t^{-2/3}$. Best fits are shown with green dashed lines.
  }
  \label{fig:odd}
\end{figure}

\textbf{Superdiffusive transport.} Next, we study the transport of the dynamical charges given by Eq.~(\ref{dyn_sym}). We compute the correlation functions $\hat C_{q_3} =\langle q_3^{(1)}(x,t) q_3^{(1)}(0,0)\rangle$ and $\hat C_{q_4} =\langle q_4^{(1)}(x,t) q_4^{(1)}(0,0)\rangle$, which  correspond to dynamically conserved charges. In Figure \ref{fig:odd}, we show the behavior of the two correlators, both of which exhibit the algebraic decay with the dynamical exponent $1/z  \approx 0.66$. Thus, the dynamical charges can also exhibit anomalous transport with an unexpected scaling function.

\clearpage
\section{Class III} \label{sec:class_3}
The main features of the models in Class III are a non-decaying correlation function of certain local observables and an exponentially increasing mean return time. 
Many models in this class are the so-called domain-wall rules with dynamics contained in small finite regions that are strictly preserved in time. These regions effectively split the system into smaller dynamical subsections that cannot interact with each other. There are special cases in which the domain walls are only partially preserved, allowing the dynamics of one subsection to influence another, albeit at a slower rate. Notably, we observe such models to exist independently of the number of local charges they possess.

\subsection{Subclass IIIa}
\label{sec:class_3a}
The first type of domain walls is characterized by a fixed finite number of strictly local conserved quantities with increasing support $r$. Based on this fact, the models can be called chaotic; however, they do not exhibit mixing, which explains the plateau observed in certain correlators. Examples of domain-wall models are provided by local update rules 16543278 and 14325687, which are $\mathcal{C}+\mathcal{T}$ and $\mathcal{C}\mathcal{P}+\mathcal{T}$ symmetric, respectively. The rules have 0 and 1 strictly local charges, respectively. We note that the simplest domain walls are composed of stripes of vacuum, as can be see for rule 14325687, but as can be seen for rule 16543278, they need not be as simple. 

\begin{figure}[h!]
\hspace{-0.3cm}
    \centering{
    \includegraphics[width=1.0\linewidth]{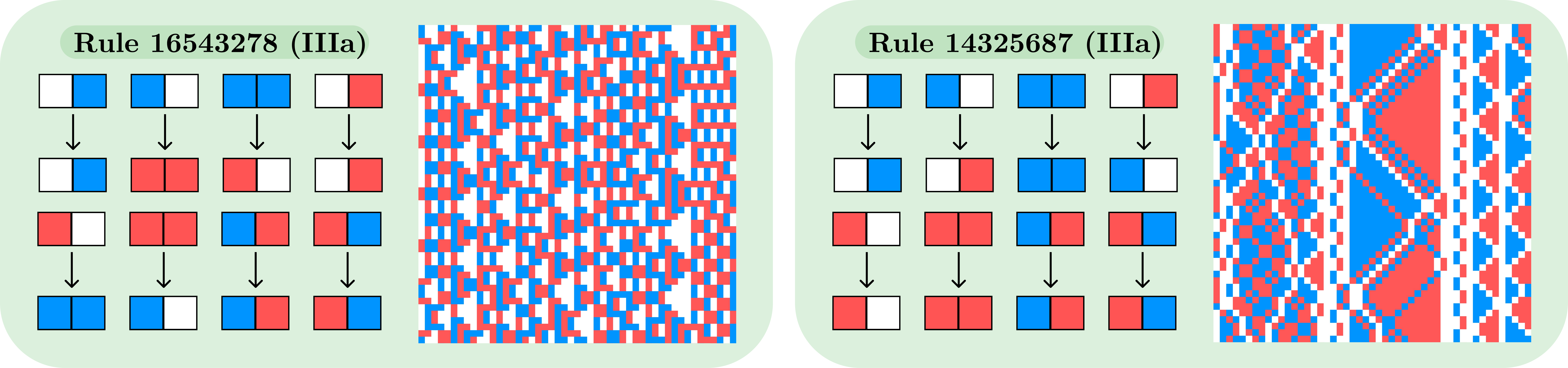}
    \label{fig:domain_wall_chaos}
    }
\end{figure}

\begin{figure}[h!]
\vspace{-20pt}
    \centering
    \includegraphics[width=1.0\linewidth]{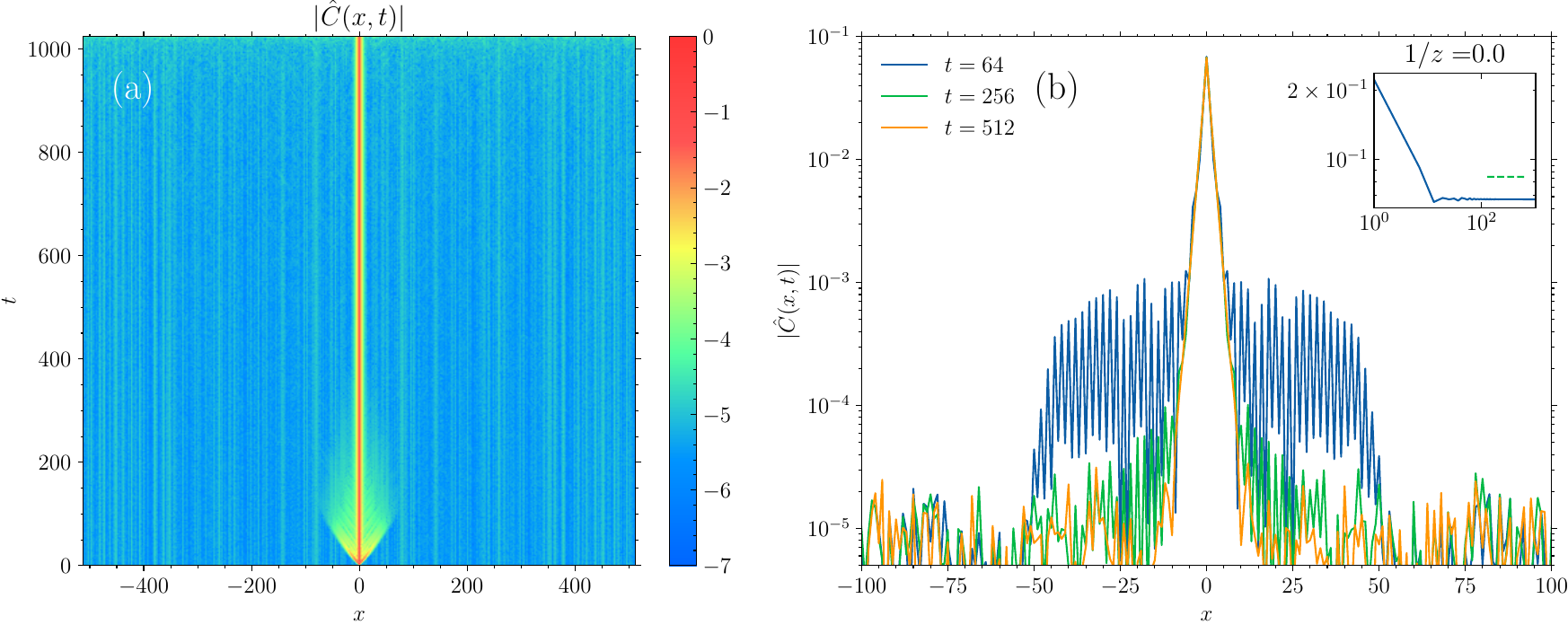}
    \caption{Correlators of the $\CC+\CT$-symmetric cellular automaton with rule 16543278. \textbf{(a)} $\hat{C}(x,t)$, \textbf{(b)} Cross-sections of $\hat{C}(x,t)$. The inset shows $\hat{C}(0,t)$.}
    \label{fig:IIIa_corr}
\end{figure}

Hereon, we focus on rule 16543278. In this model, the correlators decay to constant values and exhibit neither a drift nor spreading. This is shown in Figure~\ref{fig:IIIa_corr}. Hence, we cannot associate standard transport properties and dynamical exponents to the models in this subclass.
\\

\begin{figure}[h!]
\centerline{
    \hspace{-0.3cm}
  \includegraphics[width=0.5\linewidth]{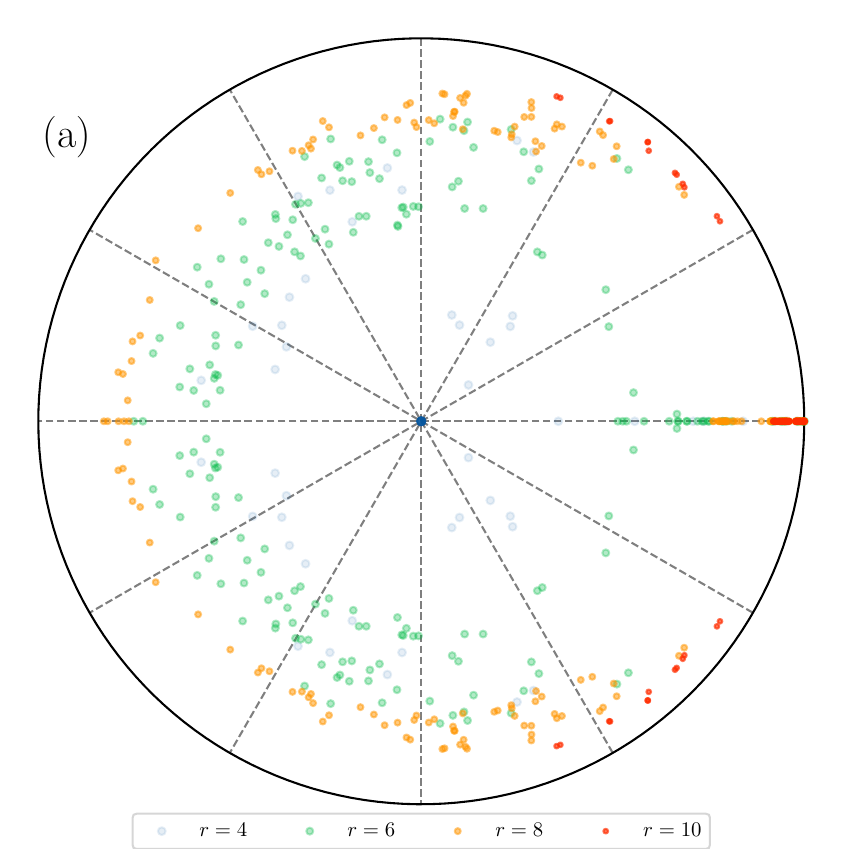}

    \hspace{-0.25cm}
   \includegraphics[width=0.5\linewidth]{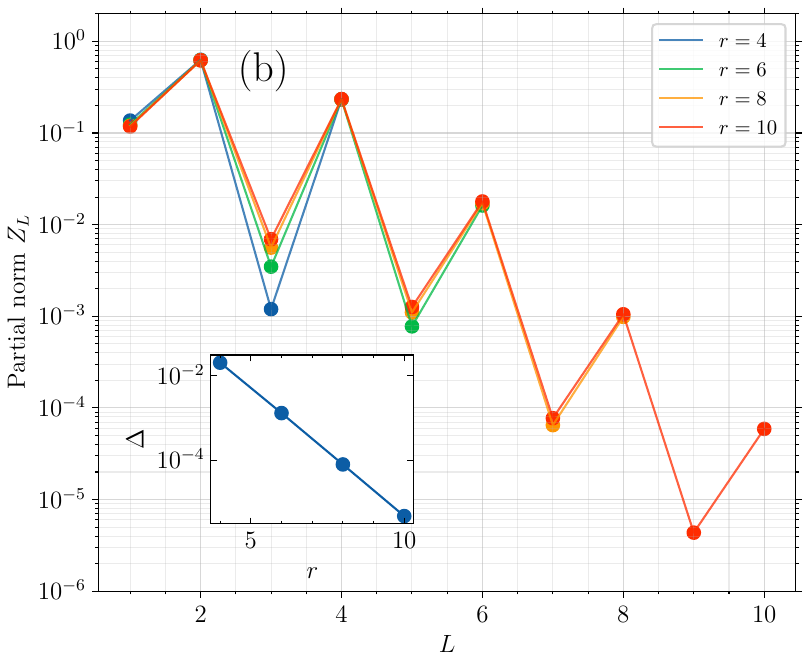}
}
  \caption{Cellular automaton with rule 16543278. \textbf{(a)} Spectrum of the transfer matrix for increasing support $r$. \textbf{(b)} Partial norms of the eigenvector associated with the gap for increasing $L$. The inset shows the exponential decay of the gap $\Delta$ with increasing support.
  }
  \label{fig:IIIa_quasilocal_I}
\end{figure}

\textbf{Quasilocal charges}. The spectrum of the transfer matrix $\mathcal T^{(r)}$ is then shown in Figure~\ref{fig:IIIa_quasilocal_I}. We verify that this rule indeed has no local conserved quantities and has an exponentially decaying gap. In addition, the partial norms of the associated leading eigenvector decay exponentially. 

\begin{figure}[h!]
\centerline{
  \includegraphics[width=0.5\linewidth]{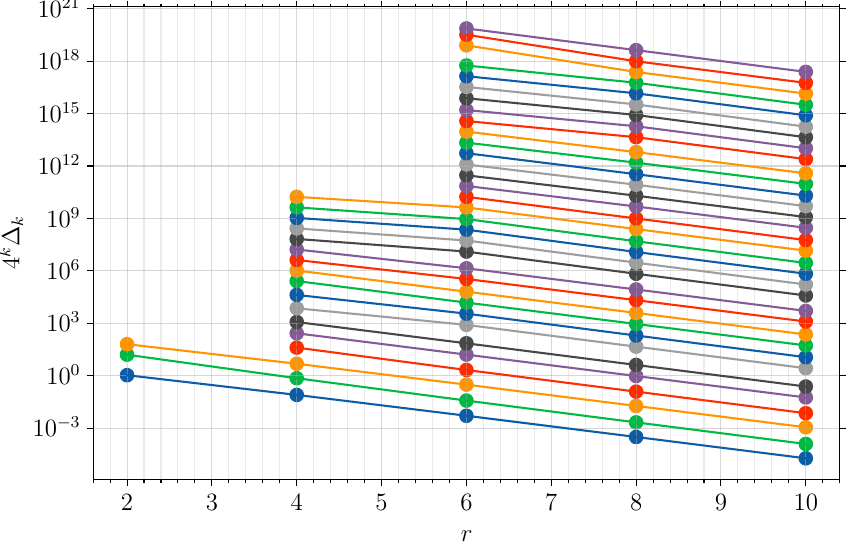}
}
  \caption{Scaling of the gaps for all real eigenvalues $\Delta_k=1-\lambda_k$ with increasing support $r$ of cellular automaton with rule 16543278. Note that the $y$ axis is rescaled by $4^k$ in order to spread the data out and make it more easily visible.
  }
  \label{fig:IIIa_manygaps}
\end{figure}

We find that this cellular automaton is exceptional in the sense that it does not only have one, but instead a large number of quasilocal conserved quantities. In Figure~\ref{fig:IIIa_manygaps}, we demonstrate how the gap scales with the support $r$ for all real eigenvalues of $\mathcal{T}^{(r)}$. Finding an exact count of the quasilocal charges appears to be difficult without exactly diagonalizing $\mathcal{T}^{(r)}$, which can only be done for small support sizes. However, our findings strongly suggest that all real eigenvalues converge exponentially to $\lambda = 1$ as support increases. This behavior leads us to propose that the (super)integrability of this model may arise solely from quasilocal charges. This is a phenomenon, which has so far never been observed in integrable spin chains.

\subsection{Subclass IIIb}

\begin{figure}[h!]
    \centering{
    \includegraphics[width=0.6\linewidth]{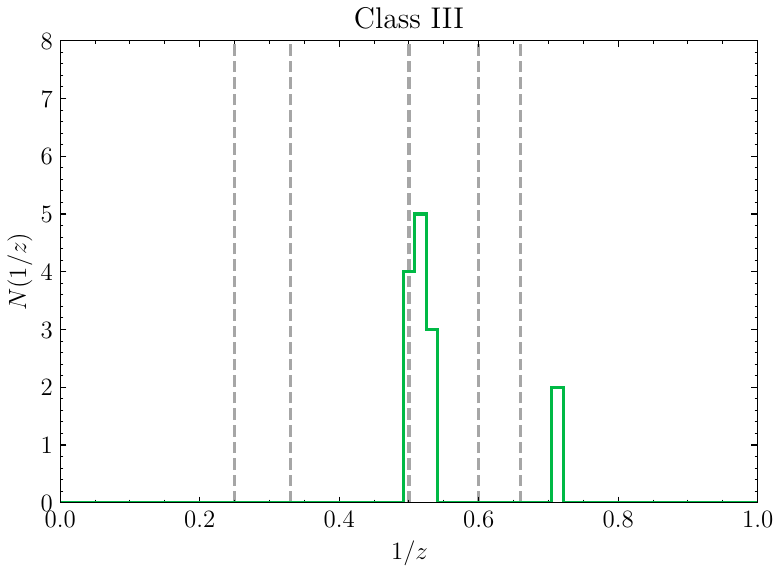}
        \caption{ Histogram of all dynamical exponents $1/z_C$ found in Class III. Importantly, all of these result correspond to models in Subclass IIIb. The vertical dashed lines, from left to right, represent the values of $1/z_C \in \{\frac{1}{4},\frac{1}{3}, \frac{1}{2}, \frac{3}{5}, \frac{2}{3}\}$. 
        }
    \label{fig:dynamical_exponents_class_III}
    }
\end{figure}

Models in Subclass IIIb have an exponential number of charges and can therefore be considered as superintegrable. For certain models (which are $\mathcal{CPT}$ symmetric), the correlator $C(t)$ decays algebraically, while $D(t)$ exhibits a constant plateau in the thermodynamic limit for all rules. The histogram of all dynamical exponents is shown again in Figure~\ref{fig:dynamical_exponents_class_III}, with most models exhibiting values close to $1/z_C=1/2$. In these models, discussing transport and drawing conclusions from correlators with support 1 turns out to be difficult. The reason is that most of the charges in this class are \textit{gliders} \cite{Borsi}, i.e., their densities $q$ are rigidly translating either left or right, $\mathbb U q = \mathbb T^{\pm 2} q$.  
\\

\textbf{Diffusive correlations.} We first consider a diffusive $\CC\CP\CT$-symmetric cellular automaton with rule 34671528. 

\begin{figure}[h!]
\centering
\includegraphics[width=0.55\linewidth]{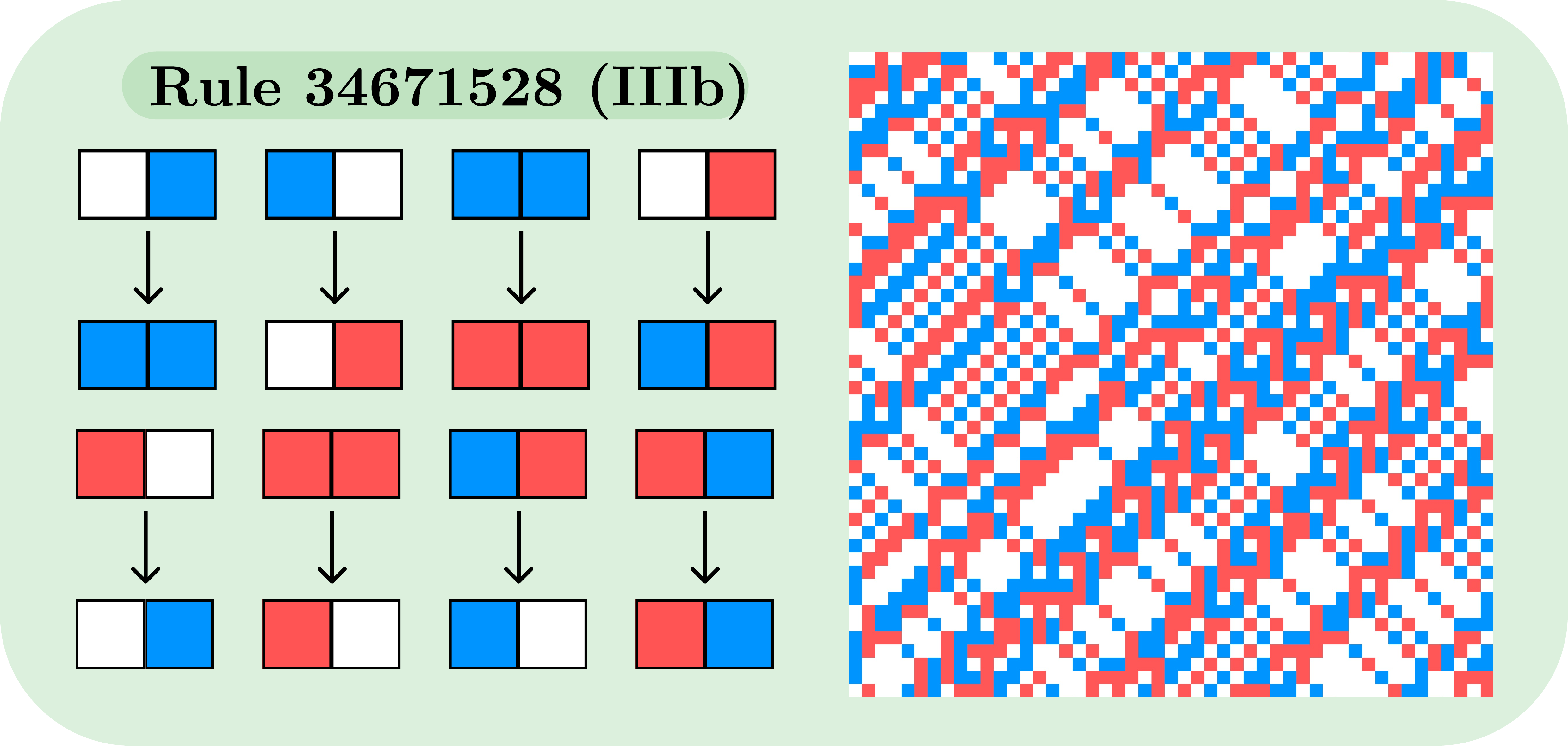}
  \label{fig:IIIc_diffusion_rule}
\end{figure}

\begin{figure}[h!]
\centering
\includegraphics[width=1.0\linewidth]{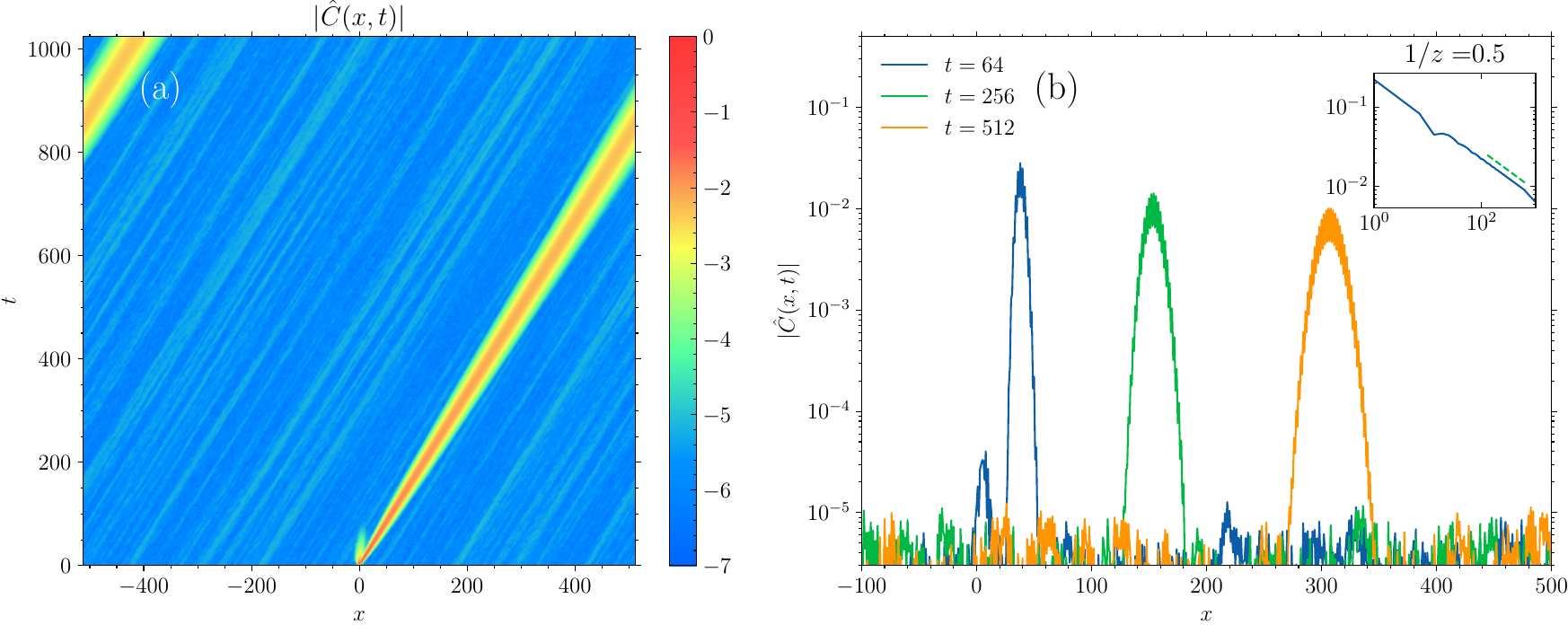}
  \caption{Correlation functions for the $\mathcal{CPT}$-symmetric cellular automaton with rule 34671528. \textbf{(a)} $\hat{C}(x,t)$, \textbf{(b)} Cross-sections of $\hat{C}(x,t)$ for increasing times $t$. The inset shows $C(t) \sim t^{-1/z_C}$, where the dashed green line shows the best fit with $1/z_C = 0.5$.}
  \label{fig:corr_34671528}
\end{figure}

The system exhibits an algebraic decay of the correlator $C(x,t)$ (see Figure~\ref{fig:corr_34671528}), While $D(x,t)$ does not decay to zero. Importantly, however, this rule does not conserve $\sum_x~[+]_x-[-]_x$, so the correlation function $C(x,t)$ is not directly related to transport. We propose that the algebraic decay arises from the overlap between range-1 observables and charges of larger support, a point we will examine further below. 

\newpage

The number of local charges for this rule, increasing with $r$ in steps of 2, is given here. 

\begin{table}[h!]
\centering
\begin{tabular}{|p{3.5cm}|p{0.8cm}|p{0.8cm}|p{0.8cm}|p{0.8cm}|p{0.8cm}|p{0.8cm}|}
\hline
Range $r$& 2 & 4 & 6 & 8 & 10 & 12 \\ \hline
Number of Charges & 1 & 2 & 6 & 10 & 18 & 34 \\ \hline
\end{tabular}
\end{table}
 
One finds that the simplest charge of range 2 is given by the local density $[20]_x$. At range 4, the two charges are translations of local densities $[2000]_x$ and $[2020]_x$. At every higher range, there is a set of charges, which span all possible combinations of $[20]$ and $[00]$ conspiring to give a basis element of support $r$, of which there are $2^{(\frac{r}{2} - 1)}$. All charges constructed in this way are gliders. The densities thus remain perfectly correlated at later times, but they have a constant drift. We propose that the diffusive-like decay of the correlation function for the observable $[+]-[-]$ cannot come from overlaps with gliders, but from other charges, which are of range 6 or higher. These charges can be found relatively easily, as long as one recognizes that the space of gliders is completely independent of them. We list the range-6 charges which are not gliders of range 6 here:
\begin{equation}
    \begin{split}
    q^{(6)}_1 =  &\,\,-2[1] +8[01]+ 2[12]-2[11] + 2[011] + 2[021] + 4[111] -[0211] +[0221]  
    \\&  +[0112]  -2[0212] + [0122]+ 2[1112] + 2[1122]+ 2[02111]-2[02211]
    \\&      -[022112]+ [021112]  + [021122]  -[022122],
    \end{split}
\end{equation}
\begin{equation}
    \begin{split}
    q^{(6)}_2 = &\,\,+ 8[101]+4[0202]+ 4[121] +4 [0222]+ 4[1022] +4[1012] +2[1212]+ 2[1222]\\& +4 [02101]+2[02121]-4[02201]-2[02221]+ 2[021012]  -2[022012] 
    \\& +[021212]    -[022212] + 2[021022]-2[022022] + [021222] -[022222].
    \end{split}
\end{equation}
The charge $q_1^{(6)}$ has non-zero overlap with the observable $[1]\equiv [+]-[-]$, thus, the transport of this charge contributes to the decay of the correlation function $\hat{C}(x,t)$. Studying the transport of this charge would therefore involve eliminating the drift velocity from the correlation functions of range-1 observables, which could exhibit diffusion-like decay. This is beyond the scope of the present work. We postpone such explorations until a later time.
\\

\newpage
\textbf{Anomalous correlations}. In addition to models with diffusive correlations, this class of models also contains a $\mathcal{CPT}$-symmetric rule 23658471 (and its equivalent 26457318), which exhibits an algebraic decay of $C(t)$.

\begin{figure}[h!]
\centering
\includegraphics[width=0.55\linewidth]{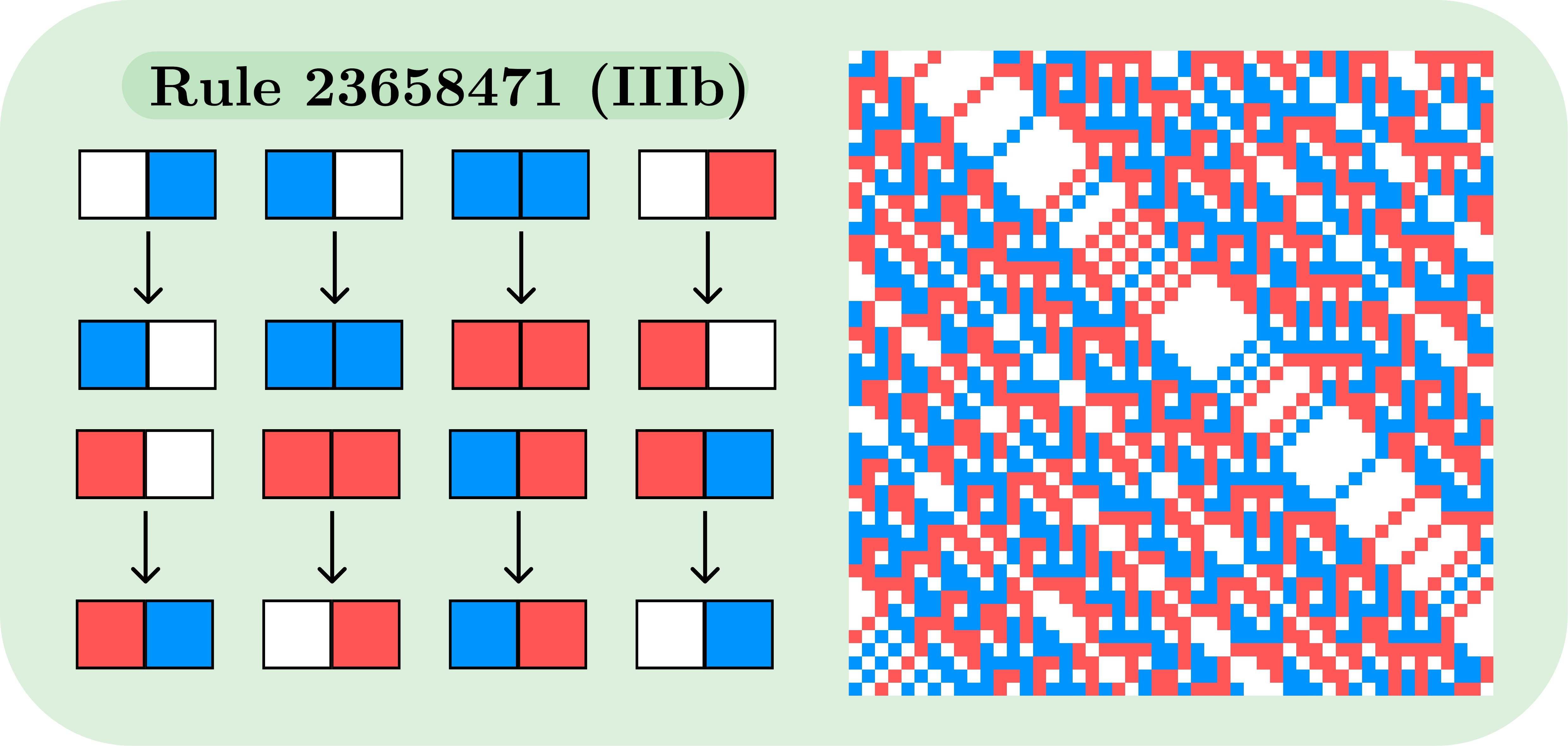}
  \label{fig:IIIc_anomalous_rule}
\end{figure}

\begin{figure}[h!]
\centering
\includegraphics[width=1.0\linewidth]{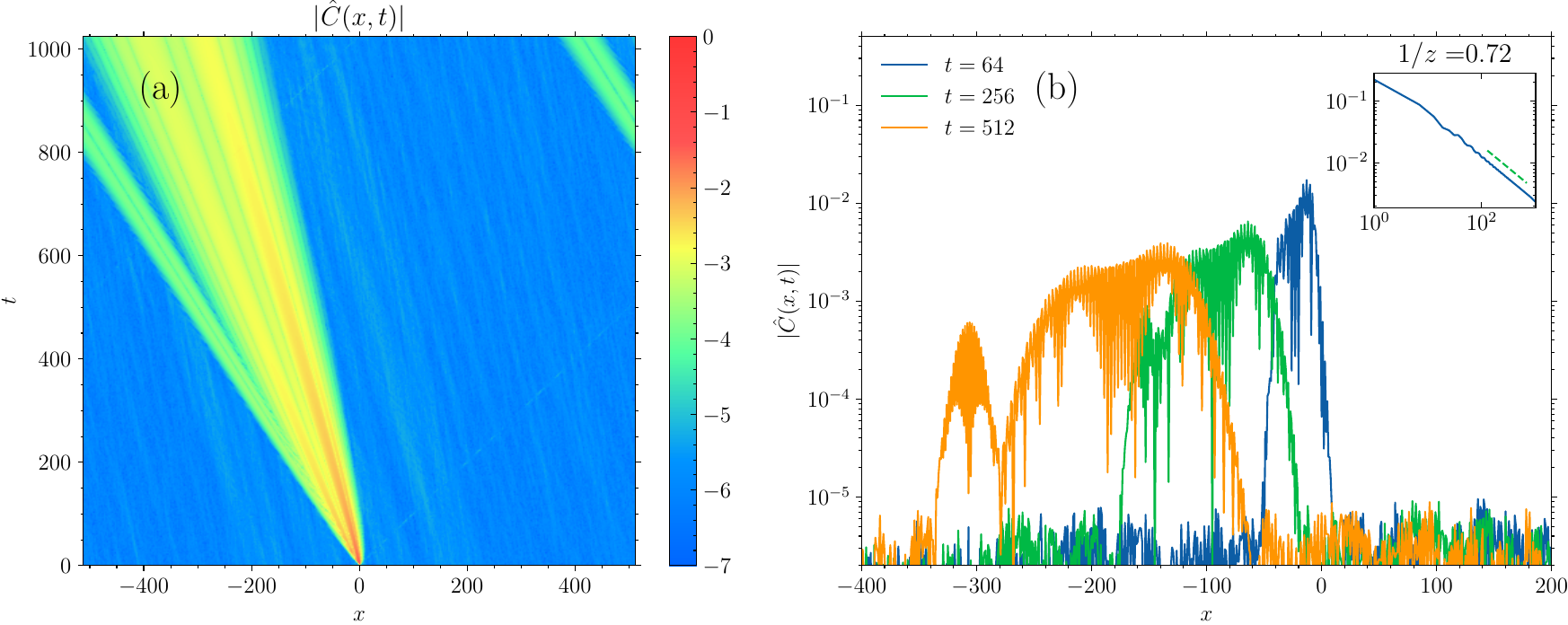}
  \caption{Correlation functions of the $\mathcal{CPT}$-symmetric cellular automaton with rule 23658471. \textbf{(a)} $\hat{C}(x,t)$, \textbf{(b)} Cross-sections of $\hat{C}(x,t)$ at increasing times $t$. The inset shows $C(t) \sim t^{1/z}$ with the dashed green line showing the best fit $1/z=0.72$.}
  \label{fig:corr_23658471}
\end{figure}

Similar to the previous example, this rule contains gliders, with the first one having the density $[20]$. In this case, there is only one non-glider charge of range 6:
\begin{equation}\label{eq:charge_6}
    \begin{split}
    q^{(6)}_3 = &\,\,4[202]+ 4[222]+8 [0101] + 4[0121] +4[2201]+ 4[2101]  + 2 [2121]   + 2[2221]
    \\ &   + 2[01212]- 4[01022] -2 [01222] + 2 [21012] +4 [01012]+ 2 [22012] 
    \\
    & + [21212] + [22212]  - 2[21022] - 2[22022] - [21222] - [22222].
    \end{split}
\end{equation}

\begin{figure}[h!]
\centerline{
  \includegraphics[width=0.8\linewidth]{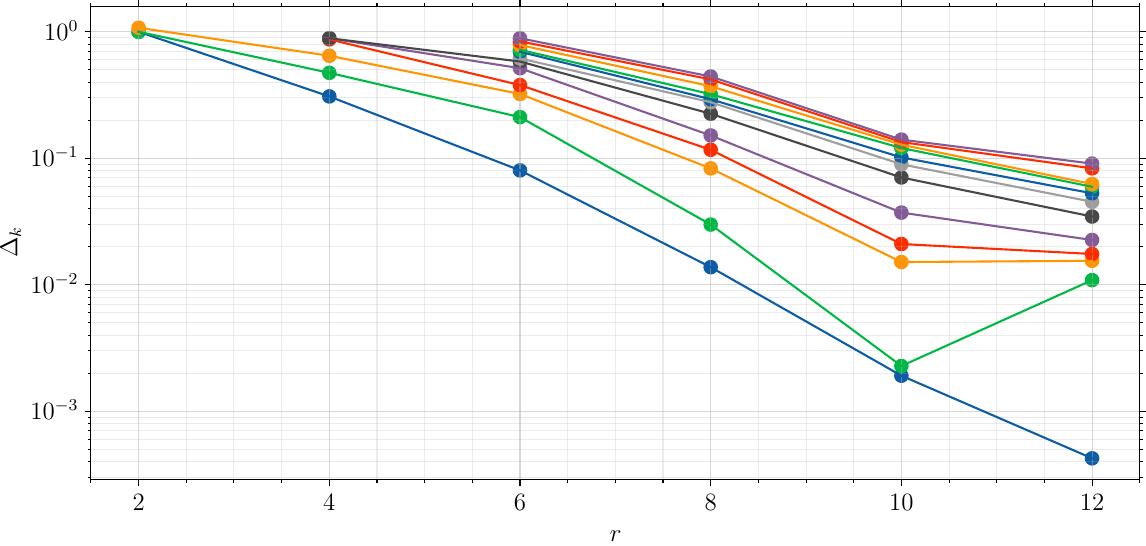}
}
  \caption{Scaling of the largest 12 real eigenvalues $\Delta_k$ with increasing support $r$ of cellular automaton with rule 23658471 up to $r=12$. We observe one eigenvalue which approaches the unit circle exponentially.}
  \label{fig:manygaps_23658471}
\end{figure}

\noindent
However, this charge does not overlap with the observable $[+]-[-]$, the correlator of which shows anomalous decay. We examined the charges of this superintegrable model up to range 12 and did not find any non-glider charges other than $q_3^{(6)}$ from Eq.~\eqref{eq:charge_6}. Thus, charges up to this range cannot explain the anomalous decay. This suggests that the observable in $\hat C(x,t)$ may have an overlap with one of the quasilocal charges of the model. In Figure~\ref{fig:manygaps_23658471}, we show the scaling of the largest 12 real eigenvalues of the transfer matrix as a function of increasing range. We observe that one of the eigenvalues indeed approaches the unit value exponentially. The anomalous correlation of the observable $[+]-[-]$ can therefore be a signature of the anomalous transport of one of the quasilocal charges with which it overlaps. However, details of this analysis will also be left to future works.

\newpage
\textbf{Dual reversible models.} Finally, we can identify a set of superintegrable, dual-reversible models that exhibit exponential decay of range-1 observables. These models have charges only of support 4 or higher, and these charges are gliders that undergo ballistic transport. Moreover, these charges do not overlap with the local range-1 observables, for which we computed the correlators, leading to their exponential decay. 

As an example, we consider the $\mathcal{C}\mathcal{P}\mathcal{T}$-symmetric rule 57638241.

\begin{figure}[h!]
\centering
\includegraphics[width=0.55\linewidth]{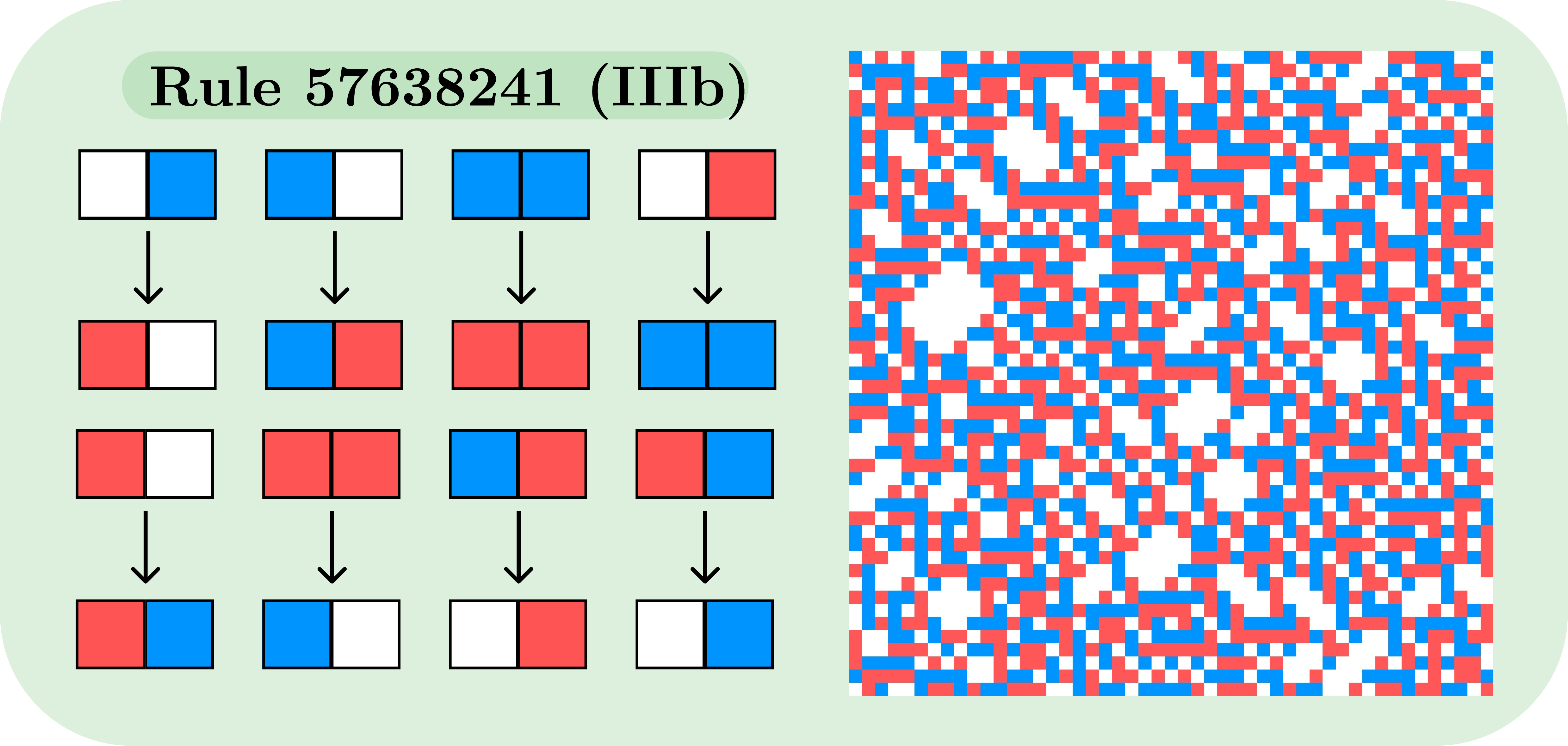}
  \label{fig:IIIc_dual_reversible_rule}
\end{figure}
The number of local charges for this rule, increasing with range in steps of 2, is given here.

\begin{table}[h!]
\centering
\begin{tabular}{|p{3.5cm}|p{0.8cm}|p{0.8cm}|p{0.8cm}|p{0.8cm}|p{0.8cm}|p{0.8cm}|}
\hline
Range $r$& 2 & 4 & 6 & 8 & 10 & 12 \\ \hline
Number of Charges & 0 & 1 & 2 & 4 & 8 & 17 \\ \hline
\end{tabular}
\end{table}

The first conserved quantity has range 3 and has the following density:
\begin{equation}
    q^{(4)} =  4[101] + 3[211]  -[121] + 3[112]  -[222].
\end{equation}
This charge is a glider, so it exhibits ballistic transport. Its density does not contain terms of lower range; in particular, it does not overlap with the observables of range 1. 

We checked the charges up to range $r=12$ and we found the same property: the charges were gliders with desities which did not contain terms of lower support size.
This consistent with a analogous property for dual unitary 
quantum circuits~\cite{masanes}. Due to the special structure of the charges, their exponential number does not lead to slow decay of range-1 observables. However, a generic observable will overlap with one of the gliders, leading to the decay of its correlator to a nonzero constant. Thus, we categorize this set of dual reversible rules as belonging to Class III, Subclass IIIb.

\newpage
\section{Class IV}\label{sec:class_4}
The models in Class IV have a mean orbit length which grows as a power-law, and their slowest correlator decays to a constant (mean) value. Interestingly, we find two subclasses of such models. The RCA in Subclass IVa have a constant number of strictly local conserved quantities (which in our case is always zero). This indicates that there could be non-local charges that constrain the dynamics. The second category of models, Subclass IVb, exhibits an exponential number of conserved charges with increasing support size $r$, indicating superintegrability. Superintegrable rules that satisfy the Yang-Baxter equation were explored in Ref.~\cite{Gambor}, where it was found that three-color rules display return times that grow polynomially, scaling as $\sim L^2$. Interestingly, when additional colors are introduced, the return times increase at a faster rate. For instance, in the case of four colors, an example of a superintegrable rule was identified with exponentially increasing return times.

Among our RCA, we can identify all set-theoretic Yang-Baxter integrable models. All such cases can be reduced to only four nonequivalent rules, which are neither free nor trivial. The first is the extensively studied hardcore charged gas model \cite{ziga_rule,ziga_rule2,ziga_rule3,ziga_rule4}, denoted in our notation as rule 21354678. This rule is known to exhibit a quadratically increasing mean return time $\sim L^2$, a result previously established in Ref.~\cite{gopalakrishnan2018operator}. The three remaining set-theoretic Yang-Baxter integrable rules 12354687, 12387654, and 54321678 also demonstrate mean return times that scale as $\sim L^2$. In addition, we identify superintegrable rules with return times that scale faster, such as $\sim L^3$ and $\sim L^4$. These cases must fall outside the scope of the framework discussed in Ref.~\cite{Gambor}.

\begin{figure}[h!]
    \centering{
    \includegraphics[width=0.6\linewidth]{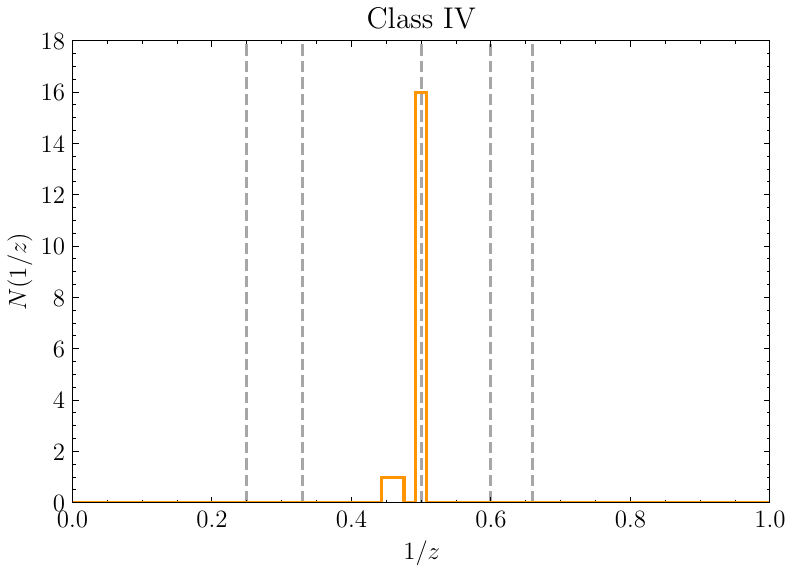}
        \caption{Histogram of all dynamical exponents $1/z_C$ and $1/z_D$ for class IV. The vertical dashed lines, from left to right, represent the values of $1/z \in \{\frac{1}{4},\frac{1}{3}, \frac{1}{2}, \frac{3}{5}, \frac{2}{3}\}$. 
       }
       \label{fig:dynamical_exponents_class_IV}
    }
\end{figure}

Despite the fact that the correlation functions decay to a non-zero constant, some of the rules exhibit a nontrivial power-law decay of correlation functions, which we show in Figure~\ref{fig:dynamical_exponents_class_IV}. Charges, which do not exhibit ballistic transport due to gliders, undergo diffusive transport with the dynamical exponent $z=1/2$.

\newpage
\subsection{Subclass IVa}
\begin{figure}[h!]
    \hspace*{-0.55cm}
    \centering{
    \includegraphics[width=1.0\linewidth]{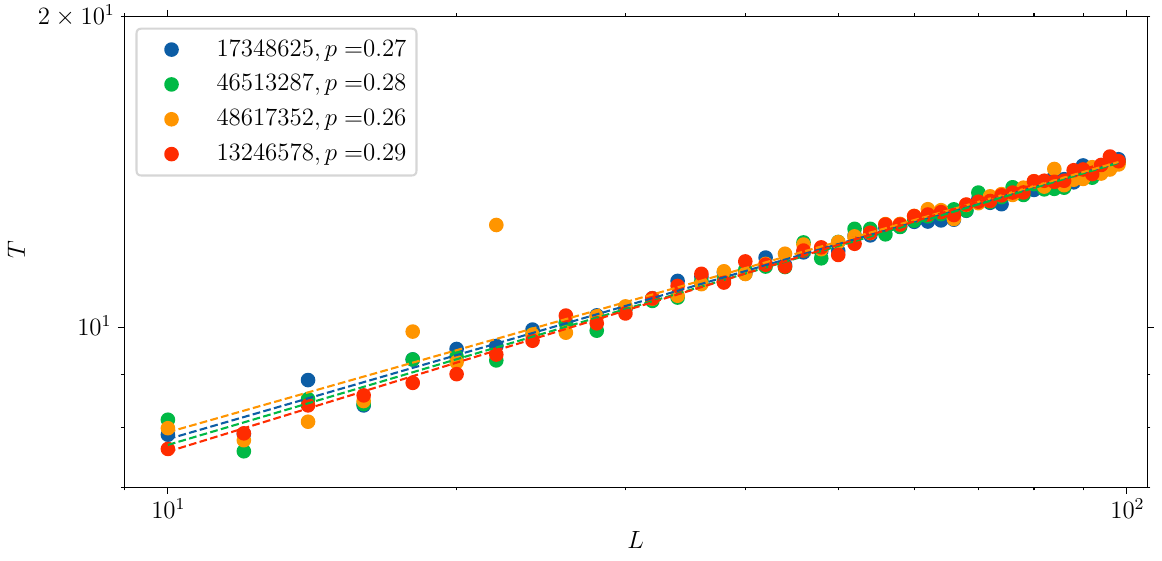}
        \caption{Examples of mean return times for Class IVa cellular automata with rules 17348625, 46513287, 48617352, 13246578, all possessing the $\mathcal{C}+\mathcal{T}$ symmetry. These rules exhibit some of the slowest increasing mean return times that scale as $\sim L^p$ with the smallest $p = 0.26$.
        }
        \label{fig:TL_IVa}
    }
\end{figure}

The few rules in this subclass exhibit a complete lack of local conserved quantities; however, their mean return times follow a sublinear power-law scaling, which is slower than that of the models in Class IVb. This is shown in Figure~\ref{fig:TL_IVa}.

This suggests that a factor other than an extensive number of local conserved quantities must be responsible for this strong constraint on the length of orbits. We propose that the presence of quasilocal conserved quantities, and more generally, dynamical symmetries, is what is responsible for this behavior. 

As an example of a cellular automaton in Subclass IVa, here, we consider rule 17348625.

\begin{figure}[h!]
\centering
\includegraphics[width=0.55\linewidth]{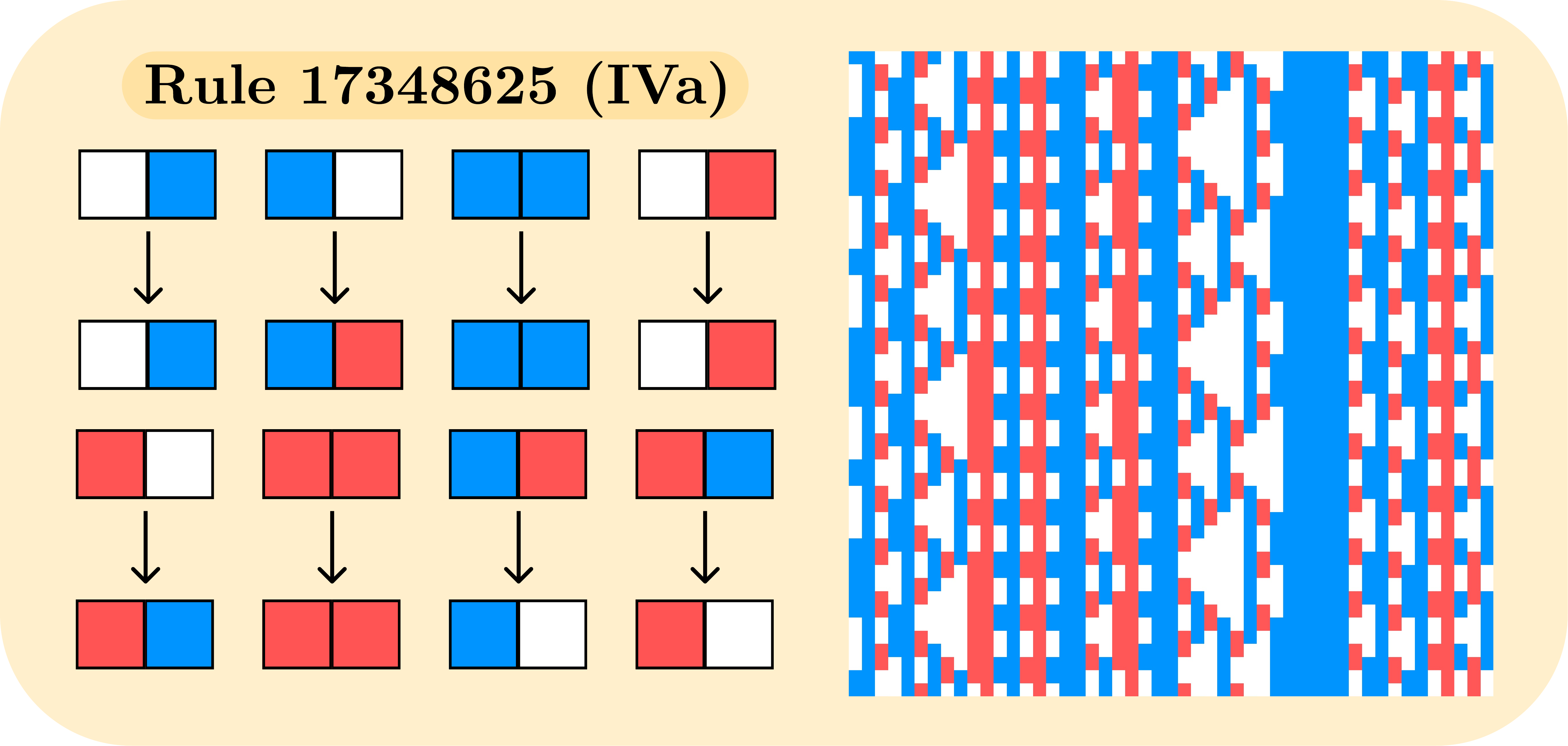}
  \label{fig:IVa_rule}
\end{figure}

\begin{figure}[h!]
\centerline{
    \hspace{-0.3cm}
  \includegraphics[width=0.45\linewidth]{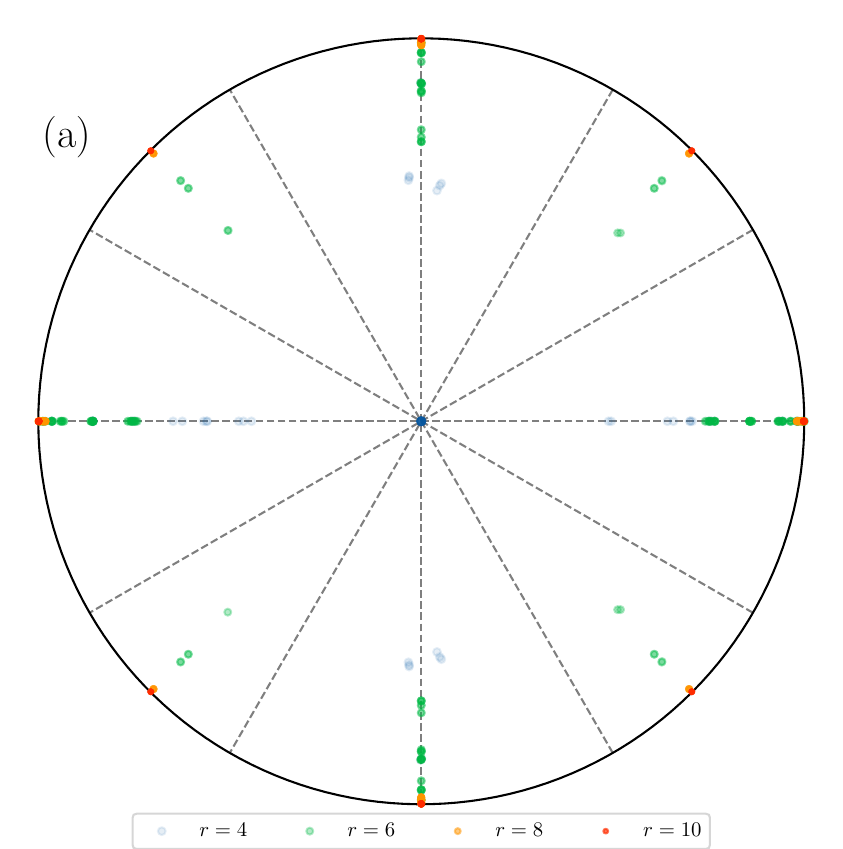}
  
    \hspace{-0.25cm}
   \includegraphics[width=0.5\linewidth]{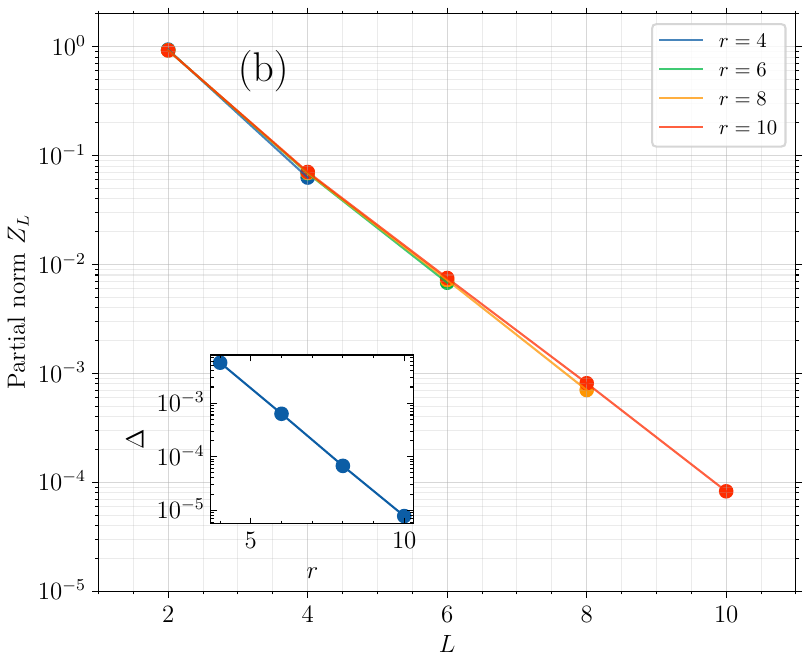}
}
  \caption{\textbf{(a)} Spectrum of the transfer matrix of the cellular automaton with rule 17348625 for increasing support $r$. \textbf{(b)} Partial norms of the eigenvector associated with the gap for increasing $L$. The partial norms at odd $L$ are 0. The inset shows the exponential decay of the gap $\Delta$ with increasing support in semilog scale.
  }
  \label{fig:IVa_example}
\end{figure}

Next, we calculate the spectrum of the transfer matrix for increasing support $r$. The results are shown in the left panel of  Figure~\ref{fig:IVa_example}. We find that, when $r$ is increased, there is an extensive number of eigenvalues that exponentially approach the unit circle, specifically the points $ e^{i \frac{j\pi}{4}}$ with $j=0,1,\ldots,7$. In the right panel of Figure~\ref{fig:IVa_example}, we show that the partial norms of the leading (smallest gap) eigenvector decay exponentially for even $L$, and that the gap closes exponentially.

\begin{figure}[h!]
\centerline{
  \includegraphics[width=0.5\linewidth]{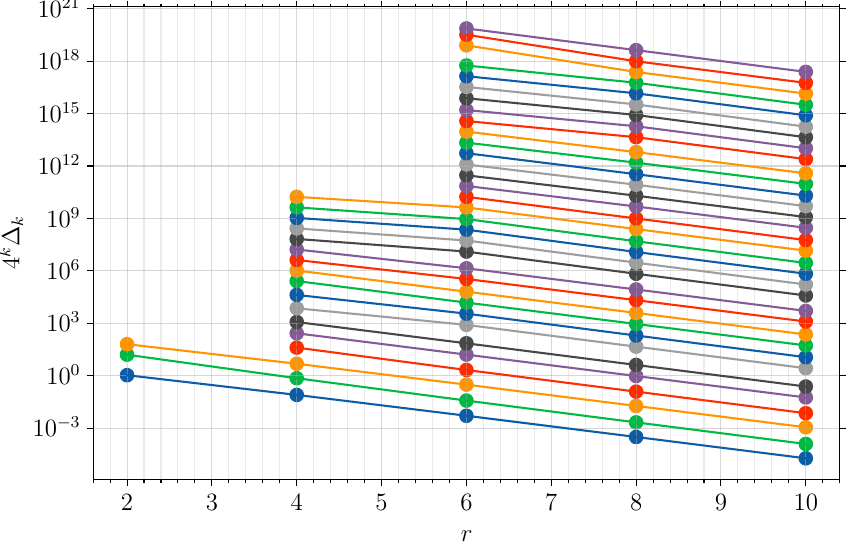}
}
  \caption{Scaling of all real eigenvalues $\Delta_k$ with increasing support $r$ of cellular automaton with rule 17348625. Note that the $y$ axis is rescaled by $4^k$ in order to spread the data out and make it more easily visible.
  }
  \label{fig:IVa_manygaps}
\end{figure}

\newpage
We also check the scaling of all real eigenvalues and confirm that they all exponentially approach the unit circle with increased support $r$. See Figure~\ref{fig:IVa_manygaps}.

\subsection{Subclass IVb}

\textbf{Trivial and free rules.} We call the simplest rules in Subclass IVb \textit{trivial} due to their constant return time. The spectrum of a trivial map is highly degenerate and each state is mapped back to itself within a few periods, regardless of system size $L$. The constant mean return times we find are 1 or 3 time steps. Moreover, all of the rules have correlation functions that decay to a nonzero constant. 

Next, the simplest nontrivial rules exhibit a linearly increasing mean return time $T(L) \propto L$. We call these rules \textit{free} because quasiparticles of each color propagate with a constant velocity and do not scatter with quasiparticles of the same or any other color. However, sometimes quasiparticles of
different colors can flip colors when scattering but do not experience any time shifts. All such rules also have constant correlators. 

Here, we shown an example of the dynamics of a free cellular automaton with rule 31256487 and a trivial automaton with rule 65413287.
\\

\begin{figure}[h!]
\centering
  \includegraphics[width=1.0\linewidth]{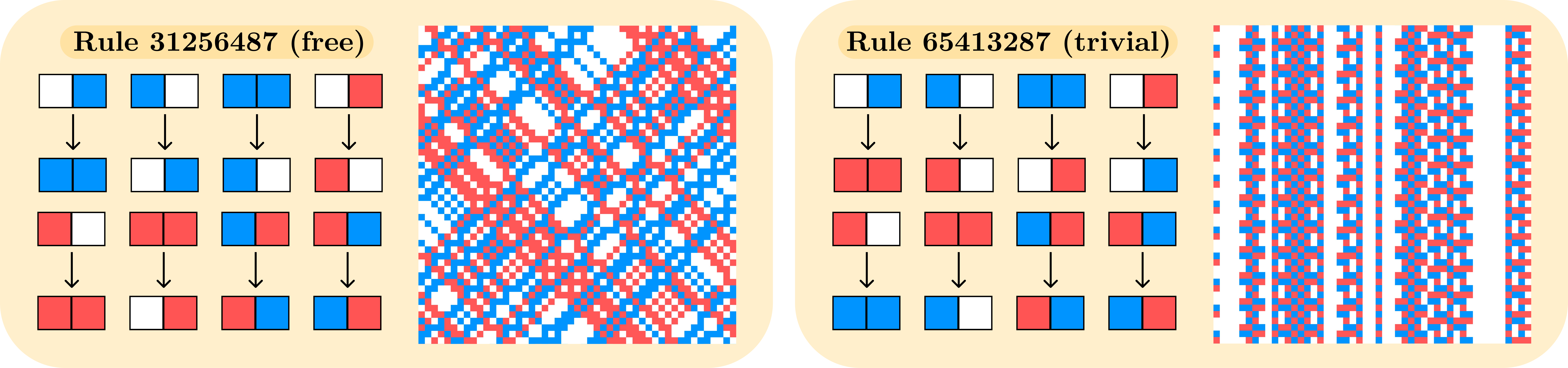}
  \label{fig:free_trivial_dynamics}
\end{figure}

\textbf{Models with polynomial return times.} Physically more interesting models in Class IV also exist that exhibit a mean return time which increases faster than linearly. Here, we discuss three such models, each with a different polynomial growth of the mean return time. Along with the hardcore charged gas, for which the mean return time scales as $L^2$, we consider a case for which it scales as $\sim L^3$ and one with $\sim L^4$.  The latter two examples possess fewer charges at the same support $r$ and might present interesting 
future playgrounds for finding exactly solvable models.

Firstly, the hardcore charged gas with rule 21354678 and a quadratically increasing mean return time $T(L) \sim L^2$ is shown here.

\begin{figure}[h!]
\centering
\includegraphics[width=0.55\linewidth]{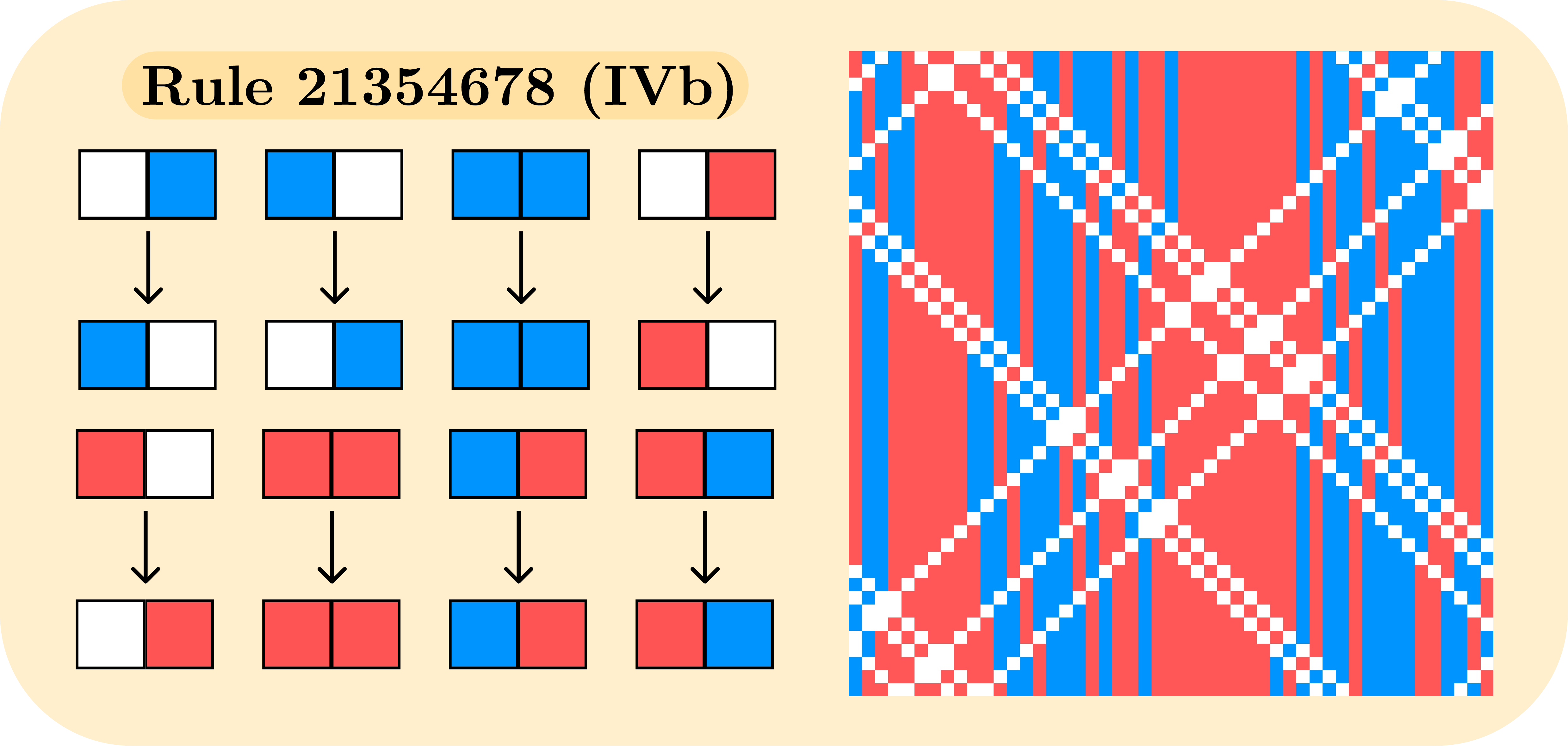}
  \label{fig:IV1_rule}
\end{figure}

Secondly, we show the cellular automaton with rule 34261578 that has a cubically increasing return time $T(L) \sim L^3$.

\begin{figure}[h!]
\centering
\includegraphics[width=0.55\linewidth]{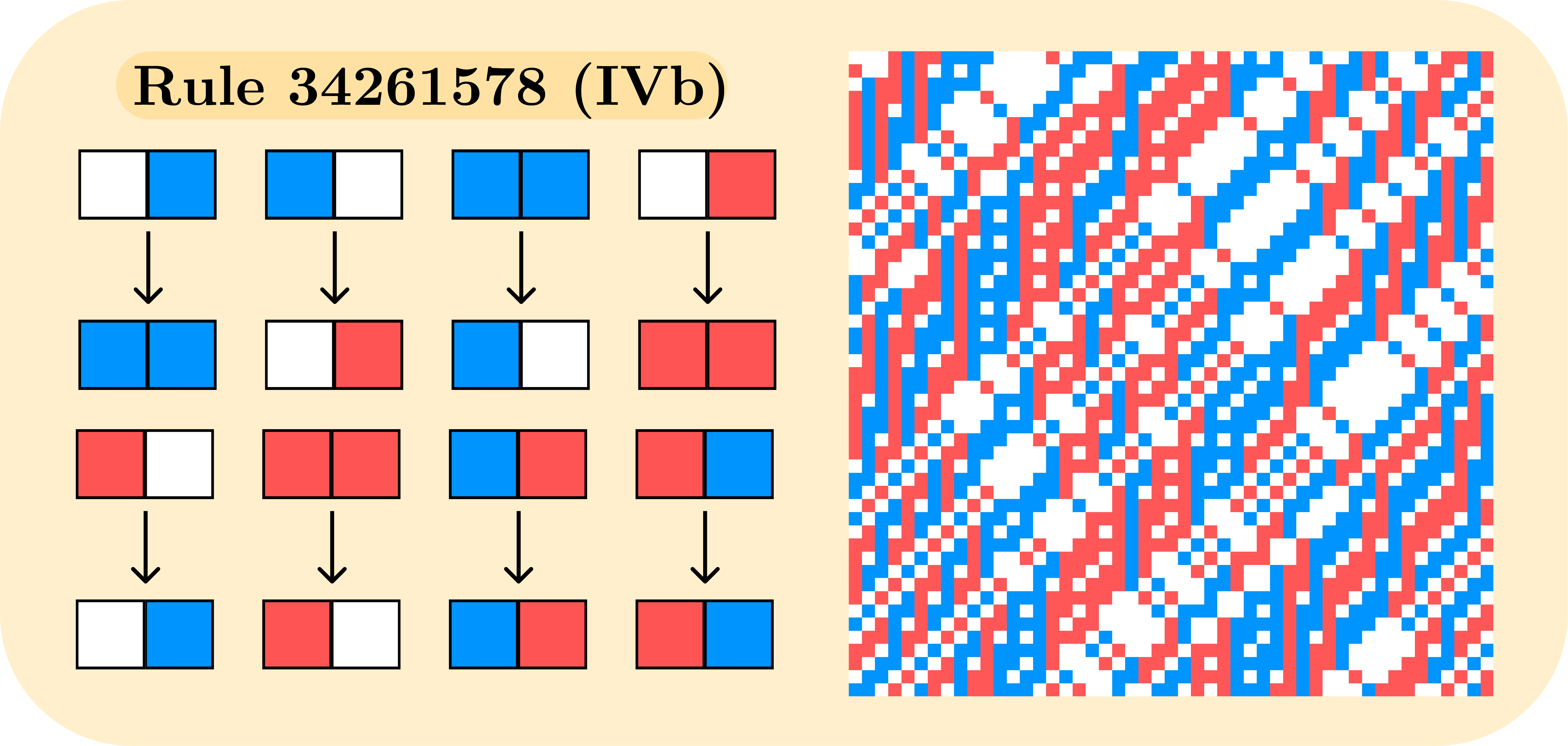}
  \label{fig:IV2_rule}
\end{figure}

Thirdly, we also present a model with rule 23651478 that exhibits a mean return time scaling as $T(L) \sim L^4$.

\begin{figure}[h!]
\centering
\includegraphics[width=0.55\linewidth]{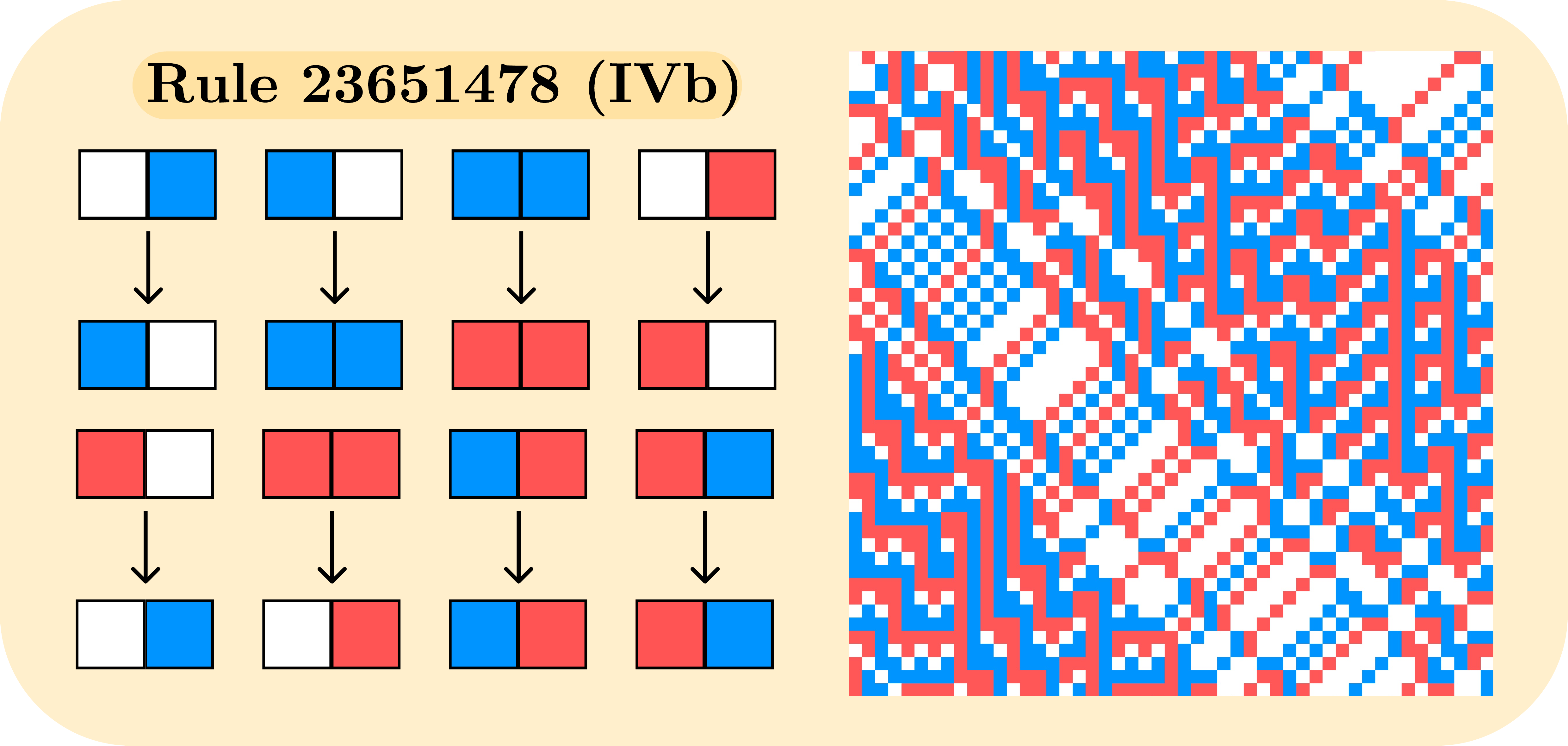}
  \label{fig:IV3_rule}
\end{figure}

\begin{figure}[h!]
\centering
  \includegraphics[width=1.0\linewidth]{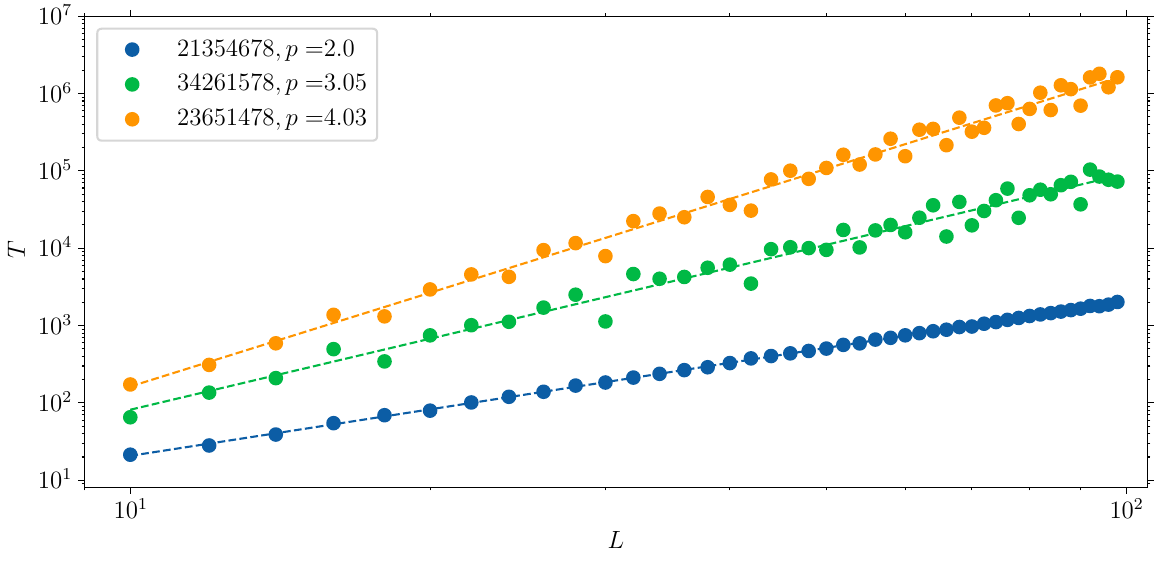}
  \caption{Scaling of the mean return time $T(L) \sim L^p$ for three examples of RCA with their rules and the best fits to $p$ stated in the legend.
  }
  \label{fig:poly_rules_snapshots}
\end{figure}

\newpage
The mean return times for the three RCA with different rates of polynomial growth are shown in Figure~\ref{fig:poly_rules_snapshots}. 

Finally, we note that these models exhibit polynomial decay of at least one correlator with a dynamical exponent $1/z_{C,D} = 1/2$.\footnote{The exception are some $L^2$ rules where the correlator can decay exponentially to a small constant value. An example is the automaton with rule 24651387.}

\newpage

\section{Conclusions and outlook}

This work attempts to propose an ergodic hierarchy of discrete state-space many-body locally interacting reversible dynamical systems. We focus on a family of three-state RCA with local brickwork rules. Postulating an invariant state (also called the vacuum state, or vacancy), we can consider the remaining two states as $\pm$ charge and focus on the rules with simple transformation properties under action of time reversal, charge conjugation and spatial reflection. We find a rich and diverse set of dynamical behaviors which allow us to propose a classification of RCA models into four types (classes). We use four empirical dynamical ergodicity indicators that are suitable for numerical computation, specifically, the mean return times, the volume scaling of the number of conserved quantities, Ruelle-Pollicott resonances, and correlation functions. Different RCA are then ordered in a specific manner from chaotic rules in Class I to superintegrable rules in Class IV.

A key theme explored in the paper is the role of mean return or recurrence times in identifying fragmentation in configuration space. While scaling of return times is broadly indicative of strong fragmentation, our results demonstrate that mean return times do not seem to be a reliable measure of an apparent complexity of an automaton. This highlights the necessity of complementary metrics such as counting local conserved charges and decay of local correlation functions.

In addition to strictly local conserved quantities, we also identify quasilocal charges in many systems. These charges are characterized by a spectral gap $\Delta$ of the corresponding dynamical transfer matrix which decreases exponentially when the maximal support size of the observables considered ($r$) increases. Additionally, the coefficients of the eigenvector associated to a quasilocal charge exhibits exponential decay with support size of the basis observable, hinting at its convergence in the thermodynamic limit. 

We find a variety of transport phenomena, mostly, but not only with the dynamical exponent $z=2$. They mainly appear in Classes III and IV, and with some also in Class II. Some of these seem to suggest normal diffusive behavior, while others show exceptional (exotic) dynamical structure factors. On top of this we find a plethora of rules with anomalous transport $1 > 1/z \neq 1/2$. We identify possible examples of superdiffusive transport $1/z=2/3$ with an unknown (non-KPZ) scaling function. Furthermore, we find a clean example of subdiffusive transport with $1/z=1/3$.

Interestingly, our investigation also uncovered numerous models of RCA that exhibit integrable-like behavior characterized by an extensive number (proportional to the volume) of conserved quantities that do not conform to the set-theoretic Yang-Baxter equation. The question remains whether broader algebraic integrability criteria exist for such systems and how they conceptually differ from `conventional' integrability results.

The results here present the motivation for further theoretical and numerical efforts focused on specific models, which might present themselves as interesting to the reader. Broader connections to classical or quantum integrability, statistical mechanics and dynamical systems are possible but the development of general mathematical methods seems to be a necessary next step.
\\\\
We close by highlighting a few promising further directions of research:
\begin{itemize}
    \item A natural extension of our work is to investigate the robustness of our results under perturbations that disrupt the deterministic nature of dynamics of our models. We propose to introduce stochasticity, which maintains the ability to efficiently simulate correlation functions using Monte Carlo techniques. Adding a dissipative contribution into our deterministic dynamics would enable us to study the dissipative form factor, which has gained significant interest recently \cite{Can2019,Li2021,Shivam2023}. Within the framework of open cellular automata, an interesting direction would be to explore the relationship between ergodicity properties and anomalous relaxation along the ideas of Refs.~\cite{Yoshimura2024,Lucas_PR}.

    \item It would be interesting to analyze other measures of chaos which are widely used to distinguish integrable systems from ergodic ones. From the classical point of view, analyzing Lyapunov exponents and topological entropies for models in diferent classes could provide new insights into the classification proposed in thisfferent classes could provide new insights into the classification proposed in this paper. If we treat our systems as quantum cellular automata (i.e., considering Floquet circuits with three local quantum states that map product states to product states), then we can analyze the known measures of quantum chaos. Namely, the out-of-time-ordered correlators \cite{Larkin1969QuasiclassicalMI}, the spectral form factor \cite{SFF_semiclass}, the Loschmit echo \cite{Gorin2006}, Krylov complexity \cite{Krylov}, Hilbert space geometry \cite{GeometryRustem} and pole skipping \cite{Grozdanov:2017ajz,Blake:2017ris} are all worth considering as they would undoubtedly help in making our classification more comprehensive.

    \item Several RCA models with volume extensive scaling of number of local conserved charges suggests that algebraic integrability
    techniques should be developed to treat the dynamics and the spectral problem for these systems.
\end{itemize}

\section*{Acknowledgements}
We would like to thank Luca Delacr\'{e}taz, Matthew Dodelson, Urban Duh, Enej Ilievski, Yusuf Kasim and Luka Paljk for useful discussions. We acknowledge funding from European Research Council (ERC) through Advanced grant QUEST (Grant Agreement No. 101096208, T.P. and R.S.), as well as the Slovenian Research and Innovation agency (ARIS) through the Program P1-0402 (All authors) and Grants N1-0219 (T.P.), N1-0368 (T.P.), N1-0245 (S.G.), J7-60121 (S.G.), and by the STFC Ernest Rutherford Fellowship ST/T00388X/1 (S.G.).

\appendix

\section{Numerical algorithm for computing conserved quantities} \label{sec:numerical_charges}

To find conserved quantities in our RCA, we follow the procedure of Refs.~\cite{Klobas_2022, Prosen_2007} and implement it numerically. Refs.~\cite{Urban,Znidaric2024} considered a generalization of the algorithm to operators and observables without translational invariance and with dependence on momentum.

\subsection{Basis in the space of extensive local observables}
Here, we describe the way to define the basis in the vector space $\mathcal{A}_{\mathbb{T}}^{(r)}$. First, recall the definition of the basis elements of vector space $\mathcal{A}^{(r)}_x$ of local observables:
\begin{equation}\label{basis_alpha}
   [\alpha_1, \alpha_2 \l \alpha_r]_x = [\alpha_1]_x [\alpha_2]_{j+1} ... [\alpha_r]_{x+r-1},~~~~~ [\alpha]_x (s) = \delta_{\alpha, s_x},
\end{equation}
\noindent
where $\alpha_i \in \{ +,-,\varnothing \}$. The observable $[\alpha]_x :X^L \rightarrow \mathbb{C}$ has the physical meaning of a number of $\alpha$ particles at site $x$. 

The local observables add up to a unit element of the algebra $[+]+[-]+[\varnothing]=\mathds{1}$. We want to change the basis, such that the identity is one of the basis elements. We change the basis to an orthogonal one in the following way
\begin{equation}
    [0] = [\emptyset] + [+] + [-], \quad [1] = [+] - [-], \quad [2] = -2[\emptyset] + [+] + [-],
\end{equation}
where $[0] \equiv \mathds{1}$. The transformation can be expressed as
\begin{equation}
    \begin{bmatrix}
        [0] \\
        [1] \\
        [2] \\
    \end{bmatrix}
    =
    R
     \begin{bmatrix}
        [\emptyset] \\
        [+] \\
        [-] \\
    \end{bmatrix},
    \quad 
    R \equiv 
    \begin{bmatrix}
        1 & 0 & -2 \\
        1 & 1 & 1 \\
        1 & -1 & 1 \\
    \end{bmatrix}.
\end{equation}

Next, we wish to build a map between the space of local observables and extensive local observables. The difficulty is that two local densities can give the same extensive observable after translations over the lattice. For example, $q_1=[0022]$ and $q_2=[2200]$ are different elements of $\mathcal{A}^{(4)}_1$, however, the extensive observables corresponding to them $Q_1=\sum_j\mathbb{T}^{2j}q_1$ and $Q_2=\sum_j\mathbb{T}^{2j}q_2$ are equivalent, $Q_1 \equiv Q_2$. 

Thus, in order to find densities $[\alpha_1,...\alpha_r]$ of observables, that form a basis in $\mathcal{A}_{\mathbb{T}}^{(r)}$, we exclude ones with two or more zeros on the left (except for the identity observable). The dimension  of the vector space is  then $\dim\left( \mathcal{A}_{\mathbb{T}}^{(r)} \right) =3^r- 3^{r-2}+1$ and the basis elements are
\begin{equation}
  A= \sum_j \mathbb{T}^{2j} a,~~~~ a=[a_1 a_2...a_r] :~~~~a_i\in\{0,1,2\},\quad a_1+a_2\neq 0
\end{equation}
and the identity observable. We define the projector of the density to this basis, which removes $2 m$ zeros from the the beginning of string to the end
\begin{equation}
    P_r [\underbrace{00...}_{2m}  a_1 a_2... a_k] = [a_1 a_2... a_k \underbrace{00...}_{r-2m}],\quad a_1+a_2\neq 0.
\end{equation}

 The vector space $\mathcal{A}_{\mathbb{T}}^{(r)}\setminus \mathcal{A}_{\mathbb{T}}^{(r-1)}$ has basis elements with the  densities $a=[a_1 a_2...a_r], \text{ where}~~(a_1+a_2)\cdot a_r\neq 0$; the condition $a_r\neq 0$ was imposed for observable not to be in the subspace $\mathcal{A}_{\mathbb{T}}^{(r-1)}$.

\subsection{Numerical algorithm}
\noindent
In Eq.~\eqref{eq:evolution}, we defined the evolution  operator $\mathbb{U}$. We can find its $3^L \times 3^L$ dimensional matrix representation (the propagator) $\mathcal{U}$ in the basis of $\mathcal{A}$ given by Eq.~\eqref{basis_alpha}. Note that this matrix can be factorized into a product of two-site propagators $\mathcal{U}_{x,x+1}$, which act only on the subspace $\mathcal{A}^{(2)}_x$ of $\mathcal{A}$ in the following way:
\begin{equation}\label{even_odd_op}
    \mathcal{U} = \underbrace{ \mathcal{U}_{23} \mathcal{U}_{45} \dots \mathcal{U}_{r,1} }_{\mathcal{U}_{\mathrm{even}}} \underbrace{\ \mathcal{U}_{12} \mathcal{U}_{34} \dots \mathcal{U}_{r-1,r}   }_{\mathcal{U}_{\mathrm{odd}}}.
\end{equation}
\noindent
We have additionally defined the matrix representations of the even and odd step.
Notice, that the in the matrix form $\mathcal{U}=\mathcal{V}^{-1}$ and $\mathcal{U}_{x,x+1}=\left(U_{x,x+1}\right)^{-1}$ due to the definition of evolution of observables Eq.~\eqref{eq:evolution}.

Since we are interested in the evolution of extensive observables, we change the basis from $\{+,-,\varnothing\}$ to $\{ 0,1,2\}$. Then, local propagator in new basis can be expressed as
\begin{equation}
    V_{x,x+1} := (R \otimes R)^{-1} \mathcal{U}_{x,x+1} (R \otimes R).
\end{equation}
In an analogous way to Eq.~\eqref{even_odd_op}, we can build the full evolution matrix $V=V_{even}\cdot V_{odd}$ from products of two-site propagators $V_{x,x+1}$. Notice that due to the local properties of the propagator, the evolution operator $\mathbb{U}$ maps $\mathcal{A}_{\mathbb{T}}^{(r)}$ to $\mathcal{A}_{\mathbb{T}}^{(r+3)}$, and the evolution of extensive local observables can be determined by the evolution of local densities.

The numerical construction of the transfer matrix $\mathcal{T}^{(r)} :\mathcal{A}_{\mathbb{T}}^{(r)}\rightarrow\mathcal{A}_{\mathbb{T}}^{(r)}$ can be done in the following steps: 
\begin{enumerate}
    \item Consider a basis translationally invariant observable $A$ with local density $a=[a_1 a_2...a_r]$, where $a_i\in \{0,1,2\}$ and $a_1+a_2\neq 0$. 
    \item The density of the evolved observable $B=\mathbb{U}A$ will have the following form:
    \begin{equation}
        b=\sum\limits_{b_i} t_{b_1,...b_{r+2}} [0 b_1 ...b_{r+2}],
     \end{equation}
     where 
     \begin{equation} t_{b_1...b_{r+2}}=\sum\limits_{s_i}V^{b_1,b_2}_{0,s_1}V^{b_3,b_4}_{s_2,s_3}...V^{b_{r+1},b_{r+2}}_{s_{r},0} \cdot V^{s_1,s_2}_{a_1,a_2} ...V^{s_{r-1},s_r}_{a_{r-1},a_r}.
    \end{equation}
    \item Next, we need to project the density $b$ to the unique basis of $\mathcal{A}^{(r+3)}_{\mathbb{T}}$. We get another vector 
    \begin{equation}
        b'=P_{r+3}b=\sum\limits_{b_i} t'_{b_1, ...,b_{r+3}}[b_1... b_{r+3}].
    \end{equation}
    \item 
    Finally, we truncate $b'$ to the subspace of observables of range $r$. Namely, we apply $P_{\mathcal{A}^{(r)}_{\mathbb{T}}}[b_1~...~b_{r}b_{r+1}b_{r+2}b_{r+3}]= [b_1~...~b_{r}]$ if $b_{r+1}=b_{r+2}=b_{r+3}=0$ and $0$ otherwise. This determines the component of the transfer matrix $\left(\mathcal{T}^{(r)}\right)^{b_1,...,b_r}_{a_1,...a_r}=t'_{b_1, ...,b_r,0,0,0}$.
\end{enumerate}

Repeating these steps over all basis densities $[a_1...a_r]$ of $\mathcal{A}_{\mathbb{T}}^{(r)}$, we find the matrix representation of the operator $\mathcal{T}^{(r)}$.

\noindent

\section{Detailed properties of 3-state RCA}
\label{app:uc}
In this appendix, we present a table of compiled selected data, which is useful for finding interesting models. We note that all the computed data is available at the \href{https://doi.org/10.5281/zenodo.17117721}{Zenodo repository}. The rules are ordered according to different classes they belong to and symmetries they exhibit. For each rule, we state the following:

\begin{itemize}
    \item How the mean return time scales with system size $L$ --- either exponentially $\sim e^{\kappa L}$ or as a power-law $\sim L^p$ --- along with the parameter $\kappa$ or $p$.
    \item How the correlation function $C$ (defined in Eq.~\eqref{eq:correlation}) behaves and how it scales with parameter $p_C$, which is either the exponential rate $\alpha$ or the power-law $1/z_C$, or constant (the long-time mean of $C$).
    \item How the correlation function $D$ (defined in Eq.~\eqref{eq:correlation}) behaves and how it scales with parameter $p_D$, which is either the exponential rate $\beta$ or the power-law $1/z_D$, or constant (the long-time mean of $D$).
    \item The number of local conserved charges found, starting with support $r=2$ and increasing it by $2$ every step.
\end{itemize}

We note that the dual reversible rules are distinguished by a star $^*$ after the group label. The East rules are distinguished by a superscript $E$ after the group label.

We also note that all floating point data is printed to three digits in precision, because we believe the error of the fit underestimates the actual systematic error of the fitting procedure. Where the estimated fitting error is larger or equal to three digits, the inaccurate digit is written in parentheses.

Finally, the extended long-time simulations done with systems of size $2^{20}$ are used for the subdiffusive rules 46532178, 16435278, 14625387, 45678321, 45321678, 25614378 and 12654378. The results are presented in Appendix~\ref{AppD}.

\setlength{\LTleft}{\fill}
\setlength{\LTright}{\fill}
\scriptsize
\begin{longtable}{|c|c|c|c|c|c|c|c|c|c|}
\hline
Rule & Group & Class & Return & $\kappa / p$ & C & $1/z_C$ & D & $1/z_D$ & $|Q|$ \\
\hline
\endfirsthead
\hline
Rule & Group & Class & Return & $\kappa / p$ & C & $1/z_C$ & D & $1/z_D$ & $|Q|$ \\
\hline
\endhead
\rowcolor{ClassI!10}
54783162 & $\mathcal{ CPT } $ & I & exp & 0.54(1) & exp & 0.048 & exp & 0.041 & 0 0 0 0 0 0 \\
\thinline
\rowcolor{ClassI!25}
34678521 & $\mathcal{ CPT } $ & I & exp & 0.55(3) & exp & 0.086 & exp & 0.153 & 0 0 0 0 0 \\
\thinline
\rowcolor{ClassI!10}
14735268 & $\mathcal{ CPT } $ & I & exp & 0.55(1) & exp & 0.086 & exp & 0.021 & 0 0 0 0 0 \\
\thinline
\rowcolor{ClassI!25}
45628731 & $\mathcal{ CPT } $ & I & exp & 0.55(2) & exp & 0.086 & exp & 0.022 & 0 0 0 0 0 \\
\thinline
\rowcolor{ClassI!10}
45723168 & $\mathcal{ CPT } $ & I & exp & 0.54(1) & exp & 0.066 & exp & 0.023 & 0 0 0 0 0 \\
\thinline
\rowcolor{ClassI!25}
58673124 & $\mathcal{ CPT } $ & I & exp & 0.548 & exp & 0.092 & exp & 0.074 & 0 0 0 0 0 \\
\thinline
\rowcolor{ClassI!10}
56481372 & $\mathcal{ CPT } $ & I & exp & 0.54(2) & exp & 0.025 & exp & 0.114 & 0 0 0 0 0 \\
\thinline
\rowcolor{ClassI!25}
14685372 & $\mathcal{ CPT } $ & I & exp & 0.54(1) & exp & 0.114 & exp & 0.071 & 0 0 0 0 0 \\
\thinline
\rowcolor{ClassI!10}
34612578 & $\mathcal{ CPT } $ & I & exp & 0.54(1) & exp & 0.114 & exp & 0.071 & 0 0 0 0 0 \\
\thinline
\rowcolor{ClassI!25}
16475823 & $\mathcal{ CPT } $ & I & exp & 0.55(2) & exp & 0.075 & exp & 0.049 & 0 0 0 0 0 0 \\
\thinline
\rowcolor{ClassI!10}
17685342 & $\mathcal{ CPT } $ & I & exp & 0.54(1) & exp & 0.14 & exp & 0.074 & 0 0 0 0 0 \\
\thinline
\rowcolor{ClassI!25}
37612548 & $\mathcal{ CPT } $ & I & exp & 0.54(1) & exp & 0.121 & exp & 0.084 & 0 0 0 0 0 \\
\thinline
\rowcolor{ClassI!10}
37621548 & $\mathcal{ CPT } $ & I & exp & 0.55(2) & exp & 0.025 & exp & 0.113 & 0 0 0 0 0 0 \\
\thinline
\rowcolor{ClassI!25}
45623178 & $\mathcal{ CPT } $ & I & exp & 0.54(1) & exp & 0.037 & exp & 0.017 & 0 0 0 0 0 \\
\thinline
\rowcolor{ClassI!10}
36421587 & $\mathcal{ CPT } $ & I & exp & 0.55(1) & exp & 0.121 & exp & 0.074 & 0 0 0 0 0 \\
\thinline
\rowcolor{ClassI!25}
45627318 & $\mathcal{ CPT } $ & I & exp & 0.54(1) & exp & 0.054 & exp & 0.044 & 0 0 0 0 0 \\
\thinline
\rowcolor{ClassI!10}
13725468 & $\mathcal{ CPT } $ & I & exp & 0.55(2) & exp & 0.049 & exp & 0.039 & 0 0 0 0 0 \\
\thinline
\rowcolor{ClassI!25}
56483172 & $\mathcal{ CPT } $ & I & exp & 0.54(1) & exp & 0.033 & exp & 0.081 & 0 0 0 0 0 \\
\thinline
\rowcolor{ClassI!10}
45627813 & $\mathcal{ CPT } $ & I & exp & 0.55(2) & exp & 0.049 & exp & 0.039 & 0 0 0 0 0 \\
\thinline
\rowcolor{ClassI!25}
14675328 & $\mathcal{ CPT } $ & I & exp & 0.54(2) & exp & 0.037 & exp & 0.015 & 0 0 0 0 0 \\
\thinline
\rowcolor{ClassI!10}
16485372 & $\mathcal{ CPT } $ & I & exp & 0.54(1) & exp & 0.121 & exp & 0.085 & 0 0 0 0 0 \\
\thinline
\rowcolor{ClassI!25}
13675428 & $\mathcal{ CPT } $ & I & exp & 0.55(2) & exp & 0.048 & exp & 0.05 & 0 0 0 0 0 \\
\thinline
\rowcolor{ClassI!10}
38612574 & $\mathcal{ CPT } $ & I & exp & 0.55(1) & exp & 0.094 & exp & 0.085 & 0 0 0 0 0 \\
\thinline
\rowcolor{ClassI!25}
13685472 & $\mathcal{ CPT } $ & I & exp & 0.55(1) & exp & 0.094 & exp & 0.086 & 0 0 0 0 0 \\
\thinline
\rowcolor{ClassI!10}
57681342 & $\mathcal{ CPT } $ & I & exp & 0.55(1) & exp & 0.023 & exp & 0.065 & 0 0 0 0 0 \\
\thinline
\rowcolor{ClassI!25}
14875326 & $\mathcal{ CPT } $ & I & exp & 0.54(1) & exp & 0.066 & exp & 0.023 & 0 0 0 0 0 \\
\thinline
\rowcolor{ClassI!10}
36421578 & $\mathcal{ CPT } $ & I & exp & 0.55(2) & exp & 0.023 & exp & 0.063 & 0 0 0 0 0 0 \\
\thinline
\rowcolor{ClassI!25}
16475328 & $\mathcal{ CPT } $ & I & exp & 0.55(2) & exp & 0.054 & exp & 0.017 & 0 0 0 0 0 \\
\thinline
\rowcolor{ClassI!10}
34186572 & $\mathcal{ CPT } $ & I & exp & 0.54(1) & exp & 0.04 & exp & 0.044 & 0 0 0 0 0 \\
\thinline
\rowcolor{ClassI!25}
13875426 & $\mathcal{ CPT } $ & I & exp & 0.54(2) & exp & 0.041 & exp & 0.05 & 0 0 0 0 0 \\
\thinline
\rowcolor{ClassI!10}
37628541 & $\mathcal{ CPT } $ & I & exp & 0.54(1) & exp & 0.033 & exp & 0.081 & 0 0 0 0 0 \\
\thinline
\rowcolor{ClassI!25}
56437218 & $\mathcal{ CPT } $ & I & exp & 0.55(3) & exp & 0.085 & exp & 0.152 & 0 0 0 0 0 \\
\thinline
\rowcolor{ClassI!10}
14685732 & $\mathcal{ CPT } $ & I & exp & 0.55(2) & exp & 0.171 & exp & 0.102 & 0 0 0 0 0 \\
\thinline
\rowcolor{ClassI!25}
37821546 & $\mathcal{ CPT } $ & I & exp & 0.55(1) & exp & 0.131 & exp & 0.139 & 0 0 0 0 0 \\
\thinline
\rowcolor{ClassI!10}
57683142 & $\mathcal{ CPT } $ & I & exp & 0.542 & exp & 0.028 & exp & 0.029 & 0 0 0 0 0 \\
\thinline
\rowcolor{ClassI!25}
57381642 & $\mathcal{ CPT } $ & I & exp & 0.54(1) & exp & 0.121 & exp & 0.08 & 0 0 0 0 0 \\
\thinline
\rowcolor{ClassI!10}
45628371 & $\mathcal{ CPT } $ & I & exp & 0.54(1) & exp & 0.05 & exp & 0.021 & 0 0 0 0 0 \\
\thinline
\rowcolor{ClassI!25}
45738261 & $\mathcal{ CPT } $ & I & exp & 0.54(2) & exp & 0.075 & exp & 0.05 & 0 0 0 0 0 0 \\
\thinline
\rowcolor{ClassI!10}
14675823 & $\mathcal{ CPT } $ & I & exp & 0.55(1) & exp & 0.067 & exp & 0.037 & 0 0 0 0 0 0 \\
\thinline
\rowcolor{ClassI!25}
38712564 & $\mathcal{ CPT } $ & I & exp & 0.55(2) & exp & 0.141 & exp & 0.176 & 0 0 0 0 0 \\
\thinline
\rowcolor{ClassI!10}
54637218 & $\mathcal{ CPT } $ & I & exp & 0.54(2) & exp & 0.015 & exp & 0.033 & 0 0 0 0 0 \\
\thinline
\rowcolor{ClassI!25}
16485732 & $\mathcal{ CPT } $ & I & exp & 0.54(2) & exp & 0.148 & exp & 0.142 & 0 0 0 0 0 \\
\thinline
\rowcolor{ClassI!10}
34628571 & $\mathcal{ CPT } $ & I & exp & 0.53(2) & exp & 0.015 & exp & 0.033 & 0 0 0 0 0 \\
\thinline
\rowcolor{ClassI!25}
14835276 & $\mathcal{ CPT } $ & I & exp & 0.539 & exp & 0.061 & exp & 0.019 & 0 0 0 0 0 \\
\thinline
\rowcolor{ClassI!10}
34812576 & $\mathcal{ CPT } $ & I & exp & 0.55(2) & exp & 0.171 & exp & 0.099 & 0 0 0 0 0 \\
\thinline
\rowcolor{ClassI!25}
45728361 & $\mathcal{ CPT } $ & I & exp & 0.54(1) & exp & 0.061 & exp & 0.02 & 0 0 0 0 0 \\
\thinline
\rowcolor{ClassI!10}
36471528 & $\mathcal{ CPT } $ & I & exp & 0.55(1) & exp & 0.093 & exp & 0.076 & 0 0 0 0 0 \\
\thinline
\rowcolor{ClassI!25}
45638271 & $\mathcal{ CPT } $ & I & exp & 0.55(2) & exp & 0.054 & exp & 0.016 & 0 0 0 0 0 \\
\thinline
\rowcolor{ClassI!10}
36428571 & $\mathcal{ CPT } $ & I & exp & 0.54(1) & exp & 0.029 & exp & 0.029 & 0 0 0 0 0 0 \\
\thinline
\rowcolor{ClassI!25}
37812546 & $\mathcal{ CPT } $ & I & exp & 0.54(2) & exp & 0.148 & exp & 0.142 & 0 0 0 0 0 \\
\thinline
\rowcolor{ClassI!10}
13625487 & $\mathcal{ CPT } $ & I & exp & 0.55(1) & exp & 0.16 & exp & 0.127 & 0 0 0 0 0 \\
\thinline
\rowcolor{ClassI!25}
34687512 & $\mathcal{ CPT } $ & I & exp & 0.54(1) & exp & 0.04 & exp & 0.044 & 0 0 0 0 0 \\
\thinline
\rowcolor{ClassI!10}
13785462 & $\mathcal{ CPT } $ & I & exp & 0.55(2) & exp & 0.141 & exp & 0.174 & 0 0 0 0 0 \\
\thinline
\rowcolor{ClassI!25}
45731268 & $\mathcal{ CPT } $ & I & exp & 0.54(1) & exp & 0.067 & exp & 0.036 & 0 0 0 0 0 0 \\
\thinline
\rowcolor{ClassI!10}
56481732 & $\mathcal{ CPT } $ & I & exp & 0.55(1) & exp & 0.131 & exp & 0.138 & 0 0 0 0 0 \\
\thinline
\rowcolor{ClassI!25}
14635278 & $\mathcal{ CPT } $ & I & exp & 0.54(1) & exp & 0.049 & exp & 0.02 & 0 0 0 0 0 \\
\thinline
\rowcolor{ClassI!10}
54738261 & $\mathcal{ CPT }^{*}$ & I & exp & 0.55(2) & exp & 0.265 & exp & 0.274 & 0 0 0 0 0 0 \\
\thinline
\rowcolor{ClassI!25}
56471823 & $\mathcal{ CPT }^{*}$ & I & exp & 0.55(2) & exp & 0.264 & exp & 0.268 & 0 0 0 0 0 0 \\
\thinline
\rowcolor{ClassI!10}
45783162 & $\mathcal{ CPT } $ & I & exp & 0.54(2) & exp & 0.041 & exp & 0.049 & 0 0 0 0 0 \\
\thinline
\rowcolor{ClassI!25}
45637218 & $\mathcal{ CPT } $ & I & exp & 0.54(2) & exp & 0.048 & exp & 0.051 & 0 0 0 0 0 \\
\thinline
\rowcolor{ClassI!10}
45781362 & $\mathcal{ CPT } $ & I & exp & 0.55(1) & exp & 0.046 & exp & 0.048 & 0 1 1 1 1 \\
\thinline
\hline
\rowcolor{ClassI!25}
13465278 & $\mathcal{ CP } $ & I & exp & 1.0(2) & exp & 0.102 & exp & 0.108 & 0 0 0 0 0 \\
\thinline
\rowcolor{ClassI!10}
34526187 & $\mathcal{ CP } $ & I & exp & 0.57(2) & exp & 0.047 & exp & 0.069 & 0 0 0 0 0 \\
\thinline
\rowcolor{ClassI!25}
35416278 & $\mathcal{ CP } $ & I & exp & 1.11(8) & exp & 0.18 & exp & 0.161 & 0 0 0 0 0 \\
\thinline
\rowcolor{ClassI!10}
34526178 & $\mathcal{ CP } $ & I & exp & 1.02(5) & exp & 0.141 & exp & 0.136 & 0 0 0 0 0 \\
\thinline
\hline
\rowcolor{ClassI!25}
56431278 & $\mathcal{ CP+T }^{*}$ & I & exp & 0.55(2) & exp & 0.265 & exp & 0.269 & 0 0 0 0 0 0 \\
\thinline
\rowcolor{ClassI!10}
13265487 & $\mathcal{ CP+T } $ & I & exp & 0.57(1) & exp & 0.064 & exp & 0.051 & 0 0 0 0 0 \\
\thinline
\hline
\rowcolor{ClassI!25}
45387612 & $\mathcal{ CT+P } $ & I & exp & 0.57(1) & exp & 0.065 & exp & 0.049 & 0 0 0 0 0 \\
\thinline
\rowcolor{ClassI!10}
54678321 & $\mathcal{ CT+P }^{*}$ & I & exp & 0.55(3) & exp & 0.264 & exp & 0.271 & 0 0 0 0 0 0 \\
\thinline
\hline
\rowcolor{ClassI!25}
12678354 & $\mathcal{ P } $ & I & exp & 1.11(8) & exp & 0.18 & exp & 0.156 & 0 0 0 0 0 \\
\thinline
\rowcolor{ClassI!10}
45678312 & $\mathcal{ P } $ & I & exp & 1.0(2) & exp & 0.102 & exp & 0.108 & 0 0 0 0 0 \\
\thinline
\rowcolor{ClassI!25}
54378612 & $\mathcal{ P } $ & I & exp & 0.569 & exp & 0.047 & exp & 0.07 & 0 0 0 0 0 \\
\thinline
\rowcolor{ClassI!10}
54678312 & $\mathcal{ P } $ & I & exp & 1.01(5) & exp & 0.141 & exp & 0.136 & 0 0 0 0 0 \\
\thinline
\thickline
\rowcolor{ClassIIa!25}
34176528 & $\mathcal{ CPT } $ & IIa & exp & 0.56(2) & exp & 0.017 & exp & 0.018 & 0 1 2 2 2 \\
\thinline
\rowcolor{ClassIIa!10}
56487312 & $\mathcal{ CPT } $ & IIa & exp & 0.55(2) & exp & 0.017 & exp & 0.018 & 0 1 2 2 2 \\
\thinline
\rowcolor{ClassIIa!25}
38721564 & $\mathcal{ CPT } $ & IIa & exp & 0.55(1) & exp & 0.036 & exp & 0.037 & 0 1 1 1 1 1 \\
\thinline
\rowcolor{ClassIIa!10}
13825476 & $\mathcal{ CPT } $ & IIa & exp & 0.55(1) & exp & 0.046 & exp & 0.048 & 0 1 1 1 1 \\
\thinline
\rowcolor{ClassIIa!25}
34827516 & $\mathcal{ CPT } $ & IIa & exp & 0.55(1) & exp & 0.036 & exp & 0.038 & 0 1 1 1 1 1 \\
\thinline
\rowcolor{ClassIIa!10}
45328671 & $\mathcal{ CPT } $ & IIa & exp & 0.55(1) & exp & 0.031 & exp & 0.034 & 0 1 1 1 1 \\
\thinline
\rowcolor{ClassIIa!25}
17385642 & $\mathcal{ CPT } $ & IIa & exp & 0.55(3) & exp & 0.036 & exp & 0.04 & 0 1 1 1 1 \\
\thinline
\rowcolor{ClassIIa!10}
14635287 & $\mathcal{ CPT } $ & IIa & exp & 0.54(1) & exp & 0.032 & exp & 0.034 & 0 1 1 1 1 1 \\
\thinline
\rowcolor{ClassIIa!25}
36412587 & $\mathcal{ CPT } $ & IIa & exp & 0.55(3) & exp & 0.036 & exp & 0.04 & 0 1 1 1 1 \\
\thinline
\rowcolor{ClassIIa!10}
38621574 & $\mathcal{ CPT } $ & IIa & exp & 0.55(2) & poly & 0.538 & poly & 1.126 & 0 1 1 1 1 \\
\thinline
\rowcolor{ClassIIa!25}
38627514 & $\mathcal{ CPT } $ & IIa & exp & 0.55(2) & poly & 0.537 & poly & 1.131 & 0 1 1 1 1 1 \\
\thinline
\rowcolor{ClassIIa!10}
14725863 & $\mathcal{ CPT } $ & IIa & exp & 0.56(3) & poly & 0.615 & poly & 0.695 & 1 1 1 1 1 \\
\thinline
\rowcolor{ClassIIa!25}
34267518 & $\mathcal{ CPT } $ & IIa & exp & 0.55(2) & poly & 0.537 & poly & 1.11 & 0 1 1 1 1 \\
\thinline
\rowcolor{ClassIIa!10}
14625873 & $\mathcal{ CPT } $ & IIa & exp & 0.55(1) & poly & 0.616 & poly & 0.644 & 1 1 1 1 1 \\
\thinline
\rowcolor{ClassIIa!25}
45721863 & $\mathcal{ CPT } $ & IIa & exp & 0.56(3) & poly & 0.615 & poly & 0.698 & 1 1 1 1 1 \\
\thinline
\rowcolor{ClassIIa!10}
34268571 & $\mathcal{ CPT } $ & IIa & exp & 0.54(1) & exp & 0.014 & poly & 0.65 & 0 1 1 1 1 \\
\thinline
\rowcolor{ClassIIa!25}
58637214 & $\mathcal{ CPT } $ & IIa & exp & 0.546 & exp & 0.014 & poly & 0.641 & 0 1 1 1 1 \\
\thinline
\rowcolor{ClassIIa!10}
36478521 & $\mathcal{ CPT } $ & IIa & exp & 0.7(2) & exp & 0.096 & poly & 0.535 & 0 0 1 1 1 \\
\thinline
\rowcolor{ClassIIa!25}
36487512 & $\mathcal{ CPT } $ & IIa & exp & 0.7(2) & exp & 0.094 & poly & 0.535 & 0 0 1 1 1 \\
\thinline
\rowcolor{ClassIIa!10}
67583142 & $\mathcal{ CPT } $ & IIa & exp & 0.7(2) & exp & 0.096 & poly & 0.535 & 0 0 1 1 1 \\
\thinline
\rowcolor{ClassIIa!25}
37186542 & $\mathcal{ CPT } $ & IIa & exp & 0.6(1) & exp & 0.094 & poly & 0.535 & 0 0 1 1 1 1 \\
\thinline
\rowcolor{ClassIIa!10}
45621738 & $\mathcal{ CPT } $ & IIa & exp & 0.9(8) & poly & 0.347 & poly & 0.575 & 1 2 2 2 2 \\
\thinline
\rowcolor{ClassIIa!25}
14625738 & $\mathcal{ CPT } $ & IIa & exp & 0.9(8) & poly & 0.346 & poly & 0.565 & 1 2 2 2 2 \\
\thinline
\rowcolor{ClassIIa!10}
45721368 & $\mathcal{ CPT } $ & IIa & exp & 0.55(1) & poly & 0.616 & poly & 0.66 & 1 1 1 1 1 \\
\thinline
\hline
\rowcolor{ClassIIa!25}
13465287 & $\mathcal{ CP } $ & IIa & exp & 0.56(2) & poly & 0.515 & poly & 0.53 & 0 0 0 0 0 0 \\
\thinline
\rowcolor{ClassIIa!10}
35416287 & $\mathcal{ CP } $ & IIa & exp & 0.9(8) & poly & 0.548 & poly & 0.525 & 1 1 1 1 1 \\
\thinline
\hline
\rowcolor{ClassIIa!25}
16435278 & $\mathcal{ CP+T } $ & IIa & exp & 0.87(8) & poly & 0.278 & poly & 0.265 & 0 0 0 0 0 \\
\thinline
\rowcolor{ClassIIa!10}
56431287 & $\mathcal{ CP+T } $ & IIa & exp & 0.55(1) & exp & 0.038 & poly & 0.482 & 0 0 0 0 0 \\
\thinline
\rowcolor{ClassIIa!25}
14625387 & $\mathcal{ CP+T } $ & IIa & exp & 0.89(9) & poly & 0.31(2) & poly & 0.25(3) & 1 1 1 1 1 \\
\thinline
\rowcolor{ClassIIa!10}
14625378 & $\mathcal{ CP+T } $ & IIa & exp & 0.78(6) & poly & 0.334 & poly & 0.509 & 2 2 2 2 2 \\
\thinline
\hline
\rowcolor{ClassIIa!25}
54378621 & $\mathcal{ CT+P } $ & IIa & exp & 0.56(1) & exp & 0.038 & poly & 0.522 & 0 0 0 0 0 \\
\thinline
\rowcolor{ClassIIa!10}
45678321 & $\mathcal{ CT+P } $ & IIa & exp & 0.87(8) & poly & 0.278 & poly & 0.265 & 0 0 0 0 0 \\
\thinline
\rowcolor{ClassIIa!25}
45321678 & $\mathcal{ CT+P } $ & IIa & exp & 0.89(9) & poly & 0.31(2) & poly & 0.25(3) & 1 1 1 1 1 \\
\thinline
\rowcolor{ClassIIa!10}
45621378 & $\mathcal{ CT+P } $ & IIa & exp & 0.77(5) & poly & 0.333 & poly & 0.505 & 2 2 2 2 2 \\
\thinline
\hline
\rowcolor{ClassIIa!25}
46532178 & $\mathcal{ CP+PT } $ & IIa & exp & 0.79(8) & poly & 0.268 & poly & 0.259 & 1 1 1 1 1 \\
\thinline
\rowcolor{ClassIIa!10}
35216478 & $\mathcal{ CP+PT } $ & IIa & exp & 0.57(3) & poly & 0.692 & poly & 0.68(3) & 2 3 3 3 3 \\
\thinline
\rowcolor{ClassIIa!25}
25614378 & $\mathcal{ CP+PT } $ & IIa & exp & 0.35(3) & poly & 0.27 & poly & 0.251 & 2 2 2 2 2 \\
\thinline
\hline
\rowcolor{ClassIIa!10}
12654378 & $\mathcal{ P+T } $ & IIa & exp & 0.34(3) & poly & 0.27 & poly & 0.251 & 2 2 2 2 2 \\
\thinline
\rowcolor{ClassIIa!25}
12687354 & $\mathcal{ P+T } $ & IIa & exp & 0.62(4) & poly & 0.692 & poly & 0.69(3) & 2 3 3 3 3 \\
\thinline
\thickline
\rowcolor{ClassIIb!10}
34612587 & $\mathcal{ CPT } $ & IIb & exp & 0.42(6) & poly & 0.501 & poly & 0.501 & 0 2 2 2 4 4\\
\thinline
\rowcolor{ClassIIb!25}
14385672 & $\mathcal{ CPT } $ & IIb & exp & 0.42(5) & poly & 0.501 & poly & 0.5 & 0 2 2 2 4 4 \\
\thinline
\hline
\rowcolor{ClassIIb!10}
34126587 & $\mathcal{ CP+T } $ & IIb & exp & 0.37(4) & poly & 0.5 & poly & 0.5(3) & 0 2 2 2 4 \\
\thinline
\hline
\rowcolor{ClassIIb!25}
54387612 & $\mathcal{ CT+P } $ & IIb & exp & 0.39(3) & poly & 0.501 & poly & 0.49(3) & 0 2 2 2 4 \\
\thinline
\hline
\rowcolor{ClassIIb!10}
38167452 & $\mathcal{ C+T } $ & IIb & exp & 0.39(4) & poly & 0.488 & poly & 0.521 & 0 2 2 2 4 \\
\thinline
\rowcolor{ClassIIb!25}
67438125 & $\mathcal{ C+T } $ & IIb & exp & 0.41(4) & poly & 0.488 & poly & 0.521 & 0 2 2 2 4 \\
\thinline
\hline
\rowcolor{ClassIIb!10}
26534178 & $\mathcal{ CP+PT } $ & IIb & exp & 0.4(3) & poly & 0.492 & poly & 0.48(3) & 2 4 4 4 6 \\
\thinline
\rowcolor{ClassIIb!25}
25314678 & $\mathcal{ CP+PT } $ & IIb & exp & 0.69(3) & poly & 0.567 & poly & 0.628 & 2 2 4 4 6 6 \\
\thinline
\hline
\rowcolor{ClassIIb!10}
12654387 & $\mathcal{ P+T } $ & IIb & exp & 0.69(3) & poly & 0.567 & poly & 0.638 & 2 2 4 4 6 \\
\thinline
\rowcolor{ClassIIb!25}
21678345 & $\mathcal{ P+T } $ & IIb & exp & 0.39(2) & poly & 0.492 & poly & 0.49(3) & 2 4 4 4 6 6\\
\thinline
\hline
\rowcolor{ClassIIb!10}
37268541 & $\mathcal{ PT+C } $ & IIb & exp & 0.37(2) & poly & 0.487 & poly & 0.52 & 0 2 2 2 6 6\\
\thinline
\rowcolor{ClassIIb!25}
68537214 & $\mathcal{ PT+C } $ & IIb & exp & 0.38(2) & poly & 0.487 & poly & 0.519 & 0 2 2 2 6 \\
\thinline
\thickline
\rowcolor{ClassIIc!10}
16435287 & $\mathcal{ CP+T } $ & IIc & exp & 0.36(3) & poly & 0.665 & poly & 0.663 & 0 2 2 4 4 8 \\
\thinline
\hline
\rowcolor{ClassIIc!25}
45378621 & $\mathcal{ CT+P } $ & IIc & exp & 0.36(3) & poly & 0.667 & poly & 0.666 & 0 2 2 4 4 8 \\
\thinline
\thickline
\rowcolor{ClassIIIa!10}
23564187 & $\mathcal{ CP } $ & IIIa & exp & 0.329 & const & 0.15(2) & const & 0.024 & 1 1 1 1 1 \\
\thinline
\hline
\rowcolor{ClassIIIa!25}
14325687 & $\mathcal{ CP+T } $ & IIIa & exp & 0.33(1) & const & 0.06(1) & const & 0.066 & 1 1 1 1 1 \\
\thinline
\hline
\rowcolor{ClassIIIa!10}
45321687 & $\mathcal{ CT+P } $ & IIIa & exp & 0.34(2) & const & 0.06(1) & const & 0.066 & 1 1 1 1 1 \\
\thinline
\hline
\rowcolor{ClassIIIa!25}
16543278 & $\mathcal{ C+T } $ & IIIa & exp & 0.316 & const & 0.068 & const & 0.034 & 0 0 0 0 0 \\
\thinline
\rowcolor{ClassIIIa!10}
18347652 & $\mathcal{ C+T } $ & IIIa & exp & 0.316 & const & 0.068 & const & 0.034 & 0 0 0 0 0 \\
\thinline
\rowcolor{ClassIIIa!25}
35162487 & $\mathcal{ C+T } $ & IIIa & exp & 0.316 & const & 0.068 & const & 0.034 & 0 0 0 0 0 \\
\thinline
\rowcolor{ClassIIIa!10}
47618325 & $\mathcal{ C+T } $ & IIIa & exp & 0.317 & const & 0.068 & const & 0.034 & 0 0 0 0 0 \\
\thinline
\hline
\rowcolor{ClassIIIa!25}
21378654 & $\mathcal{ P } $ & IIIa & exp & 0.327 & const & 0.15(2) & const & 0.024 & 1 1 1 1 1 \\
\thinline
\thickline
\rowcolor{ClassIIIb!10}
56471328 & $\mathcal{ CPT }^{*}$ & IIIb & exp & 1.02(4) & exp & 0.264 & exp & 0.272 & 0 0 1 2 4 8 \\
\thinline
\rowcolor{ClassIIIb!25}
54638271 & $\mathcal{ CPT }^{*}$ & IIIb & exp & 1.01(6) & exp & 0.265 & exp & 0.272 & 0 0 1 2 4 8 \\
\thinline
\rowcolor{ClassIIIb!10}
56438271 & $\mathcal{ CPT }^{*}$ & IIIb & exp & 0.92(8) & exp & 0.265 & exp & 0.269 & 0 1 2 4 9 \\
\thinline
\rowcolor{ClassIIIb!25}
56478321 & $\mathcal{ CPT }^{*}$ & IIIb & exp & 0.92(8) & exp & 0.266 & exp & 0.27(1) & 0 1 2 4 9 \\
\thinline
\rowcolor{ClassIIIb!10}
37681542 & $\mathcal{ CPT }^{*}$ & IIIb & exp & 0.9(1) & exp & 0.53(7) & exp & 0.311 & 0 1 2 4 9 \\
\thinline
\rowcolor{ClassIIIb!25}
57638241 & $\mathcal{ CPT }^{*}$ & IIIb & exp & 0.93(6) & exp & 0.265 & exp & 0.27(1) & 0  1  2  4  8 17 \\
\thinline
\rowcolor{ClassIIIb!10}
56473128 & $\mathcal{ CPT }^{*}$ & IIIb & exp & 0.93(5) & exp & 0.264 & exp & 0.271 & 0 1 2 4 8 \\
\thinline
\rowcolor{ClassIIIb!25}
36481572 & $\mathcal{ CPT }^{*}$ & IIIb & exp & 0.9(1) & exp & 0.53(7) & exp & 0.334 & 0 1 2 4 9 \\
\thinline
\rowcolor{ClassIIIb!10}
24658731 & $\mathcal{ CPT }^{*}$ & IIIb & exp & 0.45(2) & exp & 1.1 & const & 0.111 & 1  2  4  8 16 \\
\thinline
\rowcolor{ClassIIIb!25}
54628731 & $\mathcal{ CPT }^{*}$ & IIIb & exp & 0.41(2) & exp & 0.268 & const & 0.111 & 1  2  5 11 28 \\
\thinline
\rowcolor{ClassIIIb!10}
54731268 & $\mathcal{ CPT }^{*}$ & IIIb & exp & 0.41(2) & exp & 0.259 & const & 0.111 & 1  2  5 11 28 65 \\
\thinline
\rowcolor{ClassIIIb!25}
24853176 & $\mathcal{ CPT }^{*}$ & IIIb & exp & 0.45(2) & exp & 1.099 & const & 0.111 & 1  2  4  8 16 32 \\
\thinline
\rowcolor{ClassIIIb!10}
24857316 & $\mathcal{ CPT } $ & IIIb & exp & 0.36(4) & exp & 0.102 & const & 0.111 & 1  2  5  9 17 \\
\thinline
\rowcolor{ClassIIIb!25}
54781362 & $\mathcal{ CPT } $ & IIIb & exp & 0.85(0) & exp & 0.024 & const & 0.111 & 1  2  4  8 16 \\
\thinline
\rowcolor{ClassIIIb!10}
54328671 & $\mathcal{ CPT } $ & IIIb & exp & 0.85(8) & exp & 0.028 & const & 0.111 & 1  2  4  8 16 \\
\thinline
\rowcolor{ClassIIIb!25}
54623187 & $\mathcal{ CPT } $ & IIIb & exp & 0.83(7) & exp & 0.027 & const & 0.111 & 1  2  4  8 16 \\
\thinline
\rowcolor{ClassIIIb!10}
23851476 & $\mathcal{ CPT } $ & IIIb & exp & 0.37(4) & exp & 0.102 & const & 0.111 & 1  2  5  9 17 \\
\thinline
\rowcolor{ClassIIIb!25}
34721568 & $\mathcal{ CPT } $ & IIIb & exp & 0.86(0) & exp & 0.024 & const & 0.111 & 1  2  4  8 16 \\
\thinline
\rowcolor{ClassIIIb!10}
54327618 & $\mathcal{ CPT } $ & IIIb & exp & 0.92(9) & exp & 0.125 & const & 0.111 & 1  2  4  8 16 \\
\thinline
\rowcolor{ClassIIIb!25}
34621587 & $\mathcal{ CPT } $ & IIIb & exp & 0.9(9) & exp & 0.125 & const & 0.111 & 1  2  4  8 16 \\
\thinline
\rowcolor{ClassIIIb!10}
34671528 & $\mathcal{ CPT } $ & IIIb & exp & 0.78(5) & poly & 0.493 & const & 0.111 & 1  2  6 10 18 34 \\
\thinline
\rowcolor{ClassIIIb!25}
23658471 & $\mathcal{ CPT } $ & IIIb & exp & 0.26(3) & poly & 0.718 & const & 0.111 & 1  2  5  9 17 \\
\thinline
\rowcolor{ClassIIIb!10}
26457318 & $\mathcal{ CPT } $ & IIIb & exp & 0.26(3) & poly & 0.718 & const & 0.111 & 1  2  5  9 17 \\
\thinline
\rowcolor{ClassIIIb!25}
56427318 & $\mathcal{ CPT } $ & IIIb & exp & 0.77(5) & poly & 0.494 & const & 0.111 & 1  2  6 10 18 \\
\thinline
\rowcolor{ClassIIIb!10}
34871526 & $\mathcal{ CPT } $ & IIIb & exp & 0.78(6) & poly & 0.528 & const & 0.111 & 1  2  5  9 17 \\
\thinline
\rowcolor{ClassIIIb!25}
23758461 & $\mathcal{ CPT } $ & IIIb & exp & 0.26(3) & poly & 0.506 & const & 0.111 & 1  2  6 10 18 \\
\thinline
\rowcolor{ClassIIIb!10}
56427813 & $\mathcal{ CPT } $ & IIIb & exp & 0.77(6) & poly & 0.528 & const & 0.111 & 1  2  5  9 17 \\
\thinline
\rowcolor{ClassIIIb!25}
34621578 & $\mathcal{ CPT } $ & IIIb & exp & 0.8(1) & poly & 0.525 & const & 0.111 & 2  3  5  9 17 \\
\thinline
\rowcolor{ClassIIIb!10}
24657813 & $\mathcal{ CPT } $ & IIIb & exp & 0.58(3) & poly & 0.517 & const & 0.111 & 1  2  4  8 16 \\
\thinline
\rowcolor{ClassIIIb!25}
54627318 & $\mathcal{ CPT } $ & IIIb & exp & 0.8(9) & poly & 0.524 & const & 0.111 & 2  3  5  9 17 \\
\thinline
\rowcolor{ClassIIIb!10}
34821576 & $\mathcal{ CPT } $ & IIIb & exp & 0.9(1) & poly & 0.52 & const & 0.111 & 1  2  4  8 16 \\
\thinline
\rowcolor{ClassIIIb!25}
23751468 & $\mathcal{ CPT } $ & IIIb & exp & 0.58(4) & poly & 0.517 & const & 0.111 & 1  2  4  8 16 \\
\thinline
\rowcolor{ClassIIIb!10}
54627813 & $\mathcal{ CPT } $ & IIIb & exp & 0.9(1) & poly & 0.519 & const & 0.111 & 1  2  4  8 16 \\
\thinline
\rowcolor{ClassIIIb!25}
26457813 & $\mathcal{ CPT } $ & IIIb & exp & 0.26(3) & poly & 0.507 & const & 0.111 & 1  2  6 10 18 \\
\thinline
\hline
\rowcolor{ClassIIIb!10}
34126578 & $\mathcal{ CP+T } $ & IIIb & exp & 0.234 & const & 0.25(9) & const & 0.05(3) & 1  4 12 58 \\
\thinline
\rowcolor{ClassIIIb!25}
13265478 & $\mathcal{ CP+T } $ & IIIb & exp & 0.23(2) & const & 0.38(3) & const & 0.08(2) & 1  6 27 \\
\thinline
\hline
\rowcolor{ClassIIIb!10}
45687312 & $\mathcal{ CT+P } $ & IIIb & exp & 0.25(1) & const & 0.38(3) & const & 0.08(2) & 1   6  27 146 \\
\thinline
\rowcolor{ClassIIIb!25}
54687312 & $\mathcal{ CT+P } $ & IIIb & exp & 0.24(1) & const & 0.25(9) & const & 0.05(3) & 1  4 12 58 \\
\thinline
\thickline
\rowcolor{ClassIV!10}
17348625 & $\mathcal{ C+T }^{E}$ & IVa & poly & 0.27 & const & 0.51(6) & const & 0.07(6) & 0 0 0 0 \\
\thinline
\rowcolor{ClassIV!25}
46513287 & $\mathcal{ C+T } $ & IVa & poly & 0.275 & const & 0.51(6) & const & 0.07(6) & 0 0 0 0 \\
\thinline
\rowcolor{ClassIV!10}
48617352 & $\mathcal{ C+T } $ & IVa & poly & 0.26(2) & const & 0.51(6) & const & 0.07(6) & 0 0 0 0 \\
\thinline
\rowcolor{ClassIV!25}
13246578 & $\mathcal{ C+T }^{E}$ & IVa & poly & 0.286 & const & 0.51(6) & const & 0.07(6) & 0 0 0 0 \\
\thinline
\thickline
\rowcolor{ClassIV!10}
34681572 & $\mathcal{ CPT }^{*}$ & IVb & poly & 1.018 & const & 0.0 & const & 0.111 & 1  3  9 24 72 \\
\thinline
\rowcolor{ClassIV!25}
23657418 & $\mathcal{ CPT }^{*}$ & IVb & poly & 1.012 & const & 0.0 & const & 0.111 & 1  4 11 32 95 \\
\thinline
\rowcolor{ClassIV!10}
21654738 & $\mathcal{ CPT }^{*}$ & IVb & poly & 1.029 & const & 0.0 & const & 0.111 & 2   5  14  41 122 \\
\thinline
\rowcolor{ClassIV!25}
26458731 & $\mathcal{ CPT }^{*}$ & IVb & poly & 1.014 & const & 0.0 & const & 0.111 & 1  4 11 32 95 \\
\thinline
\rowcolor{ClassIV!10}
56428731 & $\mathcal{ CPT }^{*}$ & IVb & poly & 1.019 & const & 0.0 & const & 0.111 & 1  3  9 24 59 \\
\thinline
\rowcolor{ClassIV!25}
23651478 & $\mathcal{ CPT } $ & IVb & poly & 4.0(1) & poly & 0.507 & const & 0.111 & 1  2  5  9 18 \\
\thinline
\rowcolor{ClassIV!10}
24657318 & $\mathcal{ CPT } $ & IVb & poly & 4.0(1) & poly & 0.506 & const & 0.111 & 1  2  5  9 18 \\
\thinline
\rowcolor{ClassIV!25}
14375628 & $\mathcal{ CPT } $ & IVb & poly & 2.52(8) & const & 0.36(8) & const & 0.09(3) & 1   6  26 136 \\
\thinline
\rowcolor{ClassIV!10}
45612738 & $\mathcal{ CPT } $ & IVb & poly & 2.78(7) & const & 0.17(8) & const & 0.222 & 3  18 111 \\
\thinline
\rowcolor{ClassIV!25}
45623187 & $\mathcal{ CPT } $ & IVb & poly & 2.57(6) & const & 0.36(8) & const & 0.09(3) & 1   6  26 136 \\
\thinline
\rowcolor{ClassIV!10}
12645738 & $\mathcal{ CPT } $ & IVb & poly & 2.81(9) & const & 0.17(8) & const & 0.222 & 3  18 111 \\
\thinline
\rowcolor{ClassIV!25}
54628371 & $\mathcal{ CPT }^{*}$ & IVb & poly & 1.03(1) & const & 0.333 & const & 0.111 & 2   6  19  56 168 \\
\thinline
\rowcolor{ClassIV!10}
54623178 & $\mathcal{ CPT }^{*}$ & IVb & poly & 1.03(1) & const & 0.333 & const & 0.111 & 2   6  19  56 168 \\
\thinline
\rowcolor{ClassIV!25}
34781562 & $\mathcal{ CPT }^{*}$ & IVb & poly & 1.018 & const & 0.333 & const & 0.111 & 2   7  21  63 189 \\
\thinline
\rowcolor{ClassIV!10}
23857416 & $\mathcal{ CPT }^{*}$ & IVb & poly & 1.016 & const & 0.333 & const & 0.111 & 2   8  24  72 216 \\
\thinline
\rowcolor{ClassIV!25}
54621738 & $\mathcal{ CPT }^{*}$ & IVb & poly & 1.0(1) & const & 0.333 & const & 0.111 & 3   8  24  69 207 \\
\thinline
\rowcolor{ClassIV!10}
24658371 & $\mathcal{ CPT }^{*}$ & IVb & poly & 1.015 & const & 0.333 & const & 0.111 & 2   6  19  56 168 \\
\thinline
\rowcolor{ClassIV!25}
56428371 & $\mathcal{ CPT }^{*}$ & IVb & poly & 1.02(1) & const & 0.333 & const & 0.111 & 2   7  21  63 189 \\
\thinline
\rowcolor{ClassIV!10}
24753168 & $\mathcal{ CPT }^{*}$ & IVb & poly & 1.01(4) & const & 0.333 & const & 0.111 & 2   6  20  60 180 \\
\thinline
\rowcolor{ClassIV!25}
54728361 & $\mathcal{ CPT }^{*}$ & IVb & poly & 1.02(6) & const & 0.333 & const & 0.111 & 2   6  19  57 171 \\
\thinline
\rowcolor{ClassIV!10}
54723168 & $\mathcal{ CPT }^{*}$ & IVb & poly & 1.02(5) & const & 0.333 & const & 0.111 & 2   6  19  57 171 \\
\thinline
\rowcolor{ClassIV!25}
26458371 & $\mathcal{ CPT }^{*}$ & IVb & poly & 1.013 & const & 0.333 & const & 0.111 & 2   8  24  72 216 \\
\thinline
\rowcolor{ClassIV!10}
24758361 & $\mathcal{ CPT }^{*}$ & IVb & poly & 1.01(5) & const & 0.333 & const & 0.111 & 2   6  20  60 180 \\
\thinline
\rowcolor{ClassIV!25}
24651873 & $\mathcal{ CPT }^{*}$ & IVb & poly & 1.0(1) & const & 0.333 & const & 0.111 & 3   8  24  69 207 \\
\thinline
\rowcolor{ClassIV!10}
24653178 & $\mathcal{ CPT }^{*}$ & IVb & poly & 1.01 & const & 0.333 & const & 0.111 & 2   6  19  56 168 \\
\thinline
\rowcolor{ClassIV!25}
24651738 & $\mathcal{ CPT }^{*}$ & IVb & poly & 1.024 & const & 0.0 & const & 0.111 & 2   5  14  41 122 \\
\thinline
\hline
\rowcolor{ClassIV!10}
14325678 & $\mathcal{ CP+T } $ & IVb & poly & 1.24(3) & const & 0.48(6) & const & 0.12(3) & 3  19 149 \\
\thinline
\hline
\rowcolor{ClassIV!25}
45621387 & $\mathcal{ CT+P } $ & IVb & poly & 1.23(3) & const & 0.48(6) & const & 0.12(3) & 3   19  149 1181 \\
\thinline
\hline
\rowcolor{ClassIV!10}
12345687 & $\mathcal{ C+P+T } $ & IVb & poly & 2.56(8) & const & 0.23(8) & const & 0.222 & 4  21 145 \\
\thinline
\rowcolor{ClassIV!25}
12645378 & $\mathcal{ C+P+T } $ & IVb & poly & 2.6(9) & const & 0.23(8) & const & 0.222 & 4  21 145 \\
\thinline
\rowcolor{ClassIV!10}
45312687 & $\mathcal{ C+P+T } $ & IVb & poly & 2.57(8) & const & 0.23(8) & const & 0.222 & 4  21 145 \\
\thinline
\rowcolor{ClassIV!25}
45612378 & $\mathcal{ C+P+T } $ & IVb & poly & 2.6(8) & const & 0.23(8) & const & 0.222 & 4  21 145 \\
\thinline
\rowcolor{ClassIV!10}
21654387 & $\mathcal{ C+P+T } $ & IVb & poly & 1.995 & poly & 0.502 & const & 0.111 & 3  5  9 17 33 \\
\thinline
\rowcolor{ClassIV!25}
54321678 & $\mathcal{ C+P+T } $ & IVb & poly & 1.99(1) & poly & 0.501 & const & 0.111 & 3  5  9 17 33 \\
\thinline
\rowcolor{ClassIV!10}
54621387 & $\mathcal{ C+P+T } $ & IVb & poly & 1.99(1) & poly & 0.501 & const & 0.111 & 3  5  9 17 33 \\
\thinline
\rowcolor{ClassIV!25}
21354678 & $\mathcal{ C+P+T } $ & IVb & poly & 2.0(1) & poly & 0.502 & const & 0.111 & 3  5  9 17 33 \\
\thinline
\rowcolor{ClassIV!10}
21354687 & $\mathcal{ C+P+T }^{*}$ & IVb & poly & 1.033 & const & 0.333 & const & 0.111 & 4  12  36 108 324 \\
\thinline
\rowcolor{ClassIV!25}
54621378 & $\mathcal{ C+P+T }^{*}$ & IVb & poly & 1.033 & const & 0.333 & const & 0.111 & 4  12  36 108 324 \\
\thinline
\rowcolor{ClassIV!10}
54321687 & $\mathcal{ C+P+T }^{*}$ & IVb & poly & 1.033 & const & 0.0 & const & 0.111 & 2   6  18  54 162 \\
\thinline
\rowcolor{ClassIV!25}
21654378 & $\mathcal{ C+P+T }^{*}$ & IVb & poly & 1.033 & const & 0.0 & const & 0.111 & 2   6  18  54 162 \\
\thinline
\rowcolor{ClassIV!10}
12645387 & $\mathcal{ C+P+T } $ & IVb & const & 1.5 & const & 0.3(2) & const & 0.222 & 3  34 306 \\
\thinline
\rowcolor{ClassIV!25}
12345678 & $\mathcal{ C+P+T }^{E}$ & IVb & const & 0.5 & const & 0.667 & const & 0.222 & 6  72 648 \\
\thinline
\rowcolor{ClassIV!10}
45612387 & $\mathcal{ C+P+T } $ & IVb & const & 0.5 & const & 0.667 & const & 0.222 & 6  72 648 \\
\thinline
\rowcolor{ClassIV!25}
45312678 & $\mathcal{ C+P+T } $ & IVb & const & 1.5 & const & 0.3(2) & const & 0.222 & 3  34 306 \\
\thinline
\hline
\rowcolor{ClassIV!10}
15342687 & $\mathcal{ C+T } $ & IVb & poly & 2.54(8) & const & 0.1(9) & const & 0.222 & 3  18 111 \\
\thinline
\rowcolor{ClassIV!25}
15342678 & $\mathcal{ C+T } $ & IVb & const & 1.5 & const & 0.4(2) & const & 0.222 & 5   48  432 3888 \\
\thinline
\rowcolor{ClassIV!10}
15642378 & $\mathcal{ C+T } $ & IVb & poly & 2.49(7) & const & 0.1(9) & const & 0.222 & 3  18 111 \\
\thinline
\rowcolor{ClassIV!25}
15642387 & $\mathcal{ C+T } $ & IVb & const & 1.5 & const & 0.4(2) & const & 0.222 & 5   48  432 3888 \\
\thinline
\rowcolor{ClassIV!10}
15742836 & $\mathcal{ C+T } $ & IVb & const & 1.5 & const & 0.2(3) & const & 0.222 & 3   34  306 2754 \\
\thinline
\rowcolor{ClassIV!25}
15842763 & $\mathcal{ C+T } $ & IVb & const & 1.5 & const & 0.3(3) & const & 0.222 & 6   72  648 5832 \\
\thinline
\rowcolor{ClassIV!10}
21754836 & $\mathcal{ C+T } $ & IVb & poly & 2.01(1) & const & 0.0 & const & 0.111 & 2  4  8 16 32 \\
\thinline
\rowcolor{ClassIV!25}
45712836 & $\mathcal{ C+T } $ & IVb & const & 1.5 & const & 0.3(2) & const & 0.222 & 5   48  432 3888 \\
\thinline
\rowcolor{ClassIV!10}
54721836 & $\mathcal{ C+T } $ & IVb & poly & 2.007 & exp & 0.307 & const & 0.111 & 2  4  8 16 32 \\
\thinline
\rowcolor{ClassIV!25}
12745836 & $\mathcal{ C+T }^{E}$ & IVb & const & 1.5 & const & 0.3(2) & const & 0.222 & 5   48  432 3888 \\
\thinline
\hline
\rowcolor{ClassIV!10}
26534187 & $\mathcal{ CP+PT } $ & IVb & poly & 2.54(6) & const & 0.5(6) & const & 0.07(5) & 4  21 145 \\
\thinline
\rowcolor{ClassIV!25}
25614387 & $\mathcal{ CP+PT } $ & IVb & poly & 2.57(7) & const & 0.56(2) & const & 0.11(2) & 4  21 145 \\
\thinline
\rowcolor{ClassIV!10}
23164578 & $\mathcal{ CP+PT } $ & IVb & poly & 3.1(1) & poly & 0.496 & poly & 0.47(1) & 1  3  4  9 17 \\
\thinline
\rowcolor{ClassIV!25}
25314687 & $\mathcal{ CP+PT } $ & IVb & poly & 1.995 & const & 0.25 & const & 0.028 & 3  5  9 17 33 \\
\thinline
\rowcolor{ClassIV!10}
46532187 & $\mathcal{ CP+PT } $ & IVb & const & 1.5 & const & 0.57(6) & const & 0.13(6) & 3   34  306 2754 \\
\thinline
\rowcolor{ClassIV!25}
35216487 & $\mathcal{ CP+PT } $ & IVb & poly & 1.997 & const & 0.25 & const & 0.028 & 3  5  9 17 33 \\
\thinline
\rowcolor{ClassIV!10}
23164587 & $\mathcal{ CP+PT }^{*}$ & IVb & poly & 1.033 & const & 0.25 & const & 0.028 & 2   6  18  54 162 \\
\thinline
\hline
\rowcolor{ClassIV!25}
21378645 & $\mathcal{ P+T } $ & IVb & poly & 2.59(7) & const & 0.5(6) & const & 0.07(5) & 4   21  145 1077 \\
\thinline
\rowcolor{ClassIV!10}
21387654 & $\mathcal{ P+T }^{*}$ & IVb & poly & 1.033 & const & 0.25 & const & 0.028 & 2   6  18  54 162 \\
\thinline
\rowcolor{ClassIV!25}
21687354 & $\mathcal{ P+T } $ & IVb & poly & 3.1(1) & poly & 0.496 & poly & 0.46(1) & 1  3  5  9 17 \\
\thinline
\rowcolor{ClassIV!10}
12354678 & $\mathcal{ P+T } $ & IVb & poly & 2.57(8) & const & 0.56(2) & const & 0.11(2) & 4   21  145 1077 \\
\thinline
\rowcolor{ClassIV!25}
12354687 & $\mathcal{ P+T } $ & IVb & poly & 2.0(1) & const & 0.25 & const & 0.028 & 3  5  9 17 33 \\
\thinline
\rowcolor{ClassIV!10}
12378645 & $\mathcal{ P+T } $ & IVb & const & 1.5 & const & 0.57(6) & const & 0.13(6) & 3   34  306 2754 \\
\thinline
\rowcolor{ClassIV!25}
12387654 & $\mathcal{ P+T } $ & IVb & poly & 2.004 & const & 0.25 & const & 0.028 & 3  5  9 17 33 \\
\thinline
\hline
\rowcolor{ClassIV!10}
12745863 & $\mathcal{ PT+C } $ & IVb & poly & 2.54(5) & const & 0.13(8) & const & 0.222 & 3  18 111 \\
\thinline
\rowcolor{ClassIV!25}
21754863 & $\mathcal{ PT+C }^{*}$ & IVb & poly & 1.033 & const & 0.333 & const & 0.111 & 3   9  27  81 243 \\
\thinline
\rowcolor{ClassIV!10}
23156478 & $\mathcal{ PT+C } $ & IVb & poly & 3.0(2) & poly & 0.496 & const & 0.111 & 1  4  7 13 25 \\
\thinline
\rowcolor{ClassIV!25}
23156487 & $\mathcal{ PT+C }^{*}$ & IVb & poly & 1.04(1) & const & 0.333 & const & 0.111 & 2   7  21  63 189 \\
\thinline
\rowcolor{ClassIV!10}
24351678 & $\mathcal{ PT+C } $ & IVb & poly & 2.0(3) & const & 0.0 & const & 0.111 & 2  4  8 16 32 \\
\thinline
\rowcolor{ClassIV!25}
24351687 & $\mathcal{ PT+C }^{*}$ & IVb & poly & 1.033 & const & 0.333 & const & 0.111 & 3   9  27  81 243 \\
\thinline
\rowcolor{ClassIV!10}
24651378 & $\mathcal{ PT+C }^{*}$ & IVb & poly & 1.033 & const & 0.333 & const & 0.111 & 3   9  27  81 243 \\
\thinline
\rowcolor{ClassIV!25}
24651387 & $\mathcal{ PT+C } $ & IVb & poly & 2.0(3) & const & 0.0 & const & 0.111 & 2  4  8 16 32 \\
\thinline
\rowcolor{ClassIV!10}
24751863 & $\mathcal{ PT+C }^{*}$ & IVb & poly & 1.033 & const & 0.333 & const & 0.111 & 4  12  36 108 324 \\
\thinline
\rowcolor{ClassIV!25}
24851736 & $\mathcal{ PT+C }^{*}$ & IVb & poly & 1.033 & const & 0.0 & const & 0.111 & 2   6  18  54 162 \\
\thinline
\rowcolor{ClassIV!10}
26453178 & $\mathcal{ PT+C }^{*}$ & IVb & poly & 1.0(2) & const & 0.333 & const & 0.111 & 2   8  24  72 216 \\
\thinline
\rowcolor{ClassIV!25}
26453187 & $\mathcal{ PT+C } $ & IVb & poly & 3.0(2) & poly & 0.499 & const & 0.111 & 2  4  8 16 \\
\thinline
\rowcolor{ClassIV!10}
27358641 & $\mathcal{ PT+C } $ & IVb & poly & 3.0(2) & poly & 0.496 & const & 0.111 & 2  4  8 16 32 \\
\thinline
\rowcolor{ClassIV!25}
27658341 & $\mathcal{ PT+C }^{*}$ & IVb & poly & 1.0(2) & const & 0.333 & const & 0.111 & 2   8  24  72 216 \\
\thinline
\rowcolor{ClassIV!10}
28357614 & $\mathcal{ PT+C }^{*}$ & IVb & poly & 1.038 & const & 0.333 & const & 0.111 & 2   7  21  63 189 \\
\thinline
\rowcolor{ClassIV!25}
28657314 & $\mathcal{ PT+C } $ & IVb & poly & 3.0(2) & poly & 0.495 & const & 0.111 & 1  4  7 13 25 \\
\thinline
\rowcolor{ClassIV!10}
34261578 & $\mathcal{ PT+C } $ & IVb & poly & 3.0(2) & poly & 0.498 & const & 0.111 & 2  4  8 16 32 \\
\thinline
\rowcolor{ClassIV!25}
34261587 & $\mathcal{ PT+C }^{*}$ & IVb & poly & 1.0(2) & const & 0.333 & const & 0.111 & 2   8  24  72 216 \\
\thinline
\rowcolor{ClassIV!10}
45712863 & $\mathcal{ PT+C } $ & IVb & poly & 2.56(6) & const & 0.13(8) & const & 0.222 & 3  18 111 \\
\thinline
\rowcolor{ClassIV!25}
54721863 & $\mathcal{ PT+C }^{*}$ & IVb & poly & 1.033 & const & 0.333 & const & 0.111 & 3   9  27  81 243 \\
\thinline
\rowcolor{ClassIV!10}
56423178 & $\mathcal{ PT+C }^{*}$ & IVb & poly & 1.03(1) & const & 0.333 & const & 0.111 & 2   7  21  63 189 \\
\thinline
\rowcolor{ClassIV!25}
56423187 & $\mathcal{ PT+C } $ & IVb & poly & 3.0(2) & poly & 0.497 & const & 0.111 & 1  4  7 13 25 \\
\thinline
\rowcolor{ClassIV!10}
57328641 & $\mathcal{ PT+C } $ & IVb & poly & 3.0(2) & poly & 0.497 & const & 0.111 & 1  4  5 13 25 \\
\thinline
\rowcolor{ClassIV!25}
57628341 & $\mathcal{ PT+C }^{*}$ & IVb & poly & 1.04(1) & const & 0.333 & const & 0.111 & 2   7  21  63 189 \\
\thinline
\rowcolor{ClassIV!10}
58327614 & $\mathcal{ PT+C }^{*}$ & IVb & poly & 1.0(2) & const & 0.333 & const & 0.111 & 2   8  24  72 216 \\
\thinline
\rowcolor{ClassIV!25}
58627314 & $\mathcal{ PT+C } $ & IVb & poly & 3.0(2) & poly & 0.498 & const & 0.111 & 2  4  8 16 32 \\
\thinline
\thickline
\rowcolor{?!10}
38671524 & $\mathcal{ CPT } $ & ? & exp & 0.9(3) & poly & 0.514 & ? & ? & 0 0 1 1 1 \\
\thinline
\rowcolor{?!25}
36427518 & $\mathcal{ CPT } $ & ? & exp & 0.9(3) & poly & 0.514 & ? & ? & 0 0 1 1 1 \\
\thinline
\rowcolor{?!10}
23651487 & $\mathcal{ CPT } $ & ? & exp & 0.45(2) & ? & ? & const & 0.111 & 1  2  4  8 16 \\
\thinline
\rowcolor{?!25}
24357618 & $\mathcal{ CPT } $ & ? & exp & 0.45(2) & ? & ? & const & 0.111 & 1  2  4  8 16 \\
\thinline
\rowcolor{?!10}
34728561 & $\mathcal{ CPT } $ & ? & exp & 0.54(1) & exp & 0.048 & ? & ? & 0 0 0 0 0 \\
\thinline
\rowcolor{?!25}
14785362 & $\mathcal{ CPT } $ & ? & exp & 0.55(1) & exp & 0.138 & ? & ? & 0 0 0 0 0 \\
\thinline
\rowcolor{?!10}
45327618 & $\mathcal{ CPT } $ & ? & exp & 0.55(1) & exp & 0.159 & ? & ? & 0 0 0 0 0 \\
\thinline
\rowcolor{?!25}
36412578 & $\mathcal{ CPT } $ & ? & exp & 0.54(1) & exp & 0.14 & ? & ? & 0 0 0 0 0 \\
\thinline
\rowcolor{?!10}
34712568 & $\mathcal{ CPT } $ & ? & exp & 0.54(1) & exp & 0.138 & ? & ? & 0 0 0 0 0 \\
\thinline
\rowcolor{?!25}
13625478 & $\mathcal{ CPT } $ & ? & exp & 0.53(1) & exp & 0.054 & ? & ? & 0 0 0 0 0 \\
\thinline
\rowcolor{?!10}
34627518 & $\mathcal{ CPT } $ & ? & exp & 0.54(2) & poly & 0.53 & ? & ? & 0 1 1 1 1 \\
\thinline
\rowcolor{?!25}
24653187 & $\mathcal{ CPT } $ & ? & exp & 0.59(4) & ? & ? & const & 0.111 & 1  2  4  8 16 \\
\thinline
\rowcolor{?!10}
24358671 & $\mathcal{ CPT } $ & ? & exp & 0.6(5) & ? & ? & const & 0.111 & 1  2  4  8 16 \\
\thinline
\hline
\rowcolor{?!25}
23564178 & $\mathcal{ CP } $ & ? & exp & 0.98(9) & ? & ? & ? & ? & 0 0 0 0 0 0 \\
\thinline
\hline
\rowcolor{?!10}
18647352 & $\mathcal{ C+T } $ & ? & ? & ? & ? & ? & ? & ? & 0 0 0 0 0 0 \\
\thinline
\rowcolor{?!25}
16543287 & $\mathcal{ C+T } $ & ? & ? & ? & ? & ? & ? & ? & 0 0 0 0 0 0 \\
\thinline
\rowcolor{?!10}
35162478 & $\mathcal{ C+T } $ & ? & ? & ? & ? & ? & ? & ? & 0 0 0 0 0 0 \\
\thinline
\rowcolor{?!25}
17648325 & $\mathcal{ C+T } $ & ? & exp & 0.546 & exp & 0.406 & ? & ? & 0 0 0 0 0 0 \\
\thinline
\rowcolor{?!10}
46513278 & $\mathcal{ C+T } $ & ? & exp & 0.55 & exp & 0.404 & ? & ? & 0 0 0 0 0 0 \\
\thinline
\rowcolor{?!25}
47318625 & $\mathcal{ C+T }^{E}$ & ? & ? & ? & ? & ? & const & 0.0 & 0 0 0 0 0 0 \\
\thinline
\rowcolor{?!10}
48317652 & $\mathcal{ C+T } $ & ? & exp & 0.549 & exp & 0.402 & ? & ? & 0 0 0 0 0 0 \\
\thinline
\rowcolor{?!25}
68437152 & $\mathcal{ C+T } $ & ? & exp & 0.55(4) & ? & ? & ? & ? & 0 0 0 0 0 \\
\thinline
\rowcolor{?!10}
37168425 & $\mathcal{ C+T } $ & ? & exp & 0.55(4) & ? & ? & ? & ? & 0 0 0 0 0 0 \\
\thinline
\rowcolor{?!25}
13246587 & $\mathcal{ C+T } $ & ? & exp & 0.552 & exp & 0.405 & ? & ? & 0 0 0 0 0 0 \\
\thinline
\hline
\rowcolor{?!10}
12678345 & $\mathcal{ P+T } $ & ? & exp & 0.81(9) & poly & 0.405 & ? & ? & 1 1 1 1 1 \\
\thinline
\hline
\rowcolor{?!25}
12378654 & $\mathcal{ P } $ & ? & exp & 0.89(9) & ? & ? & ? & ? & 1 1 1 1 1 \\
\thinline
\rowcolor{?!10}
21678354 & $\mathcal{ P } $ & ? & exp & 0.98(9) & ? & ? & ? & ? & 0 0 0 0 0 \\
\thinline
\rowcolor{?!25}
45378612 & $\mathcal{ P } $ & ? & exp & 0.56(1) & poly & 0.526 & ? & ? & 0 0 0 0 0 \\
\thinline
\hline
\rowcolor{?!10}
38267514 & $\mathcal{ PT+C }^{*}$ & ? & exp & 0.172 & exp & 0.403 & ? & ? & 0  2  4 10 22 54 \\
\thinline
\rowcolor{?!25}
67538241 & $\mathcal{ PT+C }^{*}$ & ? & exp & 0.165 & exp & 0.398 & ? & ? & 0  2  4 10 22 54 \\
\thinline
\end{longtable}
\normalsize

The rules marked by a question mark are unclassifiable within our framework. This is either because the available computational resources are insufficient to make a clear observation, or the observables behave in an irregular way.

The first examples of these rules have their return times increase exponentially to a plateau. It is unclear what mechanism can cause the orbit length to plateau. Secondly, we find cases where the orbit length sporadically differs by orders of magnitude for small changes of the system size $L$, e.g., in models with rules 43216578, 13746825, 17248536, 47318625. They also exhibit a correlation function which decays in 1 step. Some of these could be explained as perfect maps~\cite{Gambor}. 

In other unclassifiable cases, we find for example cases for which the correlation functions do not show a clear exponentially or algebraically decay. We find rules where the correlation function looks more complicated or it decays in one step (47318625). We defer more detailed studies of all such interesting examples of RCA to future explorations.

\section{Extended simulations of subdiffusive models}\label{AppD}

A select few subdiffusive models exhibit slowly converging dynamical exponents $z$. To rule out the emergence of any new exotic universality classes, and to confirm the dynamical exponents on asymptotic scales, we performed additional simulations for systems of size $2^{20}$ and times up to $2^{21}$. This was done for the following set of rules:

\begin{enumerate}
    \item 46532178,
    \item 16435278 (and its equivalent 45678321),
    \item 14625387 (and its equivalent 45321678),
    \item 25614378,
    \item 12654378 (equivalent to 25614378).
\end{enumerate}
\noindent

\begin{figure}[h!]
\centerline{
  \includegraphics[width=1.0\linewidth]{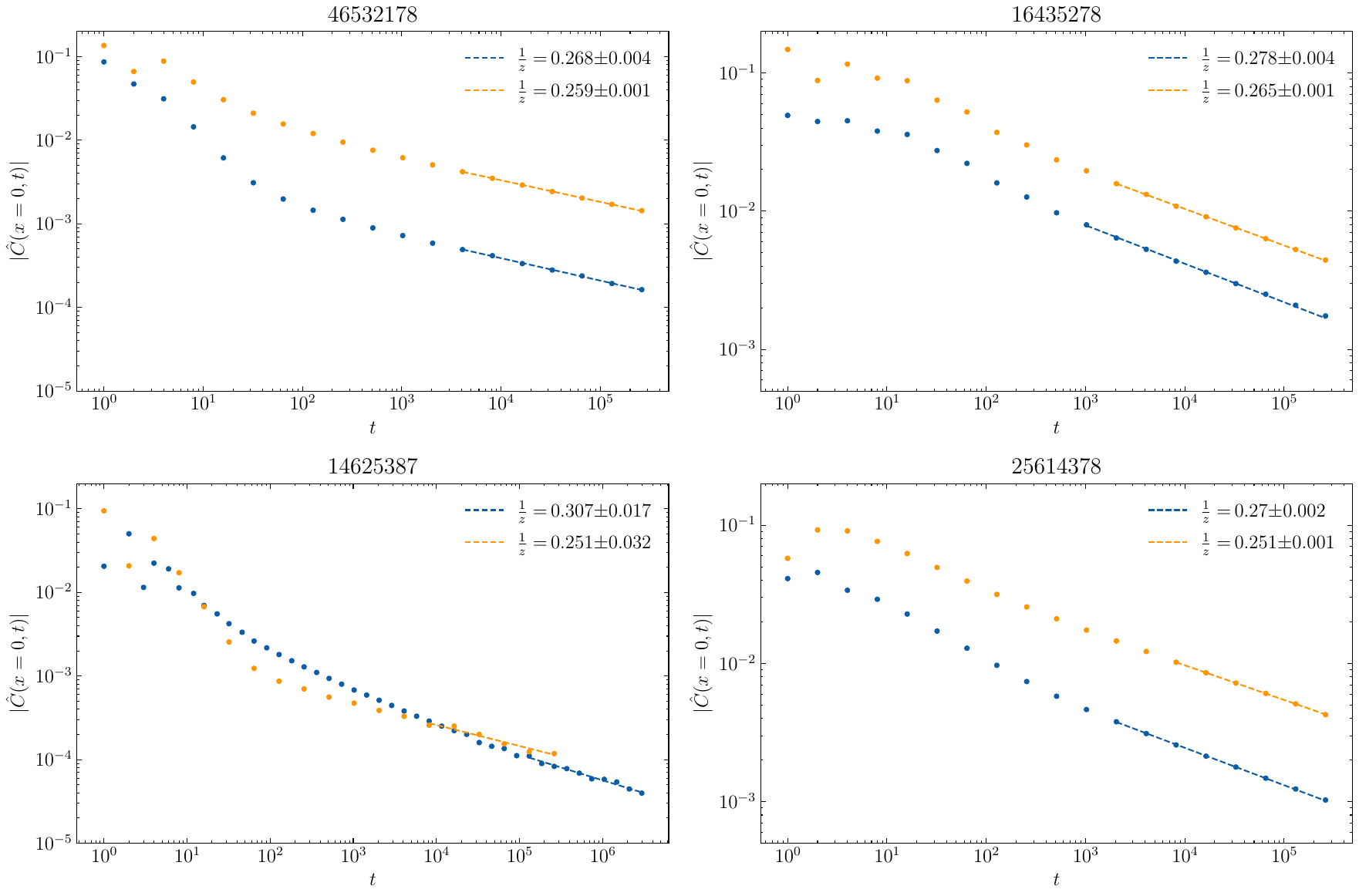}
}
  \caption{Correlators $\hat{C}(x,t)$ and $\hat{D}(x,t)$ computed for longer timescales for subdiffusive models. The blue and orange points correspond to $\hat{C}(x,t)$ and $\hat{D}(x,t)$. respectively, while the dashed lines denote power-law fits of the form $t^{-1/z}$.}
  \label{fig:subdiff_transport}
\end{figure}

We computed the correlators and examined the convergence of the dynamical exponents as a function of time. The results are presented in Figure~\ref{fig:subdiff_transport}. It appears that almost all dynamical exponents converge rather slowly to $z=4$. The exception are the cellular automata with rules 14625387 and 45321678, for which the dynamical exponents of the $\langle [\varnothing]_{xt} [\varnothing]_{00}\rangle$ correlators did not converge even at times $2^{21}$. Nevertheless, their values appear to be approaching $z=4$ as well. When available, this long-time data is used in the classification table for $z$ instead of the data extracted from the shorter-time correlators computed for other models.

\newpage

\bibliography{ref}

\begin{thebibliography}{10}
\providecommand{\url}[1]{\texttt{#1}}
\providecommand{\urlprefix}{URL }
\expandafter\ifx\csname urlstyle\endcsname\relax
  \providecommand{\doi}[1]{doi:\discretionary{}{}{}#1}\else
  \providecommand{\doi}{doi:\discretionary{}{}{}\begingroup \urlstyle{rm}\Url}\fi
\providecommand{\eprint}[2][]{\url{#2}}

\bibitem{FPUT}
E.~Fermi, J.~Pasta, S.~Ulam and M.~Tsingou,
\newblock \emph{Los alamos scientific laboratory report la-1940},
\newblock Collected Papers of Enrico Fermi \textbf{2} (1955).

\bibitem{bermanFPUT}
G.~Berman and F.~Izrailev,
\newblock \emph{The fermi--pasta--ulam problem: fifty years of progress},
\newblock Chaos: An Interdisciplinary Journal of Nonlinear Science \textbf{15}(1) (2005).

\bibitem{gallavottiFPUT}
G.~Gallavotti,
\newblock \emph{The Fermi-Pasta-Ulam problem: a status report}, vol. 728,
\newblock Springer Science \& Business Media (2007).

\bibitem{FlachPRL2019}
T.~Mithun, C.~Danieli, Y.~Kati and S.~Flach,
\newblock \emph{Dynamical glass and ergodization times in classical josephson junction chains},
\newblock Physical Review Letters \textbf{122}(5) (2019),
\newblock \doi{10.1103/physrevlett.122.054102}.

\bibitem{FlachPRL2022}
M.~Malishava and S.~Flach,
\newblock \emph{Lyapunov spectrum scaling for classical many-body dynamics close to integrability},
\newblock Phys. Rev. Lett. \textbf{128}, 134102 (2022).

\bibitem{tHooft}
G.~'t~Hooft,
\newblock \emph{{The Cellular Automaton Interpretation of Quantum Mechanics}},
\newblock Springer,
\newblock \doi{10.1007/978-3-319-41285-6} (2016).

\bibitem{wolfram1983statistical}
S.~Wolfram,
\newblock \emph{Statistical mechanics of cellular automata},
\newblock Reviews of modern physics \textbf{55}(3), 601 (1983).

\bibitem{li1990structure}
W.~Li, N.~Packard \emph{et~al.},
\newblock \emph{The structure of the elementary cellular automata rule space},
\newblock Complex systems \textbf{4}(3), 281 (1990).

\bibitem{cook2004universality}
M.~Cook \emph{et~al.},
\newblock \emph{Universality in elementary cellular automata},
\newblock Complex systems \textbf{15}(1), 1 (2004).

\bibitem{Kovtun:2012rj}
P.~Kovtun,
\newblock \emph{{Lectures on hydrodynamic fluctuations in relativistic theories}},
\newblock J. Phys. A \textbf{45}, 473001 (2012),
\newblock \doi{10.1088/1751-8113/45/47/473001},
\newblock \eprint{1205.5040}.

\bibitem{Liu:2018kfw}
H.~Liu and P.~Glorioso,
\newblock \emph{{Lectures on non-equilibrium effective field theories and fluctuating hydrodynamics}},
\newblock PoS \textbf{TASI2017}, 008 (2018),
\newblock \doi{10.22323/1.305.0008},
\newblock \eprint{1805.09331}.

\bibitem{Grozdanov:2019kge}
S.~Grozdanov, P.~K. Kovtun, A.~O. Starinets and P.~Tadi\'c,
\newblock \emph{{Convergence of the Gradient Expansion in Hydrodynamics}},
\newblock Phys. Rev. Lett. \textbf{122}(25), 251601 (2019),
\newblock \doi{10.1103/PhysRevLett.122.251601},
\newblock \eprint{1904.01018}.

\bibitem{takesue}
S.~Takesue,
\newblock \emph{Reversible cellular automata and statistical mechanics},
\newblock Phys. Rev. Lett. \textbf{59}, 2499 (1987),
\newblock \doi{10.1103/PhysRevLett.59.2499}.

\bibitem{bobenko}
A.~Bobenko, M.~Bordemann, C.~Gunn and U.~Pinkall,
\newblock \emph{On two integrable cellular automata},
\newblock Communications in mathematical physics \textbf{158}(1), 127 (1993).

\bibitem{RCA54review}
B.~Bu{\v{c}}a, K.~Klobas and T.~Prosen,
\newblock \emph{Rule 54: Exactly solvable model of nonequilibrium statistical mechanics},
\newblock Journal of Statistical Mechanics: Theory and Experiment \textbf{2021}(7), 074001 (2021).

\bibitem{Klobas_2022}
K.~Klobas and T.~Prosen,
\newblock \emph{On two reversible cellular automata with two particle species},
\newblock Journal of Physics A: Mathematical and Theoretical \textbf{55}(9), 094003 (2022),
\newblock \doi{10.1088/1751-8121/ac3ebc}.

\bibitem{Gambor}
T.~Gombor and B.~Pozsgay,
\newblock \emph{{Superintegrable cellular automata and dual unitary gates from Yang-Baxter maps}},
\newblock SciPost Phys. \textbf{12}, 102 (2022),
\newblock \doi{10.21468/SciPostPhys.12.3.102}.

\bibitem{East}
B.~Bertini, C.~De~Fazio, J.~P. Garrahan and K.~Klobas,
\newblock \emph{Exact quench dynamics of the floquet quantum east model at the deterministic point},
\newblock Phys. Rev. Lett. \textbf{132}, 120402 (2024),
\newblock \doi{10.1103/PhysRevLett.132.120402}.

\bibitem{kim2025circuitssimpleplatformemergence}
S.~W.~P. Kim, F.~Hübner, J.~P. Garrahan and B.~Doyon,
\newblock \emph{Circuits as a simple platform for the emergence of hydrodynamics in deterministic chaotic many-body systems},
\newblock arXiv preprint arXiv:2503.08788  (2025),
\newblock \doi{10.48550/arXiv.2503.08788}.

\bibitem{ziga_rule}
Žiga Krajnik, J.~Schmidt, V.~Pasquier, T.~Prosen and E.~Ilievski,
\newblock \emph{Universal anomalous fluctuations in charged single-file systems},
\newblock Phys. Rev. Res. \textbf{6}, 013260 (2024),
\newblock \doi{10.1103/PhysRevResearch.6.013260}.

\bibitem{ziga_rule2}
M.~Medenjak, K.~Klobas and T.~Prosen,
\newblock \emph{Diffusion in deterministic interacting lattice systems},
\newblock Phys. Rev. Lett. \textbf{119}, 110603 (2017),
\newblock \doi{10.1103/PhysRevLett.119.110603}.

\bibitem{ziga_rule3}
K.~Klobas, M.~Medenjak and T.~Prosen,
\newblock \emph{Exactly solvable deterministic lattice model of crossover between ballistic and diffusive transport},
\newblock Journal of Statistical Mechanics: Theory and Experiment \textbf{2018}(12), 123202 (2018),
\newblock \doi{10.1088/1742-5468/aae853}.

\bibitem{ziga_rule4}
M.~Medenjak, V.~Popkov, T.~Prosen, E.~Ragoucy and M.~Vanicat,
\newblock \emph{{Two-species hardcore reversible cellular automaton: matrix ansatz for dynamics and nonequilibrium stationary state}},
\newblock SciPost Phys. \textbf{6}, 074 (2019),
\newblock \doi{10.21468/SciPostPhys.6.6.074}.

\bibitem{moudgalya2022quantum}
S.~Moudgalya, B.~A. Bernevig and N.~Regnault,
\newblock \emph{Quantum many-body scars and hilbert space fragmentation: a review of exact results},
\newblock Reports on Progress in Physics \textbf{85}(8), 086501 (2022).

\bibitem{sala2020ergodicity}
P.~Sala, T.~Rakovszky, R.~Verresen, M.~Knap and F.~Pollmann,
\newblock \emph{Ergodicity breaking arising from hilbert space fragmentation in dipole-conserving hamiltonians},
\newblock Physical Review X \textbf{10}(1), 011047 (2020).

\bibitem{khemaniPRB}
V.~Khemani, M.~Hermele and R.~Nandkishore,
\newblock \emph{Localization from hilbert space shattering: From theory to physical realizations},
\newblock Phys. Rev. B \textbf{101}, 174204 (2020),
\newblock \doi{10.1103/PhysRevB.101.174204}.

\bibitem{zadnik2021folded}
L.~Zadnik and M.~Fagotti,
\newblock \emph{The folded spin-1/2 xxz model: I. diagonalisation, jamming, and ground state properties},
\newblock SciPost Physics Core \textbf{4}(2), 010 (2021).

\bibitem{ilievski2016quasilocal}
E.~Ilievski, M.~Medenjak, T.~Prosen and L.~Zadnik,
\newblock \emph{Quasilocal charges in integrable lattice systems},
\newblock Journal of Statistical Mechanics: Theory and Experiment \textbf{2016}(6), 064008 (2016).

\bibitem{Prosen_2004}
T.~Prosen,
\newblock \emph{Ruelle resonances in kicked quantum spin chain},
\newblock Physica D: Nonlinear Phenomena \textbf{187}(1-4), 244 (2004),
\newblock \doi{10.1016/j.physd.2003.09.017}.

\bibitem{Prosen_2007}
T.~Prosen,
\newblock \emph{Chaos and complexity of quantum motion},
\newblock Journal of Physics A: Mathematical and Theoretical \textbf{40}(28), 7881 (2007),
\newblock \doi{10.1088/1751-8113/40/28/S02}.

\bibitem{dynamical_lenart}
M.~Medenjak, T.~Prosen and L.~Zadnik,
\newblock \emph{{Rigorous bounds on dynamical response functions and time-translation symmetry breaking}},
\newblock SciPost Phys. \textbf{9}, 003 (2020),
\newblock \doi{10.21468/SciPostPhys.9.1.003}.

\bibitem{mendl}
C.~B. Mendl and H.~Spohn,
\newblock \emph{Dynamic correlators of fermi-pasta-ulam chains and nonlinear fluctuating hydrodynamics},
\newblock Phys. Rev. Lett. \textbf{111}, 230601 (2013),
\newblock \doi{10.1103/PhysRevLett.111.230601}.

\bibitem{spohn2014}
H.~Spohn,
\newblock \emph{Nonlinear fluctuating hydrodynamics for anharmonic chains},
\newblock Journal of Statistical Physics \textbf{154}(5), 1191–1227 (2014),
\newblock \doi{10.1007/s10955-014-0933-y}.

\bibitem{krajnik2020kardar}
{\v{Z}}.~Krajnik and T.~Prosen,
\newblock \emph{Kardar--parisi--zhang physics in integrable rotationally symmetric dynamics on discrete space--time lattice},
\newblock Journal of Statistical Physics \textbf{179}(1), 110 (2020).

\bibitem{ilievski2021superuniversality}
E.~Ilievski, J.~De~Nardis, S.~Gopalakrishnan, R.~Vasseur and B.~Ware,
\newblock \emph{Superuniversality of superdiffusion},
\newblock Physical Review X \textbf{11}(3), 031023 (2021).

\bibitem{krajnik2024dynamical}
{\v{Z}}.~Krajnik, J.~Schmidt, E.~Ilievski and T.~Prosen,
\newblock \emph{Dynamical criticality of magnetization transfer in integrable spin chains},
\newblock Physical review letters \textbf{132}(1), 017101 (2024).

\bibitem{takeuchi2024partial}
K.~A. Takeuchi, K.~Takasan, O.~Busani, P.~L. Ferrari, R.~Vasseur and J.~De~Nardis,
\newblock \emph{Partial yet definite emergence of the kardar-parisi-zhang class in isotropic spin chains},
\newblock arXiv preprint arXiv:2406.07150  (2024).

\bibitem{ljubotina2017spin}
M.~Ljubotina, M.~{\v{Z}}nidari{\v{c}} and T.~Prosen,
\newblock \emph{Spin diffusion from an inhomogeneous quench in an integrable system},
\newblock Nature communications \textbf{8}(1), 16117 (2017).

\bibitem{medenjak}
E.~Ilievski, J.~De~Nardis, M.~Medenjak and T.~Prosen,
\newblock \emph{Superdiffusion in one-dimensional quantum lattice models},
\newblock Phys. Rev. Lett. \textbf{121}, 230602 (2018),
\newblock \doi{10.1103/PhysRevLett.121.230602}.

\bibitem{MBL}
D.~A. Abanin, E.~Altman, I.~Bloch and M.~Serbyn,
\newblock \emph{Colloquium: Many-body localization, thermalization, and entanglement},
\newblock Rev. Mod. Phys. \textbf{91}, 021001 (2019),
\newblock \doi{10.1103/RevModPhys.91.021001}.

\bibitem{de2024absence}
W.~De~Roeck, L.~Giacomin, F.~Huveneers and O.~Prosniak,
\newblock \emph{Absence of normal heat conduction in strongly disordered interacting quantum chains},
\newblock arXiv preprint arXiv:2408.04338  (2024).

\bibitem{Brighi2023}
P.~Brighi, M.~Ljubotina and M.~Serbyn,
\newblock \emph{{Hilbert space fragmentation and slow dynamics in particle-conserving quantum East models}},
\newblock SciPost Phys. \textbf{15}, 093 (2023),
\newblock \doi{10.21468/SciPostPhys.15.3.093}.

\bibitem{Vasseur2021}
H.~Singh, B.~A. Ware, R.~Vasseur and A.~J. Friedman,
\newblock \emph{Subdiffusion and many-body quantum chaos with kinetic constraints},
\newblock Phys. Rev. Lett. \textbf{127}, 230602 (2021),
\newblock \doi{10.1103/PhysRevLett.127.230602}.

\bibitem{Rannou}
F.~Rannou,
\newblock \emph{Numerical study of discrete plane area-preserving mappings},
\newblock Astronomy and Astrophysics \textbf{31}, 289 (1974).

\bibitem{TomazProsen_2002}
T.~Prosen,
\newblock \emph{Ruelle resonances in quantum many-body dynamics},
\newblock Journal of Physics A: Mathematical and General \textbf{35}(48), L737 (2002),
\newblock \doi{10.1088/0305-4470/35/48/102}.

\bibitem{Marko_PR}
M.~\ifmmode \check{Z}\else \v{Z}\fi{}nidari\ifmmode~\check{c}\else \v{c}\fi{},
\newblock \emph{Momentum-dependent quantum ruelle-pollicott resonances in translationally invariant many-body systems},
\newblock Phys. Rev. E \textbf{110}, 054204 (2024),
\newblock \doi{10.1103/PhysRevE.110.054204}.

\bibitem{zhang2024thermalizationratesquantumruellepollicott}
C.~Zhang, L.~Nie and C.~von Keyserlingk,
\newblock \emph{Thermalization rates and quantum ruelle-pollicott resonances: insights from operator hydrodynamics} (2024), \eprint{2409.17251}.

\bibitem{jacoby2024spectralgapslocalquantum}
J.~A. Jacoby, D.~A. Huse and S.~Gopalakrishnan,
\newblock \emph{Spectral gaps of local quantum channels in the weak-dissipation limit} (2024), \eprint{2409.17238}.

\bibitem{TakashiMori}
T.~Mori,
\newblock \emph{Liouvillian-gap analysis of open quantum many-body systems in the weak dissipation limit},
\newblock Phys. Rev. B \textbf{109}, 064311 (2024),
\newblock \doi{10.1103/PhysRevB.109.064311}.

\bibitem{TomazEnej}
T.~Prosen and E.~Ilievski,
\newblock \emph{Families of quasilocal conservation laws and quantum spin transport},
\newblock Phys. Rev. Lett. \textbf{111}, 057203 (2013),
\newblock \doi{10.1103/PhysRevLett.111.057203}.

\bibitem{Enej2015}
E.~Ilievski, J.~De~Nardis, B.~Wouters, J.-S. Caux, F.~H.~L. Essler and T.~Prosen,
\newblock \emph{Complete generalized gibbs ensembles in an interacting theory},
\newblock Phys. Rev. Lett. \textbf{115}, 157201 (2015),
\newblock \doi{10.1103/PhysRevLett.115.157201}.

\bibitem{prahofer2004exact}
M.~Pr{\"a}hofer and H.~Spohn,
\newblock \emph{Exact scaling functions for one-dimensional stationary kpz growth},
\newblock Journal of statistical physics \textbf{115}(1), 255 (2004).

\bibitem{PollmannSub}
J.~Feldmeier, P.~Sala, G.~De~Tomasi, F.~Pollmann and M.~Knap,
\newblock \emph{Anomalous diffusion in dipole- and higher-moment-conserving systems},
\newblock Phys. Rev. Lett. \textbf{125}, 245303 (2020),
\newblock \doi{10.1103/PhysRevLett.125.245303}.

\bibitem{SubStrong}
P.~Zhang,
\newblock \emph{Subdiffusion in strongly tilted lattice systems},
\newblock Phys. Rev. Res. \textbf{2}, 033129 (2020),
\newblock \doi{10.1103/PhysRevResearch.2.033129}.

\bibitem{Nandy2024}
S.~Nandy, J.~Herbrych, Z.~Lenarčič, A.~Głódkowski, P.~Prelovšek and M.~Mierzejewski,
\newblock \emph{Emergent dipole moment conservation and subdiffusion in tilted chains},
\newblock Physical Review B \textbf{109}(11) (2024),
\newblock \doi{10.1103/physrevb.109.115120}.

\bibitem{DeRoeck2020}
W.~De~Roeck, F.~Huveneers and S.~Olla,
\newblock \emph{Subdiffusion in one-dimensional hamiltonian chains with sparse interactions},
\newblock Journal of Statistical Physics \textbf{180}(1–6), 678–698 (2020),
\newblock \doi{10.1007/s10955-020-02496-1}.

\bibitem{Singh2021}
H.~Singh, B.~A. Ware, R.~Vasseur and A.~J. Friedman,
\newblock \emph{Subdiffusion and many-body quantum chaos with kinetic constraints},
\newblock Physical Review Letters \textbf{127}(23) (2021),
\newblock \doi{10.1103/physrevlett.127.230602}.

\bibitem{DeNardis2022}
J.~De~Nardis, S.~Gopalakrishnan, R.~Vasseur and B.~Ware,
\newblock \emph{Subdiffusive hydrodynamics of nearly integrable anisotropic spin chains},
\newblock Proceedings of the National Academy of Sciences \textbf{119}(34) (2022),
\newblock \doi{10.1073/pnas.2202823119}.

\bibitem{Borsi}
M.~Borsi and B.~Pozsgay,
\newblock \emph{Construction and the ergodicity properties of dual unitary quantum circuits},
\newblock Phys. Rev. B \textbf{106}, 014302 (2022),
\newblock \doi{10.1103/PhysRevB.106.014302}.

\bibitem{masanes}
T.~Holden-Dye, L.~Masanes and A.~Pal,
\newblock \emph{Fundamental charges for dual-unitary circuits},
\newblock Quantum \textbf{9}, 1615 (2025),
\newblock \doi{10.22331/q-2025-01-30-1615}.

\bibitem{gopalakrishnan2018operator}
S.~Gopalakrishnan,
\newblock \emph{Operator growth and eigenstate entanglement in an interacting integrable floquet system},
\newblock Physical Review B \textbf{98}(6), 060302 (2018).

\bibitem{Can2019}
T.~Can,
\newblock \emph{Random lindblad dynamics},
\newblock Journal of Physics A: Mathematical and Theoretical \textbf{52}(48), 485302 (2019),
\newblock \doi{10.1088/1751-8121/ab4d26}.

\bibitem{Li2021}
J.~Li, T.~Prosen and A.~Chan,
\newblock \emph{Spectral statistics of non-hermitian matrices and dissipative quantum chaos},
\newblock Physical Review Letters \textbf{127}(17) (2021),
\newblock \doi{10.1103/physrevlett.127.170602}.

\bibitem{Shivam2023}
S.~Shivam, A.~De~Luca, D.~A. Huse and A.~Chan,
\newblock \emph{Many-body quantum chaos and emergence of ginibre ensemble},
\newblock Physical Review Letters \textbf{130}(14) (2023),
\newblock \doi{10.1103/physrevlett.130.140403}.

\bibitem{Yoshimura2024}
T.~Yoshimura and L.~Sá,
\newblock \emph{Robustness of quantum chaos and anomalous relaxation in open quantum circuits},
\newblock Nature Communications \textbf{15}(1) (2024),
\newblock \doi{10.1038/s41467-024-54164-7}.

\bibitem{Lucas_PR}
T.~Yoshimura and L.~S\'a,
\newblock \emph{Theory of irreversibility in quantum many-body systems},
\newblock Phys. Rev. E \textbf{111}, 064135 (2025),
\newblock \doi{10.1103/82f6-qdyd}.

\bibitem{Larkin1969QuasiclassicalMI}
A.~I. Larkin and Y.~N. Ovchinnikov,
\newblock \emph{Quasiclassical method in the theory of superconductivity},
\newblock Journal of Experimental and Theoretical Physics  (1969).

\bibitem{SFF_semiclass}
S.~M\"uller, S.~Heusler, P.~Braun, F.~Haake and A.~Altland,
\newblock \emph{Semiclassical foundation of universality in quantum chaos},
\newblock Phys. Rev. Lett. \textbf{93}, 014103 (2004),
\newblock \doi{10.1103/PhysRevLett.93.014103}.

\bibitem{Gorin2006}
T.~Gorin, T.~Prosen, T.~H. Seligman and M.~Žnidarič,
\newblock \emph{Dynamics of loschmidt echoes and fidelity decay},
\newblock Physics Reports \textbf{435}(2–5), 33–156 (2006),
\newblock \doi{10.1016/j.physrep.2006.09.003}.

\bibitem{Krylov}
D.~E. Parker, X.~Cao, A.~Avdoshkin, T.~Scaffidi and E.~Altman,
\newblock \emph{A universal operator growth hypothesis},
\newblock Phys. Rev. X \textbf{9}, 041017 (2019),
\newblock \doi{10.1103/PhysRevX.9.041017}.

\bibitem{GeometryRustem}
R.~Sharipov, A.~Tiutiakina, A.~Gorsky, V.~Gritsev and A.~Polkovnikov,
\newblock \emph{Hilbert space geometry and quantum chaos},
\newblock arXiv preprint arXiv:2411.11968  (2024),
\newblock \doi{10.48550/arxiv.2411.11968}.

\bibitem{Grozdanov:2017ajz}
S.~Grozdanov, K.~Schalm and V.~Scopelliti,
\newblock \emph{{Black hole scrambling from hydrodynamics}},
\newblock Phys. Rev. Lett. \textbf{120}(23), 231601 (2018),
\newblock \doi{10.1103/PhysRevLett.120.231601},
\newblock \eprint{1710.00921}.

\bibitem{Blake:2017ris}
M.~Blake, H.~Lee and H.~Liu,
\newblock \emph{{A quantum hydrodynamical description for scrambling and many-body chaos}},
\newblock JHEP \textbf{10}, 127 (2018),
\newblock \doi{10.1007/JHEP10(2018)127},
\newblock \eprint{1801.00010}.

\bibitem{Urban}
M.~\ifmmode \check{Z}\else \v{Z}\fi{}nidari\ifmmode~\check{c}\else \v{c}\fi{}, U.~Duh and L.~Zadnik,
\newblock \emph{Integrability is generic in homogeneous u(1)-invariant nearest-neighbor qubit circuits},
\newblock Phys. Rev. B \textbf{112}, L020302 (2025),
\newblock \doi{10.1103/tqy8-ynpd}.

\bibitem{Znidaric2024}
M.~Žnidarič,
\newblock \emph{Momentum-dependent quantum ruelle-pollicott resonances in translationally invariant many-body systems},
\newblock Physical Review E \textbf{110}(5) (2024),
\newblock \doi{10.1103/physreve.110.054204}.

\end{thebibliography}

\end{document}